\documentclass[12pt]{elsarticle}
\pdfoutput=1
\usepackage[ruled]{algorithm2e}
\renewcommand{\algorithmcfname}{ALGORITHM}
\SetAlFnt{\small}
\SetAlCapFnt{\small}
\SetAlCapNameFnt{\small}
\SetAlCapHSkip{0pt}
\IncMargin{-\parindent}

\usepackage{amsmath}
\usepackage[T1]{fontenc}
\usepackage[utf8]{inputenc}
\usepackage{amsmath}
\usepackage{graphicx}
\usepackage{amssymb} 
\usepackage{esint}
\usepackage{graphicx}

\usepackage{mathrsfs}

\journal{ar\!Xiv.org for the record.
	}

\begin{document}

\newtheorem{theorem}{Theorem}[section]
\newtheorem{lem}[theorem]{Lemma}
\newtheorem{cor}[theorem]{Corollary}
\newtheorem{mth}[theorem]{The Main Theorem}
\newtheorem{mapth}[theorem]{The Mapping Theorem}
\newtheorem{mlm}[theorem]{The Main Lemma}
\newtheorem{remlm}[theorem]{The ${\tt RemoveMax} $ Invertibility Lemma}
\newtheorem{racrlm}[theorem]{The ${\tt RemoveAll} $ Cost Lemma}
\newtheorem{rmcrlm}[theorem]{The ${\tt RemoveMax} $ Cost Lemma}
\newtheorem{losslm}[theorem]{Credit Loss Characterization Lemma}
\newtheorem{pqlm}[theorem]{The pq Lemma}
\newtheorem{fixlm}[theorem]{Subheap Repair Lemma}
\newtheorem{unlm}[theorem]{The Uniqueness Lemma}
\newtheorem{complem}[theorem]{Competition Lemma}
\newtheorem{diaglem}[theorem]{Diagram Lemma}
\newtheorem{fundlem}[theorem]{Fundamental Lemma}
\newtheorem{charlem}[theorem]{Worst-case Heap Characterization Lemma}
\newtheorem{sLBlm}[theorem]{$ \sum \lambda $ Lower Bound Lemma}
\newtheorem{dth}[theorem]{Decomposition Theorem}
\newtheorem{sith}[theorem]{Singularity Theorem}
\newtheorem{1LBth}[theorem]{The 1$ ^{\mbox{st}} $ Lower Bound Theorem}
\newtheorem{2LBth}[theorem]{The Lower Bound Theorem}
\newtheorem{UBth}[theorem]{The Upper Bound Theorem}
\newtheorem{1oth}[theorem]{The 1$ ^{st} $ Optimality Theorem}
\newtheorem{2oth}[theorem]{The 2$ ^{nd} $ Optimality Theorem}
\newtheorem{3oth}[theorem]{The 3$ ^{rd} $ Optimality Theorem}
\newtheorem{cropt}[theorem]{Optimality Criterion}
\newtheorem{hyp}[theorem]{Hypothesis}
\newtheorem{example}[theorem]{Example}
\newtheorem{theorem1}{Theorem}[subsection]
\newtheorem{lemma1}[theorem1]{Lemma}
\newtheorem{claim1}[theorem1]{Claim}
\newtheorem{corollary1}[theorem1]{Corollary}
\newtheorem{proposition1}[theorem1]{Proposition}
\newtheorem{problem1}[theorem1]{Problem}
\newtheorem{example1}{Example}[subsection]
\newtheorem{conjecture1}[theorem1]{Conjecture}
\newtheorem{remark1}[theorem1]{Remark}

\pagestyle{myheadings}
\markboth{M.A.Suchenek}{M. A. Suchenek:  A Complete Worst-Case Analysis of Heapsort {\tt (MS)}}

\begin{frontmatter}

\title{A Complete Worst-Case Analysis of Heapsort with Experimental Verification of Its Results \\ {\tt \small A manuscript (MS) intended for future journal publication}\tnoteref{label1} }
\tnotetext[label1]{\copyright 2015 Marek A. Suchenek.}
\author{MAREK A. SUCHENEK\fnref{label2}}
\fntext[label2]{I would like to thank Dr. Mohsen Beheshti, Chair of the Department of Computer Science, CSUDH, and  Dr. Neil Siegel,
                Vice President \& Chief Engineer, Technology \& Engineering,                Northrop Grumman Information Systems,
                Northrop Grumman Corporation, for the support I received while working on this paper.
                I am greatly indebted to my long-time mentor, Dr. Victor W. Marek, Professor of Computer Science at University of Kentucky, Lexington, for 37 years of his guidance and help.}


\address{California State University Dominguez Hills,
Department of Computer Science,
1000 E. Victoria St., Carson, CA 90747, USA,
  \texttt{Suchenek@csudh.edu}}

\begin{abstract}



\underline{Main results}.
A rigorous proof that  the number 
of comparisons of keys performed in the worst case by ${\tt Heapsort}$ on any array of size $N \geq 2$ is equal to:
\[
2 (N-1)\, ( \,  \lg \frac{N-1}{2}  +\varepsilon  \, ) - 2s_2(N) - e_2(N) + \min (\lfloor \lg (N-1) \rfloor, 2) + 6 + c, \]
where $ \varepsilon $, given by:
\[\varepsilon = 1 + \lceil \lg \, (N-1) \rceil -  \lg \, (N-1)
- 2^{\lceil \lg \, (N-1) \rceil -  \lg \, (N-1)} ,\]
is a function of $ N $ 
with the minimum value 0 and and the {supremum} value
\[\delta = 1 - \lg e + \lg \lg e \approx 0.0860713320559342 ,\]
$s_2(N)$ is the sum of all digits of the binary representation of $N$, $e_2(N)$ is the exponent of $2$ in the prime factorization of $N$, and $ c $ is a binary function on the set of integers defined by:
\begin{equation} \nonumber
c = 
\left\{ \begin{array}{ll}
1 \mbox{ if } \;  N \leq 2 ^{\lceil \lg N \rceil} - 4   \\ \\
0 \mbox{ otherwise,}
\end{array} \right.
\end{equation}
is presented.

\smallskip

An algorithm that generates worst-case input arrays of any size $ N \geq 2 $ for  ${\tt Heapsort}$ is offered. The algorithm has been implemented in Java,  runs in $O( N \log N )$ time, and  allows for precise experimental verification of the above formula.

\underline{\sf Significance}. The worst-case behavior of  ${\tt Heapsort}$ has escaped mathematically
 precise characterization for almost five decades now. This paper fills that important gap. The exactness of the derived number of comparisons of keys performed by ${\tt Heapsort}$ in the worst case, as opposed to merely big-oh or $ \sim $ asymptotic
 approximation or bound, allows for direct and definite experimental verification of its correctness.

\end{abstract}


\end{frontmatter}

\newpage

\tableofcontents

\section*{Introduction}

\begin{quote}
\textit{Many of those who could not
	figure it out exactly were 
	quick to dismiss the importance of precisely knowing it. But there are some significant advantages of knowing the exact value as opposed to its approximation. For one, it can be 
	conclusively
	verified by means of a direct experiment.}
\end{quote}

Some researchers tend to believe that undergraduate Computer Science is not an area for intellectually challenging and interesting problems. Some even go as far as to dismiss pursuit of their solutions as \textit{pedagogy}. But such a prejudice appears, well, prejudice, as there do exist questions that belong to undergraduate CS, yet they have been declared difficult to  answer even by some renowned scholars. Although the deceitful simplicity of some of the solutions of problems that were once considered hard might prompt a skeptic to entrench himself in his dismissiveness, it
should not puzzle those sympathetic to the $ P \neq N P  $ conjecture, one of the consequences of which stipulates the existence of hard to solve problems whose solutions are straightforward to verify as soon as their witnesses have been found.

\medskip

Take, for instance, $ {\tt Heapsort} $ invented by Williams \cite{wil:heap} and enhanced by Floyd \cite{flo:heap}. This specimen of elegance and simplicity, and a classic that has been taught across curricula of virtually every ABET-accredited Computer Science program, does belong to to undergraduate CS,
yet it apparently has resisted attempts of some seasoned researchers to accurately characterize its worst-case performance.
  In this article, I will use rather elementary mathematics to bring the worst-case analysis of $ {\tt Heapsort} $ to the point that one could consider 
complete.

\medskip

The analysis I present here is not particularly short\footnote{The 
rigorous proof of the main result is
 rather lengthy, particularly when compared to the short and elegant algebraic derivation in \cite{suc:elem} of a similar result for the Floyd's heap-construction program that constitutes the first (and faster) half of the $ {\tt Heapsort} $.}. 
This does not come as a total surprise, taking into account almost five decades that passed without its completion. 
Although Shaffer and Sedgewick declared long time ago that their paper \cite{shased:heap} ``essentially complete[d] the analysis of [$ {\tt Heapsort} $],'' they also admitted that ``there [was] another quantity that contribute[d] to the leading term of the running time that require[d] more intricate arguments,'' for which they had ``little specific information about the distribution beyond what [was] implied by [their] asymptotic results.'' Clearly, they did not attempt to derive the exact formula for the number of comparisons of keys that $ {\tt Heapsort} $ performs, which remained unknown for 49 years.

\medskip

So, here it is: for every natural number $N \geq 2$, the number 
of comparisons of keys performed in the worst case by the ${\tt Heapsort}$ on any array of size $N$ is equal to:
\[
2 (N-1)\, ( \,  \lg \frac{N-1}{2}  +\varepsilon  \, ) - 2s_2(N) - e_2(N) + \min (\lfloor \lg (N-1) \rfloor, 2) + 6 + c, \]
where $ \varepsilon $, given by:
\[\varepsilon = 1 + \theta
- 2^{\theta} \mbox{ and } \theta =  \lceil \lg \, (N-1) \rceil -  \lg \, (N-1),\]
is a continuous function (visualized on Figure~\ref{fig:epsilon} page~\pageref{fig:epsilon}) of $ N $ 
on the set of reals $ >1 $, 
with the minimum value 0 and and the maximum (\textit{supremum}, if 
$ N $
is restricted to integers) value
\[\delta = 1 - \lg e + \lg \lg e \approx 0.0860713320559342 ,\]
$s_2(N)$ is the sum of all digits of the binary representation of $N$, $e_2(N)$ is the exponent of $2$ in the prime factorization of $N$, and $ c $ is a binary function\footnote{Algebraically, $ 1 - c = \lceil \lg (N+4)  \rceil - \lceil \lg N  \rceil $.}
 (visualized on Figure~\ref{fig:c3} page~\pageref{fig:c3}) on the set of integers defined by:
\begin{equation} \nonumber
c = 
\left\{ \begin{array}{ll}
1 \mbox{ if } \;  N \leq 2 ^{\lceil \lg N \rceil} - 4   \\ \\
0 \mbox{ otherwise}.
\end{array} \right.
\end{equation}
Moreover, if $ N \geq 5 $ then the above formula simplifies to:
\begin{equation} \nonumber
2 (N-1)\, ( \,  \lg \frac{N-1}{2}  +\varepsilon  \, ) - 2s_2(N) - e_2(N)  + 8 + c. 
\end{equation}
\medskip

The method I chose for my derivation the above formula could be characterized mostly as a \textit{brute force} approach, with some 
subtler 
inductive arguments without which the \textit{brute force} alone would not accomplish much. Nevertheless, the elementary algebra involved in it seems well-worth studying in its own right as it also applies to other problems that are related to sorting and to finite binary trees.


\section{An overview}  \label{Overview}

The main subject of my analysis is the number $  C_{\tt Heapsort} ^{\tt max}(N)  $ of comparisons of keys that Williams' \textit{vanilla} $ {\tt Heapsort} $ with Floyd's improvement performs in the worst case while sorting an $ N $-element array of distinct integers. It consists of three major parts. The goal of the first part is to show that the said worst case number of comparisons is the sum of the respective numbers $  C_{\tt MakeHeap} ^{\tt max}(N)  $ and $  C_{\tt RemoveAll()} ^{\tt max}(N)  $ for  the heap-construction phase $ {\tt MakeHeap} $ and the heap-deconstruction phase ${\tt RemoveAll}$. The goals of the second and the third part are to derive the formulas for  $  C_{\tt MakeHeap} ^{\tt max}(N)  $ and $  C_{\tt RemoveAll()} ^{\tt max}(N)  $. Those three, once completed, yield the formula for $  C_{\tt Heapsort} ^{\tt max}(N)  $.

\medskip

The first part turned up the easiest of the three. Running $ {\tt MakeHeap} $ backwards on any given heap resulted, and provably so, in an  array that $ {\tt MakeHeap} $ would turn onto $ H $ while performing the maximal possible number of comparisons for any array of that size.

\medskip

The second part was somewhat harder; however, it
has been recently nailed down with a simple closed-form formula for $  C_{\tt MakeHeap} ^{\tt max}(N)  $ that 
had a 
succinct and elegant algebraic proof\footnote{See \cite{suc:elem} for such a proof.}.

\medskip

The third part was considerably more complicated than the other two. 
A fairly simple strategy for generating bad cases
for ${\tt RemoveAll}$ by running it backwards led to a straightforward\footnote{At least for those fluent with the kind of math that I am using in this paper.}, if a bit tedious, derivation of a 
closed-form formula for a lower bound 
of $ C_{\tt RemoveAll()} ^{\tt max}(N)  $, but that lower bound was less than the ``easy'' upper bound\footnote{$(2N-1)\lfloor \lg (N-1) \rfloor
	- 2 ^{\lfloor \lg (N-1) \rfloor +2} + 4$.} of $ C_{\tt RemoveAll()} ^{\tt max}(N)  $ I knew.
It was the demonstration that the said strategy could not be beaten, which fact allowed me to conclude that the derived lower bound was also an upper bound and yielded a proof of the closed-form formula for $ C_{\tt RemoveAll()} ^{\tt max}(N)  $,  that was surprisingly\footnote{It did look at the beginning as a simple exercise, only to turn out elusive as it kept evading my attempts to be formulated precisely.} convoluted and resistant to simplifications.

\medskip

Well, there must have been a reason why, to my best knowledge, a journal-quality proof of the worst-case formula had not been published despite the fact that it had been done, even if in a somewhat rough form,  for the special case of inputs of the size $ N = 2^{\lceil \lg N \rceil} -1 $ some 36 years ago. After finishing the said proof I think I got a pretty good idea why all those who attempted it might have given up before bringing their efforts to a conclusive end. 
\medskip

Out of several factors that made such a proof not quite a routine exercise, a flaw in the structure of the worst-case heaps for ${\tt RemoveAll}$ deserves a special mention. It turns out that the number of 
hereditary\footnote{A heap $ H $ is a \textit{hereditary} worst-case heap if, and only if, it is a worst-case heap and 
	$ H.{\tt RemoveMax} ()  $ is either empty or is a hereditary worst-case heap.} worst-case heaps for ${\tt RemoveAll}$ is finite\footnote{The fact that not all worst-case heaps are hereditary worst-case heaps follows also from the fact that the number $ C_{\tt RemoveMax()} (H)  $ of comparisons of keys that the operation $ H.{\tt RemoveMax} ()  $ performs on a worst-case heap $ H $ on $ N $ nodes is \textit{not} a function of $ N $.} (1017 to be exact), which rules out the existence of any greedy scheme of generating worst-case heaps for ${\tt RemoveAll}$ of arbitrary size (greater than 22).
 As a result, the proof of the said formula hangs on a singularity of worst-case heaps which states that 
 if $N = 2^{\lceil \lg N \rceil} - 4$ and $ H $ is a heap on $ N+1 $ nodes such that
  the execution of	 $ H.{\tt RemoveMax()}$ on $ H $ performs the maximum $C^{max}_{{\tt RemoveMax()}} (N+1)$, over all heaps on $ N+1 $ nodes, number $C_{{\tt RemoveMax()}} (H)$  of comparisons of keys then the heap produced by the execution of	 $ H.{\tt RemoveMax()}$ is not \label{oddheap}  a worst-case heap\footnote{It is a singular property, indeed, as for every $ N \neq  2^{\lceil \lg N \rceil} - 4 $, there is a worst-case heap $ H $ on $ N+1 $ nodes with $C_{{\tt RemoveMax()}} (H) = C^{max}_{{\tt RemoveMax()}} (N+1)$ such that $ H.{\tt RemoveMax()}$ is a worst-case heap.}. Laying down foundations for a demonstration of the above fact was perhaps the most tedious task in this study. 
 
\medskip

Here is a road map of the paper.

\medskip

Section~\ref{HeapsHeapsort}, page~\pageref{HeapsHeapsort} and on, lays down basic definitions and algebraic facts related to the subject matter.

\medskip

 Section~\ref{Notation}, page~\pageref{Notation} and on, introduces notation that I use in this paper, some of which may differ from the notation used by other authors.
 
 \medskip
 
Section~\ref{sec:runback}, page~\pageref{sec:runback} and on, describes a basic technique for constructing worst cases for $ {\tt Heapsort} $ and its components by running them backwards. Methods $ {\tt PullDown} $, $ {\tt unFixHeap} $, and $ {\tt unRemoveMax} $ are introduced there.

\medskip

Section~\ref{Decomp}, page~\pageref{Decomp} and on, presents a constructive proof that the exact characterization of the worst-case performance of $ {\tt Heapsort} $ may be computed as a sum of exact characterizations of the worst-case performances of its two phases: the heap-construction phase (for which a closed-form formula has been recently discovered) and the heap-deconstruction phase. This is accomplished by demonstrating that for any heap of $ N $ nodes, method $ {\tt unFixHeap }$ - a reverse of  $ {\tt FixHeap }$ - constructs an $ N $-element array that constitutes a worst-case array for the heap-construction phase. 

\bigskip

At this point, the only piece of information that is needed for the exact
characterization of the worst-case performance of the entire $ {\tt Heapsort} $ is a formula   for the worst-case performance of the heap-deconstruction phase.

 \medskip
 
Section~\ref{sec:pull}, page~\pageref{sec:pull} and on, introduces a solitaire game of Pull Downs a payoff of which is equal to the number of comparisons of keys that are needed to deconstruct a heap produced by the game. The said payoff is maximal if, and only if, the produced heap is a worst-case heap. The Section establishes some $ 1-1 $ correspondences between heaps produced by the game and various generators of the game, which provides some notational convenience needed in the remainder of the paper and assures consistency of the derived results. The above facts reduce the problem of construction of a worst-case heap for the heap-deconstruction phase of $ {\tt Heapsort} $ to the problem of finding a generator of a suitable game that yields a maximal payoff. 
 \medskip
 
 Section~\ref{sec:cremov}, page~\pageref{sec:cremov} and on, contains definitions and technical details of computations of credits for moves in the game of Pull Downs. It provides characterization of moves that yield maximal credits and evaluates losses of credits for some sequences of moves. 
\medskip
 
Section~\ref{strapay}, page~\pageref{strapay} and on, is mostly notational. It introduces the concept of a strategy and defines various forms of payoffs, including the upper-bound payoff, and losses of credit related to it.

\medskip 
 
Section~\ref{StraParWin}, page~\pageref{StraParWin} and on, introduces some special strategies for the game of Pull Downs: a sub-optimal strategy $ \mathsf{par} $ and a family of
strategies $ \mathsf{win}(N) $ (one strategy for each $ N \geq 2 $), each being optimal for given $ N $. Strategy $ \mathsf{par} $ is sub-optimal in that it loses 1 credit per level, relative to the upper-bound payoff, of the heap constructed, from the level 3 on, and is
optimal for each complete\footnote{On $ N = 2^{\lceil \lg N \rceil} - 1 $ nodes, that is.} heap it constructs.
It also maintains certain invariant property of the heaps it produces that is needed for the demonstration of  optimality of its improvements. 
For each $N \leq 2 ^{\lceil \lg N \rceil} - 4 $, the strategy $ \mathsf{win}(N) $ improves upon $ \mathsf{par} $ (by the total of 1 credit) in that, unlike $ \mathsf{par} $, it is greedy in the level $ \lfloor \lg N \rfloor $ of the heap constructed, so that it postpones the 1 credit loss in that level until no further postponement is possible\footnote{Until the move $2 ^{\lceil \lg N \rceil} - 4 $ .}. 
The payoffs for those strategies are derived there. They establish a lower bound for the worst-case behavior of the heap-deconstruction phase.

 \medskip
 
Section~\ref{ProofWin}, page~\pageref{ProofWin} and on, 
is devoted to proofs of optimalities of strategies $ \mathsf{par} $ and $ \mathsf{win}(N) $. The optimality of $ \mathsf{win}(N) $ allows me to conclude that the lower bound  derived in Section~\ref{StraParWin} and given by the formula for payoffs for $ \mathsf{win}(N) $ is an upper bound, too, thus yielding the sought-after characterization of the worst-case behavior of the heap-deconstruction phase. The mostly case-driven proof uses, at some point, the fact that no strategy can gain relative to a strategy that is optimal for complete heaps (in particular, relative to $ \mathsf{par} $ and $ \mathsf{win}(N) $) more than 1 credit an any level of the heaps that it produces.
 It allows strategies $ \mathsf{win}(N) $ that are optimal for complete heaps and greedy in the levels $\lfloor \lg N \rfloor$ of heaps they construct to collect the maximal payoffs and, therefore, be optimal. It also reduces considerably the number of cases that need to be tackled in the said proof.
 
 \medskip
 
 Subsection~\ref{subsec:comment}, page~\pageref{subsec:comment} and on, attempts to explain why the optimality proof I present in this paper is more complicated than one could expect it to be.\footnote{The singularity of worst-case heaps of size $N = 2^{\lceil \lg (N-1) \rceil} - 3$, indicated on page  \pageref{oddheap},  seems to be the culprit here.}  It derives some intuitively simple facts\footnote{The most basic of which is the Singularity Theorem~\ref{cor:parwin}, page~\pageref{cor:parwin}, stating that no worst-case heap of size $N = 2^{\lceil \lg N \rceil} - 4$ admits a lossless pull down.} that entail the optimality of  $ \mathsf{win}(N) $, and demonstrates that they are about as difficult to prove as those in Section~\ref{ProofWin}.

 \bigskip
 
 The remainder of the paper easily follows from the above. 
 
 \medskip
 
Section~\ref{W-cRemAll}, page~\pageref{W-cRemAll} and on, proves a closed-form formula for the worst-case number of comparisons of keys performed by the heap-deconstruction phase of $ {\tt Heapsort} $.

 \medskip
 
Section~\ref{W-cHeap}, page~\pageref{W-cHeap} and on,  proves two closed-form formulas for the worst-case number of comparisons of keys performed by the  $ {\tt Heapsort} $, one with function floor and one (mostly) without it.

 \medskip
 
Section~\ref{sec:s2e2}, page~\pageref{sec:s2e2} and on, analyses the behavior of the ``jumpy'' term $ 2s_2(N) + e_2(N) $ in the mentioned above formulas and offers its tight upper bound expressed by a function that is continuous on the set of reals except for $ N = 2^{\lfloor \lg N \rfloor} + 1 $.

 \medskip
 
Section~\ref{Roots}, page~\pageref{Roots} and on, comments on the origins of this article, with Subection~\ref{CompK-W}, page~\pageref{CompK-W} and on, comparing the presented results to those published in an old report by Kruskal and Weixelbaum.

 \medskip
 
\ref{Examples}
offers illustrating examples. 
\ref{sec:ex12}, page~\pageref{sec:ex12} and on, shows details of construction of a $ 12 $-element worst-case heap for the heap-deconstruction phase of $ {\tt Heapsort} $ in first 11 moves of strategy $ \mathsf{win}(15) $. 
\ref{sec:win31}, page~\pageref{sec:win31} and on, shows details of construction of the last level of a $ 31 $-element worst-case heap for the heap-deconstruction phase of $ {\tt Heapsort} $ in moves 15 through 30 of strategy $ \mathsf{win}(31) $. 
\ref{sec:appExInput}, page~\pageref{sec:appExInput} and on, shows program-generated examples of 500-element worst-case array for the heap-construction phase and   500-element worst-case heap for
the heap-deconstruction phase of $ {\tt Heapsort} $.

 \medskip
 
 \ref{Hereditary}, page~\pageref{Hereditary} and on, discusses hereditary worst-case heaps. Its findings explain why any greedy strategy must have failed why generating worst cases for the $ {\tt RemoveAll} $.

\section{Heaps and {\tt Heapsort} - a brief review} \label{HeapsHeapsort}

I am going to use extensively some standard undergraduate math of analysis of algorithms in this paper. Here is a quick reminder of some basics:
$\lg x $ is a logarithm base 2 of $ x $; $ \lfloor x \rfloor $ is the greatest integer
not greater than $ x $; 
$ \lceil x \rceil $ is the least integer 
not less than $ x $;
thus $ 2^{\lfloor \lg x \rfloor} $ is the greatest power of 2 not larger than $x$ and $ 2^{\lceil \lg x \rceil} $ is the least power of 2 not less than $x$; and $ \%$ is the \textit{remainder modulo} function defined for $ n \geq 1 $ by $ m \% n $ $ = $ $ m - n \times \lfloor \frac{m}{n} \rfloor $. 

\medskip

Here is an ubiquitous formula that comes handy while dealing with this kind of math, true for every positive integer $ n $:
\begin{equation} \label{eq:basic1}
\lfloor \lg n \rfloor  = \lceil \lg (n+1) \rceil - 1,
\end{equation}
a special case of which yields, for every positive integer $ n $:
\begin{equation} \label{eq:basic2}
\lfloor \lg (2^{n}-1) \rfloor  = n-1.
\end{equation}

\medskip

The rest of this Section contains some standard definitions and basic facts pertaining to heaps and ${\tt Heapsort}$ that those familiar with the subject may wish to omit and go directly to the next section on page~\pageref{Notation}. Unlike many other presentations, this one is prevailingly algebraic.

\medskip

A \textit{binary tree structure} is a non-empty\footnote{Allowing empty trees and heaps does not add any benefits to the presentation of this paper.} finite set $ I $ of positive integers, referred to as the index set, that is closed under positive integer division by $ 2 $ 
\footnote{This coincides with the usual mathematical-logical definition of binary tree (cf. \cite{b3:hnbk}) as a set of binary sequences of length $ < \alpha $ (where $ \alpha $ is an ordinal number) for $ \alpha \leq  \omega $ closed under truncation, taking into account that every finite binary sequence is equal to the binary representation of some integer with the leading 1 omitted; however, mine does not include the empty tree.}, 
\linebreak
under convention that $  1$ positively-integerly divided by $  2$ is equal to $ 1 $. A \textit{finite binary tree}, to which I will simply refer to as \textit{binary tree}, is a function $ T $ whose domain is $ I $. The elements of 
$ T $ (ordered pairs $ \langle i , T[i] \rangle $, that is)
are called nodes.
 For \label{assumed1-1} the sake of simplicity of presentation, I will assert that the said function is $ 1-1 $, which in more humane terms means that the tree in question is duplicate-free. In particular, $ T $ has the inverse function $ T^{-1} $. Thus $ T[i] $
 \footnote{I use the square brackets in lieu of parentheses here because of popularity of an array representation of binary trees.} 
 is the value stored in $ T $ at index $ i $ and $ T^{-1} [p] $ is the index of the value $ p $ in tree $ T $.  This $ 1-1 $-ness assertion will allow me to sometimes not distinguish between the nodes and their constituent indicies and values if it is clear from the context which of the three notions am I referring to. Since the sole purpose of this article is analysis of the worst-case number of comparisons of keys performed by the $ {\tt Heapsort} $ sorting algorithm and its components, the above assertion does not lead to a loss of generality. Moreover, since the $ {\tt Insert(x)} $ operation is not used in $ {\tt Heapsort} $, so is neither included nor discussed in this analysis, I may\footnote{And, usually, will.} assume, without loss of generality, that $ T $ is a permutation on $ I $ \footnote{Thus the inverse  $ T^{-1} $ of $ T $ is a permutation, too; for example, the inverse of $[ 8, 6, 7, 4, 5, 2, 3, 1 ] $ is $ [ 8, 6, 7, 4, 5, 2, 3, 1 ]$.}, that is, that the range of $ T $ coincides with $ I $.

\medskip

The index set of a binary tree defines the parent-children relationship between its nodes. A node $ p $ at index $ i $ is a parent of a node $ q $ at index $ j $, or - in other words - a node $ q $ at index $ j $ is a child of a node $ p $ at index $ i $ if, and only if, 
\begin{equation} \label{eq:parchild1} 
 i = \lfloor \frac{j}{2} \rfloor .
\end{equation}
An iterative application of (\ref{eq:parchild1}), taking into account that
$ \lfloor \frac{\lfloor \frac{j}{2} \rfloor}{2} \rfloor $ $ = $ $ \lfloor \frac{j}{4} \rfloor $, 
gives rise to the definition of the usual ancestry relation between nodes at indicies $ i $ (an ancestor's index) and $ j $ (a descendant index) in terms of existence of $ k \geq 0$ for which the equality
\begin{equation} \label{eq:ancestryk} 
 i = \lfloor \frac{j}{2^k} \rfloor 
\end{equation}
is satisfied.
The equation (\ref{eq:ancestryk})
yields the inequality
\begin{equation} \label{eq:ancestryk2} 
	i2^k \leq j < (i+1)2^k
\end{equation}
which implies
\begin{equation} \label{eq:ancestryk3} 
k \leq \lg \frac{j}{i} < k + \lg (1+\frac{1}{i}) \leq k+1,
\end{equation}
or
\begin{equation} \label{eq:distance}
k  =  \lfloor \lg \frac{j}{i} \rfloor . \footnote{If (\ref{eq:ancestryk}) is satisfied then $ k $ given by (\ref{eq:distance}) is the distance, that is, the length of path, from $ i $ to $ j $; note that (\ref{eq:distance}) does not imply (\ref{eq:ancestryk}).} 
\end{equation} 
Thus, for any 
$ i,j \in I $, $ i $ is the index of an ancestor of a node at index $ j $ if, and only if,
\begin{equation} \label{eq:ancestor} 
 i = \lfloor \frac{j}{2 ^{\lfloor \lg \frac{j}{i} \rfloor}} \rfloor .
\end{equation}
Plugging (\ref{eq:distance}) into (\ref{eq:ancestryk2}) yields another, equivalent to (\ref{eq:ancestor}), characterization of the ancestry relation:
\begin{equation} \label{eq:ancestor2} 
i2 ^{\lfloor \lg \frac{j}{i} \rfloor} \leq j < (i+1)2 ^{\lfloor \lg \frac{j}{i} \rfloor} .
\end{equation}

The equality (\ref{eq:parchild1}) has two solutions:
\begin{equation} \label{eq:parchild2} 
 j_0 = 2i \mbox{ and } j_1 = 2i+1.
\end{equation}
The node $ q $ is called the left child of $ p $ at index $ i $ if, and only if, its index is $ j_0 $, and  is called the right child of $ p $ if, and only if, its index is $ j_1 $.

\medskip

The root is defined as the node with no parent; thus $ 1 $ is the index of the root\footnote{Because $ \lfloor \frac{1}{2} \rfloor = 0 $ and $ 0 \notin I $.}. A leaf is defined as a node without children; thus its index $ i $ satisfies the condition 
\begin{equation} \label{eq:leaf1} 
2i \notin I \mbox{ and } 2i+1 \notin I ,
\end{equation}
where $ I $ is the 
index set of the tree.

\medskip

A \textit{path}\footnote{From the root.} in a binary tree $ T $ is a sequence of indicies of $ T $ defined by induction: $ \langle 1 \rangle $ is a path in $ T $, if $ \sigma $ is a path whose last element is $ i $ and $ j  $ is an index in $ T $ with $ i = \lfloor  \frac{j}{2} \rfloor $ then the concatenation $ \sigma ^{\frown} \langle j \rangle$ is a path in $ T $, and nothing else is a path in $ T $.
   
\medskip

\begin{figure}[h] 
\begin{center}
\includegraphics[scale=0.75]{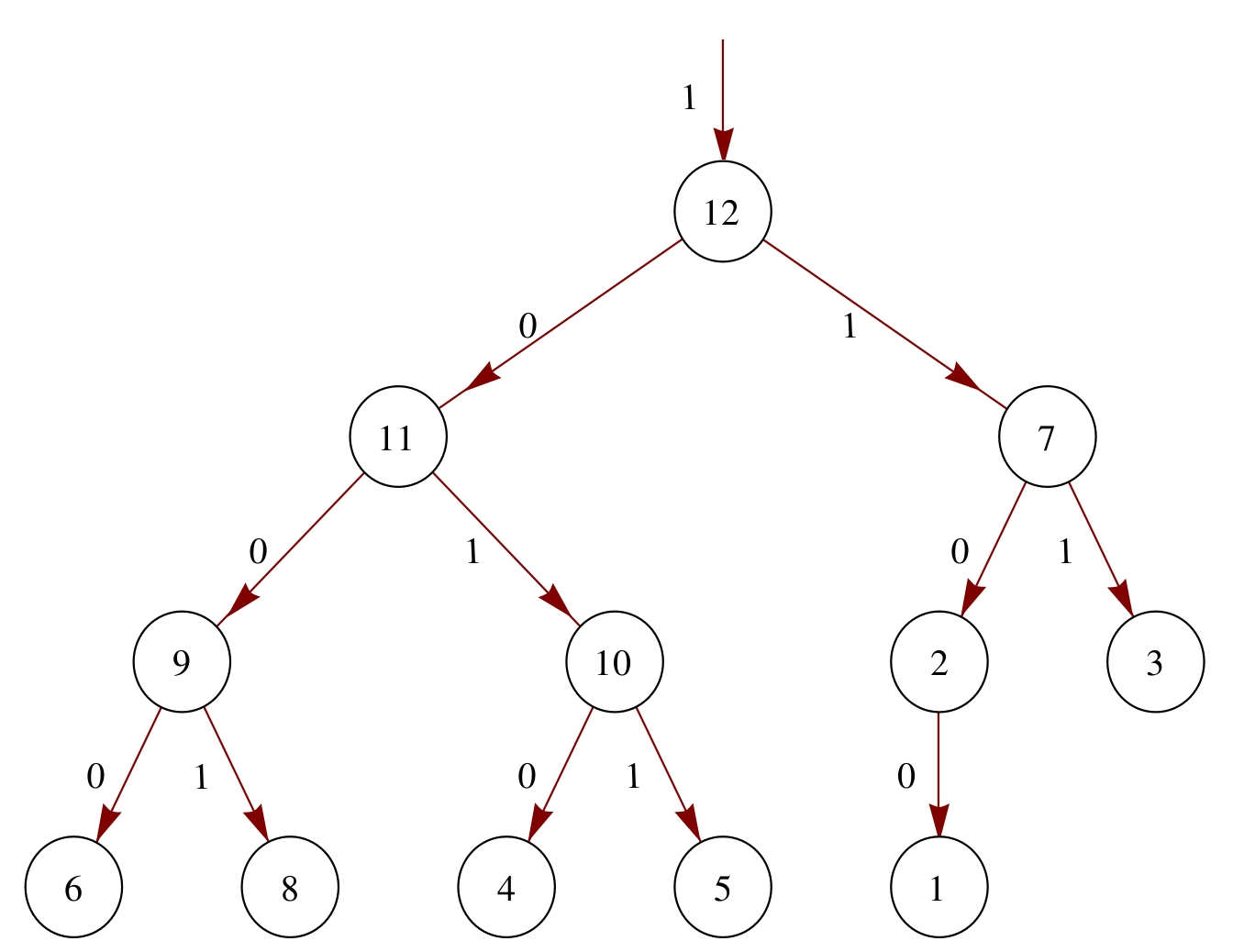} 
\end{center}
\caption{A heap of 12 nodes, with values and the navigation information shown.  \label{fig:ExHeap}}
\end{figure}

It follows that binary representations of the indices of a binary tree provide the navigation information how to get to those nodes from the root, with 0 meaning ``go to the left child'' and 1 meaning ``go to the right child'', except for the first 1 that means ``go to the root'', as it has been visualized on Figure~\ref{fig:ExHeap} page~\pageref{fig:ExHeap}. In that sense, each of those indices encodes the path (from the root) to the node at that index. Naturally, the length of such path
is one less than the number of digits in the binary representation of $ i $, that is, it is equal to 
\begin{equation} \label{eq:nodedepth} 
D_i = \lfloor \lg i \rfloor,
\end{equation}
where $ i $ is the index of the destination node (the last index in the said path). I will call $ D_i $ the depth of the node at index $ i $. Moreover, I will call a level~$ k $ of the tree the set of all its nodes (or, sometimes, indicies thereof) that have depth $ k $. Thus every node belongs to the level $ \lfloor \lg i \rfloor $, where $ i $ is the index of that node.

\medskip

In particular, the (binary representation of the) largest index $ N $ in a binary tree encodes the path (from the root of the tree) to its last node. Because that path is a longest path in the tree, its length 
\begin{equation} \label{eq:Tdepth} 
D_N = \lfloor \lg N \rfloor,
\end{equation}
also the depth of the node at index $ N $, is the depth of the entire tree. Thus the depth of the tree is 
the level number of the last non-empty level of that tree.

\medskip

Given a binary tree $ T $, \textit{the path of the largest child} is defined as the path $ \sigma $ (from the root) to a leaf such that each node in $ \sigma $, except for the root, is the largest child of its parent. One can easily conclude from the $1-1$-ness assertion that every binary tree has the unique path of the largest child. If $ \sigma $ may not go all the way down to a leaf then I will call it \textit{a} path of the largest child.

\medskip

Each node $ p $ at index $i$ in a binary tree $ T $ is the root of a subtree $ T^{(i)} $ that consists of $p$ and all its descendants, and - when treated as a separate entity - has it own index set $ I^{(i)} $. It can be computed with a help of the formula (\ref{eq:distance}) page~\pageref{eq:distance} for the length of path from $ i $ to $ j $ as
\begin{equation} \label{eq:indexsetsubtree} 
I^{(i)}  = \{  2 ^{\lfloor \lg \frac{j}{i} \rfloor} + j \, \% \, 2 ^{\lfloor \lg \frac{j}{i} \rfloor} \mid j \in I \mbox{ and } i = \lfloor \frac{j}{2 ^{\lfloor \lg \frac{j}{i} \rfloor}} \rfloor \} ,
\end{equation}
where $ I $ is the index set of $ T $ and the ancestry relation between the respective indicies $ i $ and $ j $ is given by (\ref{eq:ancestor}) page~\pageref{eq:ancestor}. 
\footnote{If $ i \geq 2 $ then $ T^{(i)} $ is not necessarily a permutation on its index set $ I^{(i)} $.}



\medskip

The \textit{height} of a node $ p $ at index $ i $ in a binary tree $ T $ is defined as the depth of the subtree $ T^{(i)} $ of  $ T $. 

\medskip

I call a binary tree a \textit{convex} binary tree if, and only if, its index set is a convex set of integers. Thus every convex binary tree on $ N $ nodes has the index set equal to 
\begin{equation} \label{eq:Iset} 
I = \{1, ... , N \}.
\end{equation}
One can conclude from (\ref{eq:indexsetsubtree}) that any subtree of a convex tree is convex.

\medskip

It follows from (\ref{eq:nodedepth}) and (\ref{eq:indexsetsubtree})
that the depth $ D_i ^N $ of a subtree $ T^{(i)} $ of a convex tree $ T $ of $ N $ nodes is equal to
\begin{equation} \nonumber 
D_i ^N = 
\max \{ \lfloor \lg ( 2 ^{\lfloor \lg \frac{j}{i} \rfloor} + j \, \% \, 2 ^{\lfloor \lg \frac{j}{i} \rfloor}) \rfloor \mid j \in I \mbox{ and } i = \lfloor \frac{j}{2 ^{\lfloor \lg \frac{j}{i} \rfloor}} \rfloor \} =
\footnote{Since $ 0 \leq j \, \% \, 2 ^{\lfloor \lg \frac{j}{i} \rfloor} < 2 ^{\lfloor \lg \frac{j}{i} \rfloor} $ so that $ 2^{\lfloor \lg \frac{j}{i} \rfloor} \leq 2 ^{\lfloor \lg \frac{j}{i} \rfloor} + j \, \% \, 2 ^{\lfloor \lg \frac{j}{i} \rfloor} < 2^{\lfloor \lg \frac{j}{i} \rfloor+1}  $ or $ {\lfloor \lg \frac{j}{i} \rfloor} \leq \lg (2 ^{\lfloor \lg \frac{j}{i} \rfloor} + j \, \% \, 2 ^{\lfloor \lg \frac{j}{i} \rfloor}) < {\lfloor \lg \frac{j}{i} \rfloor} + 1 $, that is, $ {\lfloor \lg \frac{j}{i} \rfloor} = {\lfloor \lg (2 ^{\lfloor \lg \frac{j}{i} \rfloor} + j \, \% \, 2 ^{\lfloor \lg \frac{j}{i} \rfloor}) \rfloor})$.}
\end{equation}
\begin{equation} \nonumber
= \max \{\lfloor \lg \frac{j}{i} \rfloor \mid j \in I \mbox{ and } i = \lfloor \frac{j}{2 ^{\lfloor \lg \frac{j}{i} \rfloor}} \rfloor \} =
\end{equation}
[since (\ref{eq:ancestor}) and (\ref{eq:ancestor2}) page~\pageref{eq:ancestor2} are equivalent]
\begin{equation}  \label{eq:heightbin1}
= \lfloor \lg  \frac{\max \{ j  \mid j \in I  \mbox{ and } (\ref{eq:ancestor2})\}}{i} \rfloor = \lfloor \lg  \frac{j_{\max}}{i} \rfloor.
\end{equation}
Since the maximal $ j \in I $ that satisfies (\ref{eq:ancestor2}) must also satisfy
\begin{equation} \label{eq:max_j}
i 2 ^{\lfloor \lg  \frac{j}{i}  \rfloor} \leq N < i 2 ^{\lfloor \lg  \frac{j}{i}  \rfloor + 1}
\end{equation}
(or otherwise it would not be maximal that satisfies (\ref{eq:ancestor2})), I conclude from (\ref{eq:max_j}) that
\[ {\lfloor \lg  \frac{j_{\max}}{i}  \rfloor} \leq \lg \frac{N}{i} < \lfloor \lg  \frac{j_{\max}}{i}  \rfloor + 1\]
or
\begin{equation} \label{eq:max_j2}
 {\lfloor \lg  \frac{j_{\max}}{i}  \rfloor} = \lfloor \lg \frac{N}{i} \rfloor.
 \footnote{Note that $ \lfloor \lg  {j_{\max}}  \rfloor $ may be lass than $\lfloor \lg  {N}  \rfloor$, so that the intuitively simple argument that presumes their equality is invalid.}
\end{equation}
	Applying (\ref{eq:max_j2}) to (\ref{eq:heightbin1}), I obtain one of the fundamental formulas for the precise worst-case analysis of the heap construction phase of $ {\tt Heapsort} $
\begin{equation} \label{eq:nodeheight} 
D_i ^N = \lfloor \lg \frac{N}{i} \rfloor.
\footnote{Cf. \cite{suc:elem} for a more conventional (and longer) derivation of (\ref{eq:nodeheight}). Note that (\ref{eq:nodeheight}) is not a direct concludion from (\ref{eq:distance}) as $ i $ does not have to be an ancestor of $ N $.}
\end{equation}

\medskip

If the convex binary tree in question has more than 1 node then the number of nodes in its last level is odd if, and only if, the number of nodes in the entire heap is even.

\medskip

It follows from (\ref{eq:leaf1}) and (\ref{eq:Iset}) that $ i  $ is an index of a leaf in a convex binary tree if, and only if,
\begin{equation} \label{eq:leaf2} 
2i > N ,
\end{equation}
where $ N $ is the size of the index set. 

\medskip

The \textit{leftmost descendant} $ j $ of a node at index $ i $ in a finite convex binary tree $ T $ is the first (leftmost) node in the last nonempty level of subtree $ T^{(i)} $. By virtue of (\ref{eq:ancestor2}) page~\pageref{eq:ancestor2}, (\ref{eq:max_j2}), and (\ref{eq:nodeheight}), its index $ j $ is given by this formula:
\begin{equation} \label{eq:leftmost} 
j = i \times 2^{\lfloor \lg \frac{N}{i} \rfloor}  =  i \times 2^{D_i ^N} . \, \footnote{See proof of Lemma 5.1 in \cite{suc:elem} for a conventional derivation of the equality $ j =  i \times 2^{D_i ^N} $.
	 }
\end{equation}

\medskip

Every non-empty level $ k $ in any convex binary tree, except, perhaps, the last non-empty level, is equal to $ \{ 2^k, ... , 2^{k+1}-1 \}$, while the last non-empty level $\lfloor \lg N \rfloor$ in such a tree is equal to 
$ \{ 2^{\lfloor \lg N \rfloor}, ... , N  \}$.

\medskip

Moreover, I call a \textit{complete binary tree} a convex binary tree whose index set $ I $ 
has the size $ N = 2 ^{\lfloor \lg N \rfloor+1} - 1$, that is,
\begin{equation} \label{eq:Isetconv} 
I = \{1, ... , 2 ^{\lfloor \lg N \rfloor+1} - 1 \}.
\end{equation}
Naturally, 
any subtree of a complete binary tree is a complete binary tree. 
Of course, $ N \geq 1$ is the size of the index set of a complete heap if, and only if, $ N+1 $ is a power of two, that is, $ \lfloor \lg (N+1) \rfloor $ $=$ $ \lceil \lg (N+1) \rceil $, or, by (\ref{eq:basic1}) page~\pageref{eq:basic1},
\begin{equation} \label{eq:completeCharact} 
\lfloor \lg N \rfloor =  \lfloor \lg (N+1) \rfloor - 1 .
\end{equation}
Also, the largest complete binary tree of no more than $ N $ nodes has 
\begin{equation} \label{eq:completeCharactNodes} 
M =  2 ^{\lfloor \lg (N+1) \rfloor} - 1 
\end{equation}
nodes, since $ 2 ^{\lfloor \lg (N+1) \rfloor} $ is the greatest power of 2 not greater than $ N+1 $ so that $ 2 ^{\lfloor \lg (N+1) \rfloor} $ is the greatest power of 2 minus 1 not greater than $ N $.
By (\ref{eq:basic2}) page~\pageref{eq:basic2} and (\ref{eq:Tdepth}) page~\pageref{eq:Tdepth},
the depth of such largest complete binary tree of no more than $ N $ nodes is
\begin{equation} \label{eq:completeCharactDepth} 
\lfloor \lg (2 ^{\lfloor \lg (N+1) \rfloor} - 1) \rfloor =  \lfloor \lg (N+1) \rfloor - 1. 
\end{equation}

\medskip

A heap is a convex\footnote{Some authors use adjective \textit{complete} in this context, instead. A neat definition borrowed from mathematical logic allows one to identify the set of indicies of a countable $ k $-ary tree with a set of positive integers closed under positive integer division by $ k $, with $ \frac{i}{k} $ assumed equal to $ 1 $ if $ 1 \leq i < k $. In light of such a definition, a finite \textit{convex} $ k $-ary tree $ T $ has a set of indicies $ I $ that comprises of the first $ n $ positive integers, which one could describe as \textit{complete}, although $ I $ is in fact a  \textit{convex} set of integers so the adjective \textit{convex}  appears like a better descriptor of tree $ T $. I will reserve adjective $ complete $ to binary trees with $ 2^D-1 $ nodes, where $ D $ is a positive integer. It is worth noting that in mathematical logic trees were studied long before they were used in Computer Science, using a definition of a complete $ k $-ary tree  that ours is compatible with; in particular, according to that definition (cf. \cite{b3:hnbk}, p. 381), a finite complete $ k $-ary tree of depth D has $ k^D-1 $ nodes.}, 
partially ordered binary tree.
 \textit{Partially ordered} means that every sequence of values along any path in the tree is ordered in a decreasing order.

\medskip

An example of a heap is visualized on Figure \ref{fig:ExHeap}.

\medskip

\medskip

Quite obviously, any non-empty convex binary $ H $ tree may be represented as a one-dimensional array
whose indicies range over the index set of $ H $, that is, from $ 1 $ to $ N $, the number of nodes of the tree.

\medskip

\begin{figure}[h] 
\begin{center}
\begin{tabular}{|c||c|c|c|c|c|c|c|c|c|c|c|c||}
  \hline
 index&1&2&3&4&5&6&7&8&9&10&11&12 \\
  \hline
  value & 12 & 11 & 7 & 9 & 10 & 2 & 3 & 6 & 8 & 4 & 5 & 1  \\
  \hline
\end{tabular} 
\end{center}
  \caption{Array representation of the heap of Figure \ref{fig:ExHeap}. \label{fig:ArrayHeap}}
\end{figure}

The table in Figure \ref{fig:ArrayHeap} shows an array that represents the heap of Figure \ref{fig:ExHeap} with the indices of the array shown in the top row of the table.

\medskip

 If tree  $ H $ is partially ordered then its every subtree $ H^{(i)} $, where $ i $  is in the index set of $ H $, is partially ordered as well, so a subtree of a heap is a heap\footnote{Except that it is not necessarily a permutation on its index set.}. I will call $ H^{(i)} $  a subheap of $ H $.

\medskip


\medskip

{\tt Heapsort} (see, e.g., \cite{knu:art} for its description and partial analysis) consists of two phases: {\em heap construction} and a sequence of removals from the constructed heap that I call {\em heap deconstruction}. 

\medskip

Both phases use a subroutine {\tt FixHeap} that inherits an \textit{almost heap}, defined as a heap whose root, referred to in some contexts as a \textit{patch}, may violate the {\em partially ordered tree} condition in the definition of heap, and turns it onto a heap by bubble-sorting its root into the path of the largest child. This is done by demoting the said patch down the heap while promoting the largest of its current children until the demotee reaches the level where it is not less than any of its current children, if it still has any at that level. Since each step in that process requires comparing, directly or indirectly, the demotee to its all children\footnote{Clearly, the demotee must have been compared to its largest child; knowing which child is the largest does require comparing children to each other, if there are two children, that is.}, the total number of comparisons of keys that {\tt FixHeap} performs during one call is equal to the total number of children of elements of the path of demotion that it follows\footnote{Which is \textit{a} path of the largest child.}. Because {\tt FixHeap} is the only part of {\tt Heapsort} that performs comparisons of keys, the above characterization is the point of departure of the analysis presented in this paper.

\medskip

The heap-construction phase, referred to as ${\tt MakeHeap}$ in this paper and credited to Floyd \cite{flo:heap}, inherits an array that represents a convex binary tree $H$ and rearranges it onto a heap by calling ${\tt FixHeap}$ for its parts that represent subtrees of $H$ that have been already rearranged onto almost subheaps, beginning from the one that has the last non-leaf (stored at the index $\lfloor \frac{N}{2}\rfloor$ in the array) of $H$ as the root\footnote{{\tt MakeHeap} could have begun calling {\tt FixHeap} from the last node of $H$, but this would produce the same sequence of comparisons of keys and demotions because {\tt FixHeap} does not do anything to a one-node tree.} and ending with the entire tree $H$ (the root of which is stored at index $1$). This is accomplished by the following Java statement:
\begin{equation} \label{java}
 \mbox{\tt
        for (int i = N/2; i > 0; i-{}-) FixHeap(i); }
\end{equation}
The heap-deconstruction phase, referred to as {\tt RemoveAll} in this paper, consists of $N$ calls to a subroutine  {\tt RemoveMax}. Each of these calls removes the current root of the heap, patches the resulting vacancy with the current last node of the heap, and calls ${\tt FixHeap}(1)$ in order to turn the resulting almost heap onto a heap after each removal. The removed nodes are then stored in the array $ {\tt heap} $ from the last index up in the order they were removed, which process yields an array that is sorted in an increasing order.  
This is accomplished by the following Java statement:
\begin{equation} \label{java2}
 \mbox{\tt
        for (int i = N; i > 0; i-{}-) heap[i] = RemoveMax(); }
\end{equation}

A complete code of {\tt HeapSort} may be easily found in about every standard text on Data Structures and Algorithms, or in \cite{suc:heapsort}.

\section{Notation and basic facts} \label{Notation}

I am going to measure the running time of {\tt Heapsort} and its components by the number of comparisons of keys that they perform, using the following notation.

\medskip

For any operation $ X $, $C_{X} (Y)$ denotes the number of comparisons of keys that $ X $ performs while executed on its input $ Y $. For instance,
$ C_{{\tt FixHeap}} (T^{(i)}) $, which I will also denote as $ C_{{\tt FixHeap}(i)} (T) $, is the number of comparisons of keys that the {\tt FixHeap} performs while turning the almost subheap $ T^{(i)} $ rooted at node $i 
$ of a convex binary tree $ T $ onto a heap.

\medskip

Moreover, $C_{ X}  ^{\tt max}(N)$ denotes the maximum number of comparisons of keys that $ X $ performs while executed on its any valid input of size $ N $. It is given by this formula:
\begin{equation} \label{eq:defCXMax}
C_{X} ^{\tt max} (N) = \max \{ C_{ X} (Y) \mid  Y \mbox{ is $ X $'s valid input of size }  N\} .
\end{equation}
For instance,
\begin{equation} \label{eq:defCremoveAllMax}
C_{\tt RemoveAll()} ^{\tt max} (N) = \max \{ C_{\tt RemoveAll()} (H) \mid H \mbox{ is a heap on }  N  \mbox{ nodes}\} .
\end{equation}

\medskip 

{\em Residue}\footnote{\textit{Proper residue} would be a more adequate but longer term.} of a heap $ H $ is either the heap that is the result of one applications of $ {\tt RemoveMax} $ to $ H $, or a residue of a residue of $ H $.

\medskip

$ \omega $ is the set of all non-negative integers.
$ \omega ^+$ is the set of all positive integers.
A (non-empty) sequence is a function whose domain is a convex subset of  $ \omega ^+$ that contains 1. If $ s $ is a sequence and $ N \subseteq \omega ^+ $ then $ s \! \restriction \! N  $ is the result of restricting (of the domain of $ s $) to  N. 
${\mathsf v} ^{\frown} {\mathsf w}$ is the concatenation of sequences ${\mathsf v}$ and ${\mathsf w}$.

\section{A useful trick: Running Heapsort backwards} \label{sec:runback}

It turns out that $ {\tt FixHeap} $ is invertible, and so are $ {\tt MakeHeap}$, $ {\tt RemoveMax}$, and $ {\tt RemoveAll}$. This fact allows for running the entire $ {\tt Heapsort}$ backwards in order to produce inputs that force it to follow predetermined paths of demotions within the heap. This is useful in construction of cases that establish lower bounds on its worst-case behavior. Although the transition relation for  $ {\tt Heapsort} $, a deterministic algorithm, is a function, its inverse relation is not a function. So, some extra information is required in order to execute it backwards.

\medskip

\begin{figure}[h] 
\begin{center}
\includegraphics[scale=.35]{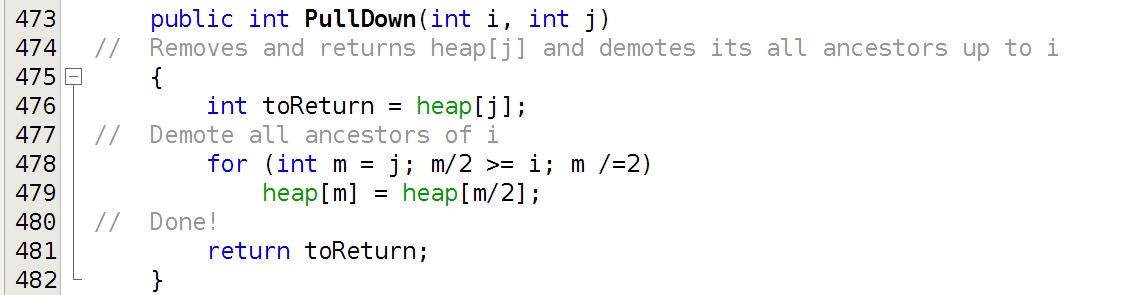} 
\end{center}
\caption{A Java code for operation $ {\tt PullDown}(i, j) $; $ i $ and $ j $ are indicies. \label{fig:PullDown} }
\end{figure}

The basic operation I will use to accomplish all the above is $ {\tt PullDown}$ whose Java code is shown on Figure~\ref{fig:PullDown} on page~\pageref{fig:PullDown}. It takes a convex binary tree $ H $ and two indicies in $H $, $ i $ and $ j $, with $ j $ a presumed descendant of $ i $,\footnote{It actually works fine even if $ j $ is not e descendant of $ i $; in such a case, it works as if $ i $ were equal to 1.} removes and returns node $ H[j] $ (instructions in lines 476 and 481 of the referenced above Java code), and demotes $ j $'s all proper ancestors in $ H $ that are descendants of $ i $ (the $ {\tt}for $-loop in lines 478 and 479 of the Java code).

\medskip

The following Subheap Repair Lemma provides a useful characterization of the implementation of the operation $ {\tt PullDown}$.

\medskip


\begin{fixlm} \label{lem:FH}
Let $ H $ be a convex binary tree of $ N \geq $ nodes, $ i \leq \lfloor \frac{N}{2} \rfloor $ be its index such that the subtree
$ H^{(i)} $ roted at $ i $ is a heap, $ j $ be a proper descendant index of $ i $ in $ H $, and $ H^{\prime} $ be the convex binary tree that is the result of executing the following Java instruction
\begin{eqnarray} \label{eq:PullDownFH}
 \mbox{\tt
        H.heap[j] = H.PullDown(i,j);}
\end{eqnarray}
on $ H $, where the Java code of method $ {\tt PullDown} $ is visualized on Figure~\ref{fig:PullDown}.

\begin{enumerate}
     \renewcommand\labelenumi{\theenumi}
     \renewcommand{\theenumi}{(\roman{enumi})}

\item \label{item:FH1} The execution of $ {\tt FixHeap} $ on $ H^{\prime} $ yields $ H $, that is,
\begin{equation} \label{eq:c1FH}
H^{\prime} . {\tt FixHeap}(i) = H .
\end{equation}

\item \label{item:FH2}
The number $ C_{{\tt FixHeap}(i)} (H^{\prime}) $ of comparisons of keys performed by the execution of $ {\tt FixHeap} $ on $ H^{\prime} $
is given by this equality:
\begin{equation} \label{eq:creditFH} 
C_{{\tt FixHeap}(i)} (H^{\prime}) = 2 (\lfloor \lg j  \rfloor - \lfloor \lg i \rfloor -1)  + \#_{\lfloor \frac{j}{2} \rfloor}^N + \#_j ^N,
\end{equation}
where 
\begin{equation} \label{eq:no_of_childrenFH}
\#_m ^N \; \footnote{$ \#_m ^N = signum (\frac{N}{m} -2) +1 $, where $ signum (x) $ is the sign of $ x $ (-1 if $ x<0 $, 0 if $ x=0 $, and 1 if $ x>0 $).}
  = 
\left\{ \begin{array}{ll}
0 \mbox{ if } 2m>N   \\ \\
1 \mbox{ if } 2m=N   \\ \\
2 \mbox{ if } 2m<N
\end{array} \right.
\end{equation}
is the number of children of the node at index $ m $ in a heap of $ N $ elements. 

\item \label{item:FH3} If $ j $ is the leftmost descendant of $ i $ in $ H $, and, therefore, in  $ H^{(i)} $, then
the number $ C_{{\tt FixHeap}(i)} (H^{\prime}) $ of comparisons of keys performed by the execution of $ {\tt FixHeap} $ on $ H^{\prime} $
is maximal, that is, it satisfies the equality 
\begin{equation} \label{eq:creditFH50} 
C_{{\tt FixHeap}(i)} (H^{\prime}) = C_{{\tt FixHeap}(i)} ^{\tt max} (N).
\end{equation}
\end{enumerate}
\end{fixlm}
{\bf Proof} Because  $H ^{(i)}$ is a heap and $ j $ is a proper descendant of $ i $ in $ H $ and, therefore, in $H ^{ (i)}$,
\begin{equation} \label{eq:FHj<j}
H[j] < H[i].
\end{equation}
Moreover, for every child $ k $ of $ j $ in $ H $ and, therefore, in $H ^{ (i)}$,
\begin{equation} \label{eq:FHchild}
H[k] < H[j].
\end{equation}
Since, as an effect of instruction (\ref{eq:PullDownFH}), $H ^{\prime}[i]$ $ = $  $ H[j] $, $H ^{\prime}[i]$ $ = $  $ H[j] $,
inequality (\ref{eq:FHj<j}) implies
\begin{equation} \label{eq:FHj<j'}
H^{\prime}[j] > H^{\prime}[i],
\end{equation}
and inequality (\ref{eq:FHchild}) implies, for every child $ k $ of $ j $ (unaffected by instruction (\ref{eq:PullDownFH})) in in $ H^{\prime} $ and, therefore, in $H ^{\prime (i)}$,
\begin{equation} \label{eq:FHchild'}
H^{\prime}[k]  < H^{\prime}[i].
\end{equation}

\label{page:pathdemotion}
Each node demoted as a result of $ {\tt H.PullDown(i,j)} $ in instruction (\ref{eq:PullDownFH})
becomes the largest child after the demotion, so that the path $ \sigma $ $ = $ $ \langle i,...,j\rangle $ of such demotions becomes a path of the largest child in the subtree $H ^{\prime (i)}$ of $H ^{\prime}$. This is a consequence of the fact that each demotee was the parent (before the demotion) of its sibling (after the demotion), except for the case when the demotee became a leaf in $ H^{\prime} $, thus it must be greater than its sibling (after the demotion), if it has any. Thus the demotions of $H ^{\prime} [i]$ done by the subsequent execution of $ {\tt FixHeap()} $  will follow $ \sigma $ and terminate after $H ^{\prime} [i]$ is demoted to index $ j $ since all children of $ j $
are, by virtue of (\ref{eq:FHchild'}), less than the demotee $H ^{\prime} [i]$. As a result, (\ref{eq:c1FH}) holds. This completes the proof of \ref{item:FH1}.

\medskip

The number of comparisons of keys performed by  $ {\tt FixHeap()} $ following a path $ \sigma $ of the largest child is equal to the total number of children of indicies in $ \sigma $. Thus,
\begin{equation} \label{eq:FH600}
 C_{{\tt FixHeap}(i)} (H^{\prime}) = \sum _{k \in \sigma} \#_k ^N = \sum _{k \in \sigma \setminus \{ \lfloor \frac{j}{2} \rfloor,j \}} \#_k ^N 
  + \#_{\lfloor \frac{j}{2} \rfloor}^N + \#_j ^N.
\end{equation}
Since all elements of $ \sigma  $, except, perhaps, for $ \lfloor \frac{j}{2} \rfloor $ and $ j $, have 2 children each, $\sum _{k \in \sigma \setminus \{ \lfloor \frac{j}{2} \rfloor,j \}} \#_k ^N $ is equal to twice the number of levels between the level of $ i $ and the level of $ \lfloor \frac{j}{2} \rfloor $ (not including the latter), that is,
\[ \sum _{k \in \sigma \setminus \{ \lfloor \frac{j}{2} \rfloor,j \}} \#_k ^N =   2(\lfloor \lg \lfloor \frac{j}{2} \rfloor \rfloor - \lfloor \lg i \rfloor) = 2(\lfloor \lg \frac{j}{2}  \rfloor - \lfloor \lg i \rfloor) =
2(\lfloor \lg j  \rfloor - \lfloor \lg i \rfloor - 1) .
\]
Thus,
\begin{equation} \label{eq:FH500}
\sum _{k \in \sigma \setminus \{ \lfloor \frac{j}{2} \rfloor,j \}} \#_k ^N = 2(\lfloor \lg j  \rfloor - \lfloor \lg i \rfloor - 1)  .
\end{equation}

Substituting (\ref{eq:FH500}) to the right-hand side of (\ref{eq:FH600}) yields (\ref{eq:creditFH}).  This completes the proof of \ref{item:FH2}.

\medskip

Clearly, the path from $ i $ to its leftmost descendant $ j $ has the total number of children in  the subtree $H ^{\prime (i)}$ of $H ^{\prime}$ at least as large as any other path in $H ^{\prime (i)}$ has. Therefore, the equality (\ref{eq:creditFH50}) holds. This completes the proof of \ref{item:FH3} and the proof of the Lemma.
\hspace*{\fill} $\Box$

\medskip

\textbf{Note}. Since $ {\tt FixHeap}(i) $ performs the maximum number of comparisons if it demotes $ i $ to the index of its leftmost descendant, it may be computed directly from the equality (\ref{eq:creditFH}) page~\pageref{eq:creditFH}, substituting $ j = i \times 2^{\lfloor \lg \frac{N}{i} \rfloor} $ given by the equality (\ref{eq:leftmost}) page~\pageref{eq:leftmost} and using the equality (\ref{eq:creditFH50}) of part \ref{item:FH3} of the Subheap Repair Lemma \ref{lem:FH} that
\begin{equation} \label{eq:CFixHeapi} 
C_{{\tt FixHeap}(i)} ^{\tt max} (N) = \lfloor \lg \frac{N}{i} \rfloor + \lfloor \lg \frac{N-1}{i} \rfloor ,
\end{equation}
but since I already did it in \cite{suc:elem}, Corollary 5.2, I refrained from redoing it here.
Also, substituting 1 for $ i $ in the equality (\ref{eq:CFixHeapi}) yields
\begin{equation} \label{eq:CFixHeap1} 
C_{{\tt FixHeap}(1)} ^{\tt max} (N) = \lfloor \lg N \rfloor + \lfloor \lg (N-1) \rfloor .
\end{equation}
\medskip

For running backwards the heap-construction phase ${\tt MakeHeap}$, 
I will use an operation $H . {\tt unFixHeap}(i) $ that takes a subheap $ H^{(i)} $ rooted at node $i \leq \lfloor \frac{N}{2} \rfloor $ of a convex binary tree $  H $ of $ N $ nodes and turns it onto an almost subheap $ H^{\prime (i)} $  of the resulting convex binary tree $ H^{\prime} $ $ = $ $ H . {\tt unFixHeap}(i) $, while leaving the remainder of $ H $ unchanged, with the following two constrains satisfied:

\begin{equation} \label{eq:Fixconstrain1}
(H . {\tt unFixHeap}(i) ) . {\tt FixHeap}(i) = H 
\end{equation}
and 
\begin{equation} \label{eq:Fixconstrain2}
C_{{\tt FixHeap}(i)} (H. {\tt unFixHeap}(i)) = 
C_{{\tt FixHeap}(i)} ^{\tt max} (N) ,
\end{equation}
where 
$C_{{\tt FixHeap}(i)} (T)$ 
is the number of comparisons of keys that the {\tt FixHeap} performs while turning the almost subheap 
rooted at node $i \leq \lfloor \frac{N}{2} \rfloor $ of a convex binary tree $ T $
 onto a heap, and $ C_{{\tt FixHeap}(i)} ^{\tt max} (N) $ is the maximum of $C_{{\tt FixHeap}(i)} (T)$ over all convex binary trees $ T $ of $ N $ nodes whose subtrees rooted at index $ i $ are almost heaps.

\medskip

\begin{figure}[h] 
\begin{center}
\includegraphics[scale=.35]{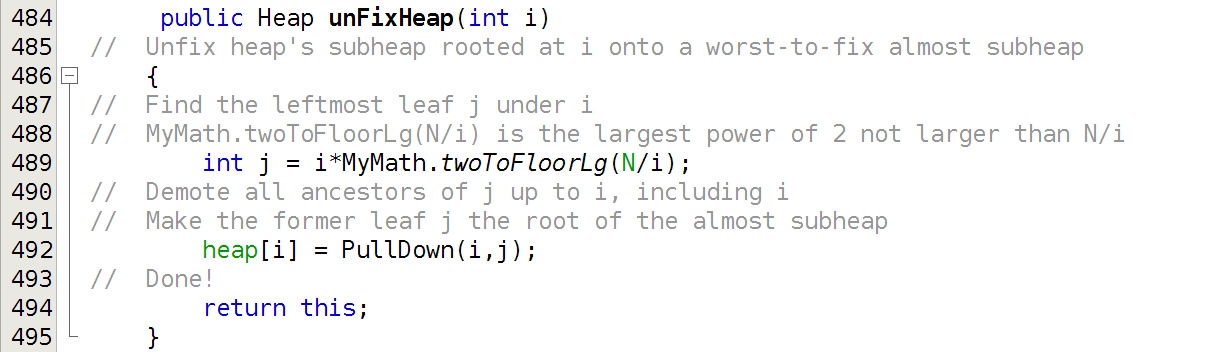} 
\end{center}
\caption{A Java code that implements operation $ {\tt unFixHeap} $. 
Instruction in line 489 computes the index of the leftmost descendant of $ i $ using the formula (\ref{eq:leftmost}).
The static method $ {\tt twoToFloorLg}(n) $ in class $ {\tt MyMath} $ computes $2 ^{\lfloor \lg n \rfloor}  $. The method $ {\tt PullDown(i,j)} $ removes and returns the node at index $ j $ and demotes its all ancestors up to and including an ancestor at index $ i $; it is shown on Fig.~\ref{fig:PullDown} on page~\pageref{fig:PullDown} and is discussed in Section~\ref{sec:pull}. \label{fig:unfix}}
\end{figure}

Operation $ {\tt unFixHeap} $ is not unique. Any one that satisfies the constrains (\ref{eq:Fixconstrain1}) and (\ref{eq:Fixconstrain2}) will do.
I am going to prove that the Java code shown on Figure~\ref{fig:unfix} implements operation $ {\tt unFixHeap} $ that satisfies those constrains.

\medskip

Let $ H $ be a convex binary tree of $ N $
nodes whose subtree $ H^{(i)} $ roted at index $ i \leq \lfloor \frac{N}{2} \rfloor$ of $ H $ is a heap, and let $ j $, given by (\ref{eq:leftmost}) page~\pageref{eq:leftmost} and computed by the instruction in line 489 of Java code shown on Figure~\ref{fig:unfix}, be the leftmost descendant of $ i $. Since $ j $ is a leaf and $ i $ is not, $ j $ is a proper descendant of $ i $, so that Subheap Repair Lemma~\ref{lem:FH} does apply.

\medskip

Application of Subheap Repair Lemma~\ref{lem:FH}~\ref{item:FH1} yields  (\ref{eq:Fixconstrain1}).  Thus constrain (\ref{eq:Fixconstrain1}) is satisfied.

\medskip

Application of Subheap Repair Lemma~\ref{lem:FH}~\ref{item:FH3} yields  (\ref{eq:Fixconstrain2}).  Thus constrain (\ref{eq:Fixconstrain2}) is satisfied.

\medskip

\begin{figure}[h] 
\begin{center}
\includegraphics[scale=.35]{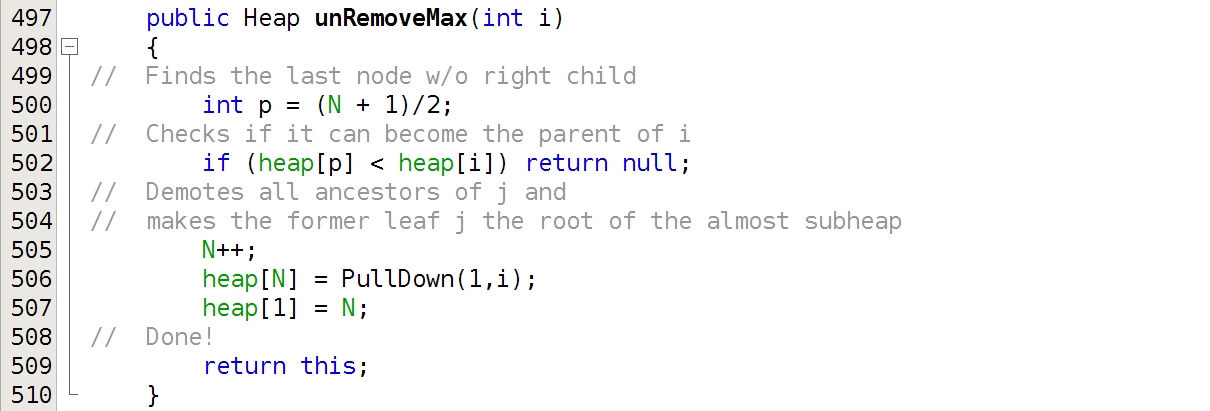} 
\end{center}
\caption{A Java code that implements operation $ {\tt unRemoveMax}(i) $.  The method $ {\tt PullDown(1,i)} $  removes and returns the node at index $ j $ and demotes its all ancestors; it  is shown on Fig.~\ref{fig:PullDown} on page~\pageref{fig:PullDown} and is discussed in Section~\ref{sec:pull}. \label{fig:unRemoveMax}}
\end{figure}

For running backwards the heap-deconstruction phase ${\tt RemoveAll}$, I will use the operation $H . {\tt unRemoveMax}(i) $, visualized on Figure~\ref{fig:unRemoveMax}, 
that undoes the effects of any given operation ${\tt RemoveMax} $ that produced a given heap $ H $. More specifically, $H . {\tt unRemoveMax}(i) $ takes a heap 
$  H $ on $ N $ nodes and an index $ i $ of its node $  H [i] $
\footnote{The  presumed \textit{patch} used by ${\tt FixHeap(1)}$ called by the ${\tt RemoveMax()} $ that
the $ {\tt unRemoveMax}(i) $ presumably undoes.} that satisfies the constrain 
\begin{equation} \label{eq:Remconstrain1}
H [i] \leq H [\lfloor \frac{N+1}{2} \rfloor] ,
\end{equation}
and produces a\footnote{\textit{The}, as the Uniqueness Lemma~\ref{lem:Uniq} page~\pageref{lem:Uniq} states.} heap 
\begin{equation} \label{eq:defHprime}
H^{\prime} = H.{\tt unRemoveMax}(i)
\end{equation}
on $ N+1 $ nodes that satisfies these two constrains:
\begin{equation} \label{eq:Remconstrain2}
H^{\prime}[N+1] = H [i] ,
\end{equation}
and
\begin{equation} \label{eq:Remconstrain3}
H^{\prime} . {\tt RemoveMax()} = H
.
\end{equation}

The constrains (\ref{eq:Remconstrain2}) and (\ref{eq:Remconstrain3}) are self-explanatory. The constrain (\ref{eq:Remconstrain1}) is an input constrain for $ {\tt unRemoveMax}(i) $ and needs a comment. It allows $ {\tt unRemoveMax}(i) $, line 506 of the Java code visualized on Figure~\ref{fig:unRemoveMax}, to make $ H[i] $ a child,  in heap $ H^{\prime} $, of the first node $ H [\lfloor \frac{N+1}{2} \rfloor] $ of heap $ H$ without the right child. This, for any $ H $ and $ i $ that satisfy the constrain (\ref{eq:Remconstrain1}), assures the existence of $ H^{\prime} $ that satisfies constrains (\ref{eq:Remconstrain2}) and (\ref{eq:Remconstrain3}). If $ H $ was produced by ${\tt RemoveMax()} $ that used $ H[i] $ as the patch then, of course, (\ref{eq:Remconstrain1}) is satisfied. In the case of $ i = \lfloor \frac{N+1}{2} \rfloor $ (the said node without the right child is the patch), this constrain reduces to a tautology $ H [\lfloor \frac{N+1}{2} \rfloor] \leq H [\lfloor \frac{N+1}{2} \rfloor] $.
In the case $ i = 2 \lfloor \frac{N+1}{2} \rfloor = N + (N \mod 2)$ (the left child of the said node without the right child is the patch), $  N $ is even and the constrain reduces to $ H [N] \leq H [\frac{N}{2}] $, true for every heap $ H$ on $ N $ nodes. These, for any $ H $, assure the existence of index $ i $ in $ H $ that satisfies the constrain (\ref{eq:Remconstrain1}).

\medskip

The $ i $ and $ H^{\prime} $, whose existences have been demonstrated above, are unique, as the following Lemma states. 

\begin{unlm} \label{lem:Uniq}.	
	
	\begin{enumerate}
		\renewcommand\labelenumi{\theenumi}
		\renewcommand{\theenumi}{(\roman{enumi})}
		
		\item \label{item:Uniq1}
		
		For every heap $ H $ on $ N  $ nodes and its every index $ i $  that satisfies the constrain (\ref{eq:Remconstrain1}), there is the unique heap  $ H^{\prime} $ that satisfies constrains (\ref{eq:Remconstrain2}) and  (\ref{eq:Remconstrain3}).
		
		\item \label{item:Uniq2}
		
		For every heap $ H^{\prime} $ on $ N \geq 2 $ nodes there is a unique index $ i \leq N -1$ in the heap $ H $ defined by (\ref{eq:Remconstrain3}) such that the constrain (\ref{eq:Remconstrain1}) and the equality (\ref{eq:defHprime}) are satisfied. 
		
	\end{enumerate}
	
\end{unlm} 
{\bf Proof}. \ref{item:Uniq1} The existence of such a heap $ H^{\prime} $ follows form the foregoing discussion. Now, suppose that heaps on $ N+1 $ nodes $ H^{\prime} $ and its substitute $ G$ satisfy (\ref{eq:defHprime}) and (\ref{eq:Remconstrain3}). Let $ \sigma $ be the path (from the root) to $ i $. Obviously, $ \sigma $ is a path in all three heaps, $ H $, $ H^{\prime} $, and $ G $. Since $ {\tt RemoveMax}() $ does not modify its explicit argument ($ H^{\prime} $ or $ G $) except for the elements at indices along path $ \sigma $, I infer that for every index $ k $ in $ H $ with $ k \notin \sigma $,
\begin{equation} \nonumber
H^{\prime}[k] = H[k] = G[k],
\end{equation}
that is,
\begin{equation} \label{eq:Unique1proof100}
H^{\prime}[k] =  G[k],
\end{equation}
Also, by virtue of constrain (\ref{eq:Remconstrain2})
\begin{equation} \label{eq:Unique1proof200}
H^{\prime}[N+1] =  G[N+1].
\end{equation}
Let $ j \in \sigma $, that is, $ j > 1 $ be an ancestor of $ i $. 
We have:

\begin{equation} \label{eq:Unique1proof300}
H^{\prime}[j] = H[j/2] = G[j].
\end{equation}
Also
\begin{equation} \label{eq:Unique1proof400}
H^{\prime}[1] = N+1 = G[1].
\end{equation}
Thus the equality (\ref{eq:Unique1proof100}) is also satisfied for all $ i \in \sigma $. Therefore,
\[ H^{\prime} = G. \]
Hence the uniqueness of $  H^{\prime} $.

\medskip

\ref{item:Uniq2} The existence of such an $ i $ follows from the foregoing discussion. The uniqueness follows from part \ref{item:Uniq1} of the Lemma (already proved) and constrain (\ref{eq:Remconstrain2}).\footnote{Recall that any heap is a 1-1 function.}
\hspace*{\fill} $\Box$

\medskip

\begin{figure}[h] 
\begin{center}
\includegraphics[scale=.25]{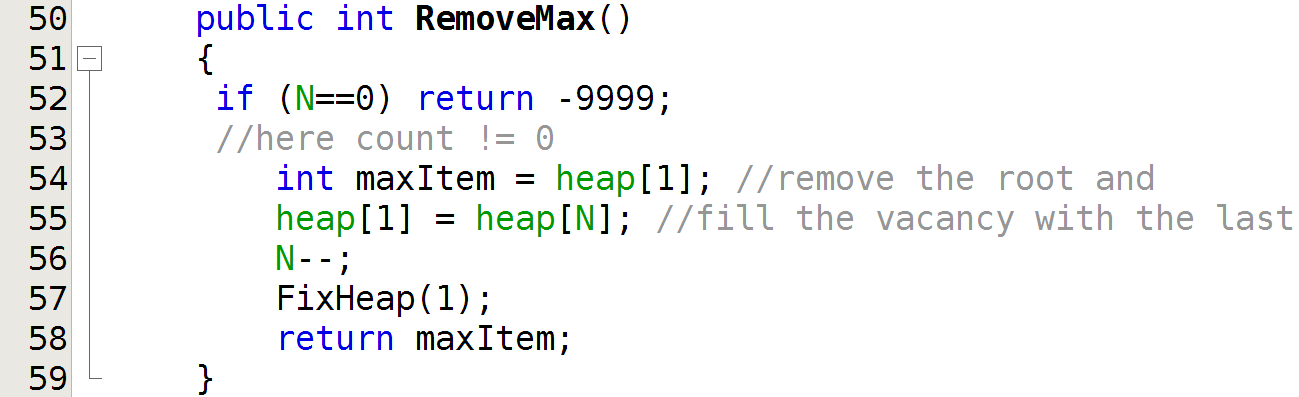} 
\end{center}
\caption{A Java code that implements operation $ {\tt RemoveMax}() $.  The method $ {\tt FixHeap(1)} $ implements the operation described at the end of Section~\ref{Overview}. \label{fig:RemoveMax}}
\end{figure}

The above Uniqueness Lemma assures that the heap  $  H^{\prime} $ postulated by constrains (\ref{eq:Remconstrain2}) and (\ref{eq:Remconstrain3}) does exist and is unique.
I am going to show that the Java code visualized on Figure~\ref{fig:unRemoveMax} produces an $ N+1 $-node heap $ H^{\prime} $ out of given $ N $-node heap $ H $ that satisfy constrains (\ref{eq:Remconstrain2}) and (\ref{eq:Remconstrain3}) if the constrain (\ref{eq:Remconstrain1}) is met; otherwise the move $ i $ is invalid and $ {\tt unRemoveMax}(i) $ returns the
$ {\tt null} $ value as a result of execution of instructions in lines 500 and 502 of the Java code of Figure~\ref{fig:unRemoveMax} page~\pageref{fig:unRemoveMax}. Without loss of generality\footnote{Since for every heap $ H $, the result of execution of $ {\tt RemoveMax}() $ on $ H $ is determined by $ H $, any two correct implementations of $ {\tt RemoveMax}() $ are functionally equivalent.}, I will use an example of implementation of  $ {\tt RemoveMax}() $
whose Java code is visualized on Figure~\ref{fig:RemoveMax}. 

\medskip

$ H^{\prime} $ is a heap because $ H $ is a heap, and instruction in line 507 of the Java code of Figure~\ref{fig:unRemoveMax} page~\pageref{fig:unRemoveMax} makes its root larger than any other node of $ H^{\prime} $, and instruction in line 506 of that code attaches a new child to the node $ p $ at index $ N $ that, by asserted constrain (\ref{eq:Remconstrain1}), is less than $ p $, and the rest of that code does not affect the ordering of the paths in $ H^{\prime} $.

\medskip

Constrain (\ref{eq:Remconstrain2}) is secured by instructions in lines 505 and 506 of the above referenced Java code.
In order to show that constrain (\ref{eq:Remconstrain3}) is met, let us look into the implementation of operation 
$ {\tt RemoveMax} $ of Figure~\ref{fig:RemoveMax}.

\medskip

Clearly, instructions in lines 55 and 56 of the Java code is shown on Figure~\ref{fig:RemoveMax} reverse the effects of the instructions in lines 505 and 503, as well as the effects of the assignment to $ {\tt heap[N]} $ in line 506 of the Java code is shown on Figure~\ref{fig:unRemoveMax}. At this point, the resulting tree is the same as if instruction
\begin{equation} \label{eq:H.Pulldown(1,i)}
\mbox{\tt
        H.heap[1] = H.PullDown(1,i);}
\end{equation}
were applied to the original heap. Therefore, the Subheap Repair Lemma \ref{lem:FH} applies (substituting 1 for $ i $ and $ i $ for $ j $), and its part \ref{item:FH1} implies that line 57
of the Java code is shown on Figure~\ref{fig:RemoveMax} restores the original heap. Thus constrain (\ref{eq:Remconstrain3}) is satisfied. As a result, $ {\tt RemoveMax}() $ and $ {\tt unRemoveMax}(i) $ are inverses to one another, as the following Lemma states.

\medskip

\begin{remlm} \label{lem:RemMax}.	
	
	\begin{enumerate}
		\renewcommand\labelenumi{\theenumi}
		\renewcommand{\theenumi}{(\roman{enumi})}
		
		\item \label{item:RemMax2}
		
		For every heap $ H $ on $ N  $ nodes and every index $ i \leq N $ that satisfies the constrain (\ref{eq:Remconstrain1}) page~\pageref{eq:Remconstrain1}, 
		\begin{equation} \label{eq:RemMax2} 
		(H.{\tt unRemoveMax}(i)) . {\tt RemoveMax()} = H .
		\end{equation}
		
				\item \label{item:RemMax1}
	
	For every heap $ H $ on $ N \geq 2 $ nodes there is a unique index $ i \leq N -1$ such that the constrain (\ref{eq:Remconstrain1}) page~\pageref{eq:Remconstrain1} is satisfied and 
	\begin{equation} \label{eq:RemMax1}
	(H.{\tt RemoveMax}()) . {\tt unRemoveMax(i)} = H .
	\end{equation}

		\end{enumerate}
	
\end{remlm} 
{\bf Proof}. \ref{item:RemMax2} Let $ H $ be a heap on $ N  $ nodes and $ H^{\prime} $ be a heap defined by (\ref{eq:defHprime}) page~\pageref{eq:defHprime}. Substituting (\ref{eq:defHprime}) to (\ref{eq:Remconstrain3}) page~\pageref{eq:Remconstrain3} yields (\ref{eq:RemMax2}).

\medskip

 \ref{item:RemMax1} Let $ H^{\prime} $ be a heap on $ N \geq 2 $ nodes, $ H $ be the heap defined by (\ref{eq:Remconstrain3}) and $ i $ be the unique index whose existence is assured by the Uniqueness Lemma~\ref{lem:Uniq}~\ref{item:Uniq2}. Combining (\ref{eq:defHprime}) and (\ref{eq:Remconstrain3}), both of which are satisfied, yields
 	\begin{equation} \label{eq:RemMax1proof900}
 	 H^{\prime} =  H . {\tt unRemoveMax(i)}  =  (H^{\prime}.{\tt RemoveMax}()) . {\tt unRemoveMax(i)},
 	\end{equation}
 	or
 \begin{equation} \label{eq:RemMax1proof9500}
 H^{\prime} =    (H^{\prime}.{\tt RemoveMax}()) . {\tt unRemoveMax(i)}.
 \end{equation}	
Since $ H^{\prime} $ was any heap on $ N \geq 2 $ nodes, (\ref{eq:RemMax1proof9500}) implies (\ref{eq:RemMax1}). 
\hspace*{\fill} $\Box$

\medskip

The following Lemma will be useful while proving correctness of construction of worst-case heaps for 
$ {\tt RemoveAll} $.

\begin{rmcrlm} \label{lem:newRemMax} 
Let $ H $ be a heap of $ N \geq 2 $ nodes and let $ H^{\prime} $ $ = $ $ H. {\tt unRemoveMax(i)} $, where $ 2 \leq i \leq N $.
The number $ C_{{\tt RemoveMax()}} (H^{\prime}) $ of comparisons of keys that the operation $ {\tt RemoveMax()} $  performs on $ H^{\prime} $ is given by this equality:
\begin{equation} \label{eq:newRemMaxcredit}
C_{{\tt RemoveMax()}} (H^{\prime}) =  2 (\lfloor \lg i  \rfloor -1)  + \#_{\lfloor \frac{i}{2} \rfloor}^N + \#_i ^N,
\end{equation}
where $\#_m ^N$, given by  (\ref{eq:no_of_childrenFH}) page~\pageref{eq:no_of_childrenFH}, 
is the number of children of the node at index $ m $ in a heap of $ N $ elements, 
with convention  $  \#_0 ^N = 0$.
\end{rmcrlm} 
{\bf Proof}.
 The only comparisons of keys within $ {\tt RemoveMax()} $ are performed by the call to $ {\tt FixHeap(1)} $ in line
57 of the Java code visualized on Figure~\ref{fig:RemoveMax}. Thus
\begin{equation} \label{eq:newRemMax100}
C_{{\tt RemoveMax()}} (H^{\prime}) = C_{{\tt FixHeap(1)}} (H^{\prime \prime}),
\end{equation}
where $ H^{\prime \prime} $ is the heap on $ N $ nodes with $ H^{\prime \prime} [k] = H^{\prime}[k] $ for $ 2 \leq k \leq N $ and $ H^{\prime \prime} [1] = H^{\prime}[N+1] $.
 Since, as noted in the discussion above, $ H. {\tt unRemoveMax(i)} $ performed actions that were comprised by (\ref{eq:H.Pulldown(1,i)}), the  Subheap Repair Lemma \ref{lem:FH} applies (substituting 1 for $ i $ and $ i $ for $ j $) and the equality (\ref{eq:creditFH}) of its part \ref{item:FH2}, taking
 into account equality (\ref{eq:newRemMax100}), yields 
(\ref{eq:newRemMaxcredit}).
\hspace*{\fill} $\Box$

\textbf{Note}. Equalities (\ref{eq:newRemMax100}) and (\ref{eq:CFixHeap1}) page \ref{eq:CFixHeap1} (the latter with substituting $ N - 1 $ for $ N $ since the almost heap sent to $ {\tt FixHeap}(1) $ has one less node than the one sent to {\tt RemoveMax}) yield for every $ N \geq 3 $:
\begin{equation} \label{eq:RemMaxmaxcredit}
C_{{\tt RemoveMax()}} ^{\tt max} (N) = \lfloor \lg (N-1) \rfloor + \lfloor \lg (N-2) \rfloor .
\end{equation}

\section{Decomposition of the worst-case analysis of ${\tt Heapsort}$} \label{Decomp}

It is easy to generate worst-case input arrays of arbitrary size for $ {\tt MakeHeap} $. A Java code presented and proved correct in Appendix A of \cite{suc:elem} does just that. 
However, efficiently\footnote{As opposed to, say, exhaustive search.} generating worst-case heaps for $ {\tt RemoveAll} $ and worst-case input arrays for the entire $ {\tt Heapsort} $, except for some special sizes, have been, to my best knowledge, unknown. Moreover, although the sum of upper bounds of program's components is an upper bound on the running time of the entire program, the converse is not necessarily true. For instance, for $ N \geq 13 $ an upper bound on the number comparisons of keys performed by $ {\tt RemoveAll} $ on an $ N $-element heap is less than the sum of any upper bounds on numbers of comparisons of keys performed by the sequence of $ N $ $ {\tt RemoveMax} $es that comprise it. And this singularity\footnote{\label{foo:singlarity}This singularity is characterized by Theorems~\ref{cor:parwin} and \ref{cor:parwinVar} page~\pageref{cor:parwin}.} is one of the reasons why the worst-case analysis of $ {\tt Heapsort} $ is not totally a routine task.

\medskip

Fortunately, the problem of generation of worst-case inputs for $ {\tt Heapsort} $ can be decomposed on two subproblems: how to, given any $ N \geq 1 $, generate a worst-case heap $ H $ of size $ N $ for $ {\tt RemoveAll} $, and how to, given a heap $ H $, generate a worst-case array for $ {\tt MakeHeap} $ that $ {\tt MakeHeap} $ converts onto $ H $. 
 Clearly, solving those subproblems in that order will result in a worst-case array $ A $ of size $ N $ for $ {\tt Heapsort} $. Careful evaluation of the number $ C_{\tt Heapsort}(A) $ of comparison of keys  performed on the resulting array will allow me to derive the exact formula for $ C^{max}_{\tt Heapsort}(N) $.

\medskip

The mentioned above decomposition property implies that any upper bound on the number comparisons of keys performed by $ {\tt Heapsort} $ on an $ N $-element array is equal to the sum of upper bounds on numbers of comparisons of keys performed by that $ {\tt MakeHeap} $ and $ {\tt RemoveAll} $ on an $ N $-element array and an $ N $-element heap, respectively. In particular, the least upper bound is equal to such a sum, thus yielding
\begin{equation} \label{eq:sum_maxes}
C^{max}_{\tt Heapsort}(N)  = C^{max}_{\tt MakeHeap}(N) + C^{max}_{\tt RemoveAll()}(N) .
\end{equation}



\medskip

\begin{figure}[h] 
\includegraphics[scale=.35]{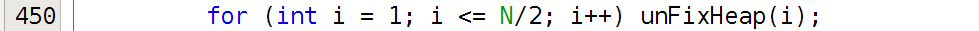} 
\caption{A Java statement that constructs a worst-case input array for ${\tt MakeHeap}$ given an output heap. The method $ {\tt unFixHeap(i)} $ is shown on Fig.~\ref{fig:unfix}.
\label{fig:mkworst}}
\end{figure}

Based on the above observations, in order to prove (\ref{eq:sum_maxes}) it suffices to demonstrate that,
given a heap $ H $, the Java loop statement visualized on Figure~\ref{fig:mkworst} constructs a worst-case input array for the ${\tt MakeHeap}$, that is, an array that the ${\tt MakeHeap}$ will convert onto $ H $ while performing the maximum possible number of comparisons. The equalities (\ref{eq:c1FH}) and (\ref{eq:creditFH50}) of the Subheap Repair Lemma~\ref{lem:FH}~\ref{item:FH1} and
\ref{item:FH3}, page~\pageref{eq:creditFH50}, via constrains (\ref{eq:Fixconstrain1}) and (\ref{eq:Fixconstrain2}), page~\pageref{eq:Fixconstrain1}, they entail, make the latter a routine exercise.

\medskip

\begin{figure}[h] 
\begin{center}
\includegraphics[scale=.5]{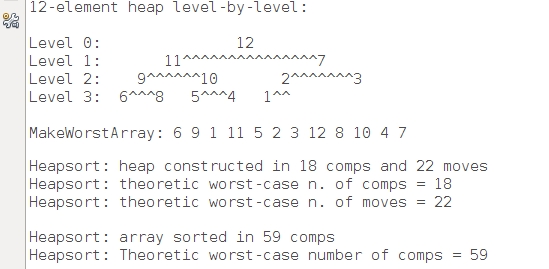} 
\end{center}
\caption{A 12-element worst-case array $ [6, 9, 1, 11, 5, 2, 3, 12, 8, 10, 4, 7] $ for the heap of Fig.~\ref{fig:ExHeap} on page~\pageref{fig:ExHeap} (a worst-case heap for the deconstruction phase of $ {\tt Heapsort} $); a fragment of output generated by my Java program containing the code of Fig.~\ref{fig:mkworst} on page~\pageref{fig:mkworst}. \label{fig:mkworstout}}
\end{figure}

Figure~\ref{fig:mkworstout} shows an output generated by my Java program containing the above code.

\medskip

First, I will prove, by induction on $ K $, that for every $ K \in \{ \lfloor \frac{N}{2} \rfloor -1 ,..., N \} $, that the following program $ P_K $:
\begin{eqnarray}
 \mbox{\tt
        for (int i = N/2; i <= K; i+{}+) unFixHeap(i);} \label{javaUnMakeHeapKMakeHeapK1} \\
\mbox{\tt
        for (int i = K; i >= N/2; i-{}-) FixHeap(i);   \hspace{0.1in}}  \label{javaUnMakeHeapKMakeHeapK2}       
\end{eqnarray}
leaves the heap $ H $, on which it is run, unchanged, that is, 
\begin{equation} \label{eq:invFHuFH}
H = H ^\prime ,
\end{equation}
where $ H ^\prime $ is the value of $ H $ after the execution of $ P_K $ on it,
and that the line (\ref{javaUnMakeHeapKMakeHeapK2}) in $ P_K $ forces the $ {\tt FixHeap} $ to perform the total of 
\begin{equation} \label{eq:UnFixFixK} 
C
(N,K) = \sum _{i=\lfloor \frac{N}{2} \rfloor} ^{K} C_{{\tt FixHeap}(i)} ^{\tt max} (N)
\end{equation}
comparisons of keys.


\medskip

If $ K = \lfloor \frac{N}{2} \rfloor -1 $ then program $ P_K $ performs no actions, so that the invariant (\ref{eq:invFHuFH}) is satisfied, and the numbers of comparisons done by the $ {\tt FixHeap} $ is 0, which is equal to the right-hand side of (\ref{eq:UnFixFixK}).

\medskip

If  $ K \in \{ \lfloor \frac{N}{2} \rfloor ,..., N \} $ then program $ P_K $ is functionally equivalent to:

\begin{eqnarray}
 \mbox{\tt
        for (int i = N/2; i <= K-1; i+{}+) unFixHeap(i);} \label{javaUnMakeHeapKMakeHeapK+1-1} \\
\mbox{\tt
         unFixHeap(K); \hspace{2.56in}} \label{javaUnMakeHeapKMakeHeapK+1-2} \\
\mbox{\tt
         FixHeap(K);  \hspace{2.73in}}  \label{javaUnMakeHeapKMakeHeapK+1-3}  \\     
\mbox{\tt
        for (int i = K-1; i >= N/2; i-{}-) FixHeap(i);  \hspace{0.1in}}  \label{javaUnMakeHeapKMakeHeapK+1-4}       
\end{eqnarray} 
By (\ref{eq:Fixconstrain1}) page~\pageref{eq:Fixconstrain1}, line (\ref{javaUnMakeHeapKMakeHeapK+1-3}) cancels out the effects of line (\ref{javaUnMakeHeapKMakeHeapK+1-2}), and by the invariant (\ref{eq:invFHuFH}) of the inductive hypothesis, line (\ref{javaUnMakeHeapKMakeHeapK+1-4}) cancels out the effects of line (\ref{javaUnMakeHeapKMakeHeapK+1-1}). So, program $ P_K $ leaves the heap $ H $ it is run on unchanged and the invariant (\ref{eq:invFHuFH}) is satisfied.

\medskip

By (\ref{eq:Fixconstrain2})  page~\pageref{eq:Fixconstrain2}, putting $ i = K $, the number of comparisons of keys performed by $ {\tt FixHeap} $ in line (\ref{javaUnMakeHeapKMakeHeapK+1-3}) is equal to
\begin{equation}
 C_{{\tt FixHeap}(K)} ^{\tt max} (N) ,
\end{equation}
and by the inductive hypothesis, taking into account that  line (\ref{javaUnMakeHeapKMakeHeapK+1-3}) cancels out the effects of line (\ref{javaUnMakeHeapKMakeHeapK+1-2}), the number of comparisons of keys performed by $ {\tt FixHeap} $ in line (\ref{javaUnMakeHeapKMakeHeapK+1-4}) is equal to
\begin{equation} \label{eq:UnFixFixKproof} 
 \sum _{i=\lfloor \frac{N}{2} \rfloor} ^{K-1} C_{{\tt FixHeap}(i)} ^{\tt max} (N),
\end{equation}
so that
\[ C
(N,K) =   C_{{\tt FixHeap}(K)} ^{\tt max} (N) +\sum _{i=\lfloor \frac{N}{2} \rfloor} ^{K-1} C_{{\tt FixHeap}(i)} ^{\tt max} (N),\]
or (\ref{eq:UnFixFixK}).
\begin{equation}  \label{eq:CNKsum}
C
(N,K)  =
\sum _{i=\lfloor \frac{N}{2} \rfloor} ^{K} C_{{\tt FixHeap}(i)} ^{\tt max} (N).
\end{equation}
This completes the inductive proof of (\ref{eq:UnFixFixK}) and the invariant (\ref{eq:invFHuFH}).

\medskip

Since for every $ N \geq 2 $, $ C
(N,N)  $ 
is a lower bound on the number $ C^{max}_{\tt MakeHeap}(N) $ of comparisons that $ {\tt MakeHeap} $ performs on any $ N $-element array and \linebreak $ \sum _{i=\lfloor \frac{N}{2} \rfloor} ^{N} C_{{\tt FixHeap}(i)} ^{\tt max} (N) $ is an upper bound on $ C^{max}_{\tt MakeHeap}(N) $ , I conclude from (\ref{eq:CNKsum}), putting $ K = N $, that for every $ N \geq 2 $, $ C
(N,N)  $ is the least upper bound on $ C^{max}_{\tt MakeHeap}(N) $, that is,
\begin{equation} 
C
(N,N)  =
C^{max}_{\tt MakeHeap}(N).
\end{equation}
Thus, for every heap $ H $, the Java loop statement visualized on Figure~\ref{fig:mkworst} constructs a worst-case input array for the ${\tt MakeHeap}$ that ${\tt MakeHeap}$ converts onto $ H $\footnote{In other words, every heap is a worst-case heap to build for the $ {\tt MakeHeap} $.}. Hence, the equality (\ref{eq:sum_maxes}) holds.


\begin{figure}[h] 
\begin{center}
\includegraphics[scale=1]{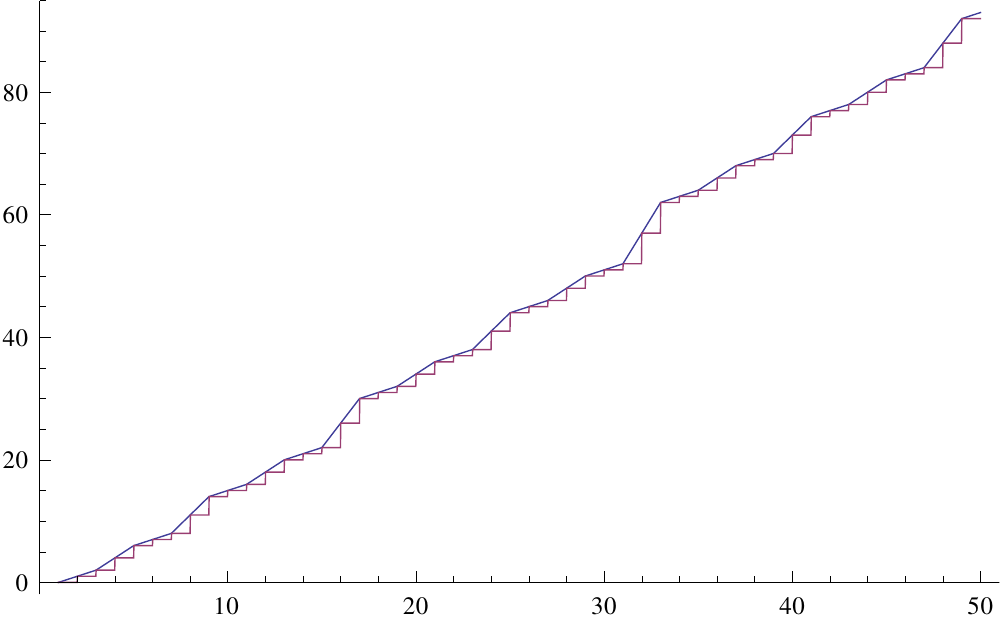} 
\end{center}
\caption{The number of comparisons of keys performed by ${\tt MakeHeap}$ in the worst case on any $ N $-element array, also known as the \textit{Sloan sequence} A092054 (cf. \cite{slo:oeis}). The lower line visualizes the right-hand side of (\ref{eq:CNKsum}) 
	for $ K = \lfloor N \rfloor $, while the upper line visualizes the linear interpolation of the right-hand side of (\ref{eq:MkHeapElem}) between integer points. \label{fig:makeHeap}}
\end{figure}
\medskip

By virtue of Theorem 7.1 in \cite{suc:elem},
\begin{equation} \label{eq:MkHeapElem}
C_{\tt MakeHeap} ^{\tt max} (N) = 2N - 2s_2(N) - e_2(N) ,
\end{equation}
where $s_2(N)$ is the sum of all binary digits of $N$ and $e_2(N)$ is the exponent of $2$ in the prime factorization of $N$.
Thus the array $ A $ produced by the Java loop statement visualized on Figure~\ref{fig:mkworst} run on any heap $ H $ of $ N $ distinct elements, when given as the input to ${\tt MakeHeap} $ forces it to perform
$ 2N - 2s_2(N) - e_2(N) $ comparisons of keys and to yield the heap $ H $.

\medskip

\noindent \textit{\textbf{Note}}. The sequence of integers
\[ 1, 2, 4, 6, 7, 8, 11, 14, 15, 16, 18, 20, 21, 22, 26, 30, 31, 32, 34, 36, 37, ... \]
 given by the formula (\ref{eq:MkHeapElem}) and visualized on Figure~\ref{fig:makeHeap} appears
 as the \textit{Sloan sequence} A092054 in \cite{slo:oeis}.

\medskip

At this point it becomes clear that all that is needed for the completion of the worst-case analysis of $ {\tt Heapsort} $ is a derivation of a formula for the worst-case number of comparisons of keys performed by the {\tt RemoveAll}. I will do just that in the sequel of this paper.

\section{The game of Pull Downs} \label{sec:pull}

Both  ${\tt Heapsort}$ and  ${\tt MakeHeap}$ have nice worst-case decomposition properties that allow to compute their worst-case numbers of comparisons of keys as sums of worst cases of their components. Unfortunately, the same cannot be said of  ${\tt RemoveAll}$. Although, for  arrays of size less than or equal to 12, the worst-case number of comparisons of keys performed by ${\tt RemoveAll}$ is equal to the sum of the worst-case numbers of comparisons of keys of the ${\tt RemoveMax}$'s that make ${\tt RemoveAll}$, it is not the case for arrays of more than 12 elements\footnote{As I have indicated at the begining of Section~\ref{Decomp} page~\pageref{Decomp}; see footnote $ ^{\ref{foo:singlarity}} $.}. In the latter case, the worst-case number of comparisons of keys is performed by ${\tt RemoveAll}$ is always less than the sum of the worst-case numbers of comparisons of keys of the ${\tt RemoveMax}$'s that make ${\tt RemoveAll}$.

\medskip

In order to construct worst-case heaps for ${\tt RemoveAll}$ and prove that they actually are worst-case heaps, I will resort to games and strategies.

\medskip

Imagine the following solitaire game played by the Player.

\medskip

\label{def:game} A \textit{game} is a sequence $ {\mathsf H} = \langle {\mathsf H}_{n+1} \mid n \in \omega \rangle $ of heaps in which $ {\mathsf H}_1 $ is the one-element heap whose only node is $ 1 $,\footnote{Formally, $ {\mathsf H}_1  = \{ ( 1,1 ) \}$, so the only node of  $ {\mathsf H}_1  $ is $ (1,1) $; naturally, it is identified by its value 1 (the second element in the pair).} and for every positive integer $ i $, heap $ {\mathsf H}_{i+1} $ is the result of a valid application ${\mathsf H}_{i}. {\tt unRemoveMax(k_i)} $ of the operation $ {\tt unRemoveMax(k_i)} $, visualized on Figure~\ref{fig:unRemoveMax} page~\pageref{fig:unRemoveMax}, to heap $ {\mathsf H}_{i} $. In particular, each $ {\mathsf H}_{i} $ is a heap on $ i $ nodes. 
Each application ${\mathsf H}_{i}. {\tt unRemoveMax(k_i)} $ is an $ i $-th \textit{move}, and I will refer to it in some contexts as a \textit{pull down}\footnote{A call $ {\tt PullDown(1,i)} $ to method $ {\tt PullDown} $, visualized on Figure~\ref{fig:PullDown} page~\pageref{fig:PullDown}, is part of method $ {\tt unRemoveMax} $; hence the name \textit{pull down}.}$ ^, $\footnote{The reason for using both \textit{moves} and \textit{pull downs} is to simplify notation. For instance, it allows using \textit{pull down} $ v $ (on a heap $ H $) in lieu of \textit{move} $ H^{-1}[v] $.}
 if it is a valid application, that is, if the instance 
$ {\mathsf H}_i [k_i] \leq {\mathsf H}_i [\lfloor \frac{i+1}{2} \rfloor ] $ of the constrain (\ref{eq:Remconstrain1}) is satisfied.  \\

\medskip

\label{def:playerPayoff}
The Player draws at random an integer $ N \geq 2$ and executes a sequence of $ N-1 $ consecutive\footnote{\label{foo:consecutive} The adjective \textit{consecutive} in this context means that each next pull down of the said sequence is applied to the heap produced by the foregoing pull down.} 
pull downs on the 1-element heap $ {\mathsf H}_1 $. The result is an $ N $-element heap $ {\mathsf H}_N $. Player's goal is to maximize the payoff for the game defined as the number $C_{\tt RemoveAll()} ^{\tt max} (N)$ of comparisons of keys that the $ {\tt Heapsort} $'s reconstruction phase ${\tt RemoveAll}$ will perform while run on the heap $ {\mathsf H}_N $.

\medskip

Given a game $ {\mathsf H} $, its every $ i $-th move ${\mathsf H}_{i}. {\tt unRemoveMax}({\mathsf k}_i) $ is determined by the index $ {\mathsf k}_i $,
and 
the game $ {\mathsf H} $ itself is determined by the sequence \linebreak $ {\mathsf k} = \langle {\mathsf k}_{n+1} \mid n \in \omega \rangle $ of indices that are the arguments of respective \linebreak ${\mathsf H}_{i}. {\tt unRemoveMax}({\mathsf k}_i) $'s. Thus there is a $ 1-1 $ function $  \mathscr{G} ({\mathsf k})$ between the sequences $ {\mathsf k} $ of valid moves and games that is defined by:
\begin{equation} \label{eq:GamesMovesFun} 
\mathscr{G} ({\mathsf k}) =  {\mathsf H}. 
\end{equation}
 Because of that, I will identify moves ${\mathsf H}_{i}. {\tt unRemoveMax}({\mathsf k}_i) $ with their arguments $ {\mathsf k}_i $. Also, 
since I assumed (on page \pageref{assumed1-1}) that any heap 
$ H $ is a $ 1 - 1 $ function,
and as such has the inverse $ H^{-1} $,
any index $ k $ in $ H $ is unambiguously identified by the node $ H[k] $ that is stored at $ k $. 
Thus the game $ {\mathsf H} $ is also determined by the sequence 
\begin{equation} \label{eq:defVK} 
{\mathsf v}   =  {\mathsf H}({\mathsf k} ) \;
\footnote{Incorporating (\ref{eq:GamesMovesFun}), $ {\mathsf H}({\mathsf k} ) $ may be written as $ {\mathsf H}(\mathscr{G}^{-1} ({\mathsf H})) $ that yields $ {\mathsf v} $ directly from $ {\mathsf H}$ without any references to $ {\mathsf k} $.}
 = \langle {\mathsf H}_{i+1}[{\mathsf k}_{i+1}] \mid i \in \omega \rangle
\end{equation}
of patches
where  $ {\mathsf k}_i $ is the argument of respective ${\mathsf H}_{i}. {\tt unRemoveMax}({\mathsf k}_i) $, simply because the sequence $ {\mathsf k} $ in (\ref{eq:GamesMovesFun}) is determined by
\begin{equation} \label{eq:defKV} 
{\mathsf k}   =  {\mathsf H}^{-1}({\mathsf v} ) =  \langle {\mathsf H^{-1}}_{i+1}[{\mathsf v}_{i+1}] \mid i \in \omega \rangle .
\end{equation}
This fact allows me to identify pull downs with \textit{patches} (the values that are being pulled down) rather than with their indices. For instance, given a heap $ H $, pull down 6 is the move 
$H. {\tt unRemoveMax}(H^{-1}[6]) $.
\medskip

\label{def:heapBySequence} 
I will also consider finite subgames  $ {\mathsf H}_{J,K} $ of game $ {\mathsf H} $ that I define as finite sequences of heaps $\langle {\mathsf H}_i \mid J \leq i \leq K \rangle $ from $ {\mathsf H} $. Clearly, any finite subgame $ {\mathsf H}_{J,K} $ is determined by the heap $ {\mathsf H}_J $ and the subsequence
$ {\mathsf k}_{J,K-1}  =  \langle {\mathsf k}_{i} \mid J \leq i \leq K-1 \rangle $ 
of $ {\mathsf k} =   \mathscr{G}^{-1} ({\mathsf H})$ 
 of valid moves.
 This gives rise to function  $ \mathscr{G}_{{\mathsf H}_J} $ defined by
\begin{equation} \label{eq:defseqheaps} 
\mathscr{G}_{{\mathsf H}_J} ( {\mathsf k}_{J,K-1} ) = \mathscr{G} ({\mathsf k})_{J,K} 
\; \footnote{The $ K-J $ moves result in a $ K-J +1$-element subgame: the original heap  $ {\mathsf H}_J $ plus the $ K-J $ heaps $ {\mathsf H}_{J+1},...,  {\mathsf H}_{K}$ created by that subgame.}.
\end{equation}
As before, given  heap $ {\mathsf H}_J $, any finite subgame $ {\mathsf H}_{J,K} $ is unambiguously determined by
the corresponding subsequence
\begin{equation} \label{eq:defVjkKjk} 
 {\mathsf v}_{J,K-1}  =   {\mathsf H}({\mathsf k}_{J,K-1} ) = {\mathsf H}({\mathsf k})_{J,K-1} = 
  \langle {\mathsf H}_{i}[ {\mathsf k}_{i}] \mid J \leq i \leq K-1 \rangle 
\end{equation} 
of pull downs applied consecutively$ ^{\ref{foo:consecutive}} $
to $ {\mathsf H}_J $, 
simply because the sequence $ {\mathsf k}_{J,K-1} $ in (\ref{eq:defseqheaps}) is determined by 
\begin{equation} \label{eq:defKjkVjk} 
 {\mathsf k}_{J,K-1}  =   {\mathsf H}^{-1}({\mathsf v}_{J,K-1} ) = {\mathsf H}^{-1}({\mathsf v})_{J,K-1} = 
  \langle {\mathsf H}^{-1}_{i}[ {\mathsf k}_{i}] \mid J \leq i \leq K-1 \rangle .
\end{equation} 
For example, if the subsequence ${\mathsf v}_{1,6}$ of pull downs is equal to 
$ \langle 1, 1, 1, 1, 2, 1 \rangle  $ then the corresponding subsequence ${\mathsf k}_{1,6}$ of moves is equal to: 
\begin{equation} \label{eq:defK16kV16} 
 {\mathsf k}_{1,6}  =   
  \langle {\mathsf H}^{-1}_{1}[1], {\mathsf H}^{-1}_{2}[1], {\mathsf H}^{-1}_{3}[1], 
  {\mathsf H}^{-1}_{4}[1], {\mathsf H}^{-1}_{5}[2], {\mathsf H}^{-1}_{6}[1] \rangle 
  = \langle 1, 2, 3, 4, 4, 5 \rangle .
\end{equation} 


 
 \medskip
 
 Given heap $ {\mathsf H}_J $, each of these subsequences will serve as a definition of the last heap $ {\mathsf H}_K $ in the subgame $ {\mathsf H}_{J,K} $. For instance, one can verify (a program can do it) that given the 1-element heap $ {\mathsf H}_1 $, the sequence of pull downs $ \langle 1, 1, 1, 1, 2, 1 \rangle  $, and the sequence of moves 
 $ \langle 1, 2, 3, 4, 4, 5 \rangle $, define the 7-element heap $ {\mathsf H}_7 $ visualized on Figure~\ref{fig:worst7}.
\medskip

Since $ {\tt RemoveMax} $ is a function\footnote{$ {\tt RemoveMax} $ is not a $ 1-1 $ function since the range of its restriction to $ N $-element heaps has a lesser cardinality than its domain.}, pull downs of two different patches $ v $ and $ w $ (for instance, pull down 1 and pull down 2) produce different heaps regardless whether they were applied to the same heap or not. For if both pull down $ v $ and pull down $ w $ produce the same heap $ H $, then the heap $ G $ that they were applied to is given by $ H. {\tt RemoveMax()}$ and, therefore, is unique, and so is the patch that was used to fill the vacancy after the largest element of $ H $ was removed. Thus, $ v = w $.  Hence, different sequences ${\mathsf v}_{J,K-1}$ of consecutive pull downs are never \textit{coalescing}\footnote{So that the graph of the game of Pull Downs is a tree.} in that they always produce different heaps $ {\mathsf H}_K $ from their residua $ {\mathsf H}_J $ no matter what the heaps $ {\mathsf H}_J $ are\footnote{As long as all moves in the said sequences are valid.}; a routine induction argument yields the proof. As a result, different games are never \textit{coalescing} as well\footnote{Thus heaps and games form trees of sequences whose property of never \textit{coalescing} implies the uniqueness of the path (from the root) to any given node.}.
\medskip

 In particular, every heap $ {\mathsf H}_K $  has the unique sequence ${\mathsf v}_{1,K-1}$ of consecutive pull downs which produce it ($ {\mathsf H}_K $, that is) from the heap $ {\mathsf H}_1 $.\footnote{The sequence $ {\mathsf v_{1,K-1}} $ can be constructed via the equality (\ref{eq:defVjkKjk}), where $ {\mathsf H_{1,K-1}} $ is the reversed sequence of consecutive residua of the heap $ {\mathsf H}_K $.}

\medskip

Thus there is a $ 1 - 1 $ correspondence between all heaps $ {\mathsf H}_K $ and sequences of valid applications of ${\mathsf H}_k . {\tt unRemoveMax}(i_k) $. In particular, the number of different heaps of size $ K $ is equal to the number of different sequences ${\mathsf v}_{1,K-1}$ of $ K-1 $ patches\footnote{Recall that not every node can serve as the $ k $th patch $  v_k$ since its index $ i = {\mathsf H}_{k}^{-1}[v_k] $ must satisfy an instance of  inequality (\ref{eq:Remconstrain1}) for $ n = k $.}.

\medskip

The above observations prove the following Theorem that will allow me to reduce the worst-case analysis of $ {\tt RemoveAll} $ to analysis of some winning strategies for the game of Pull Downs.

\begin{mapth} \label{thm:1-1heapPatch}
For every heap $ H $ of $ K \geq 2$ nodes and its residue $ \tilde{H} $ of $ J < K $ nodes, there is a  unique sequence  ${\mathsf v}_{J,K-1}$ of $ K-J $ consecutive pull downs that produce $ H $ from  $ \tilde{H} $; in particular, there is a unique sequence  ${\mathsf v}_{1,K-1}$ of $ K-1 $ consecutive pull downs that produce $ H $ from  the 1-element heap. 
\end{mapth}
{\bf Proof} follows from the above discussion.
\hspace*{\fill} $\Box$

\medskip

\label{def:creative} 
I will call the unique sequence of pull downs mentioned in Mapping 
 Theorem~\ref{thm:1-1heapPatch} 
a \textit{creative sequence} of $ H $ relative to $ \tilde{H} $ and denote it by 
$ \mathscr{S}_{\tilde{H}} (H)$. 
In the case of $ \tilde{H} $ bring the 1-element heap $ {\mathsf H}_1 $, will I call it simply a creative sequence and use notation $ \mathscr{S} (H)$ in lieu $ \mathscr{S}_{{\mathsf H}_1} (H)$.
By Mapping 
Theorem~\ref{thm:1-1heapPatch}, for any $ \tilde{H} $, $ \mathscr{S}_{\tilde{H}} (H)$ is a $ 1-1 $ function of $ H $ and, therefore, has the inverse  $ \mathscr{S}^{-1} _{\tilde{H}}$. 
\label{def:createdT}
I will denote it by  $ \mathscr{T} _{\tilde{H}}$, or, in the case of $ \tilde{H} = {\mathsf H}_1 $, by 
$ \mathscr{T} $.

\medskip

%

Thus $ \mathscr{S}_{\tilde{H}} (H)$ is the sequence $ {\mathsf v}  $ of pull downs that result in heap $ H $ when applied consecutively to heap $ \tilde{H} $, and $ \mathscr{T} _{\tilde{H}}( {\mathsf v}  )$ is the heap $ H $ created by application of the sequence $ {\mathsf v}  $ of consecutive pull downs to the  heap $ \tilde{H} $.

\medskip

It follows from the above definitions that $ \tilde{H} $ is a residue of $ H $ if, and only if, $ \mathscr{S} (\tilde{H})$ 
is an initial proper subsequence of $ \mathscr{S} (H)$ \footnote{Formally, if $ \mathscr{S} (\tilde{H}) \subseteq \mathscr{S} (H)$ and $ \mathscr{S} (\tilde{H}) \neq \mathscr{S} (H)$}.

\section{Credits for moves} \label{sec:cremov}

I define credit $ cr (i,k) $ for a 
move $ k $ applied to a heap $ H $ on $ i $ nodes
to be equal to the number $ C_{\tt RemoveMax()} (H ^{\prime}) $ of comparisons of keys that application
 of operation $ {\tt RemoveMax}() $ to a heap $ H ^{\prime} $ produced by $H. {\tt unRemoveMax}(k) $ will perform. 
 By the  {\tt RemoveMax} Cost Lemma~\ref{lem:newRemMax} page~\pageref{lem:newRemMax}, it is given by the equality (\ref{eq:newRemMaxcredit}) page~\pageref{eq:newRemMaxcredit}.
It is a function of $ k $ and the number $ i $ of nodes of $ H $.

\medskip

The equation (\ref{eq:newRemMaxcredit}) 
page \pageref{eq:newRemMaxcredit} 
can be rewritten to a human-readable form as follows, substituting $ i $ for $ N $ and $ k $ for $ i $. 
\medskip

If $ k $ is an index of a leaf with a sibling then
\[ \#_{\lfloor \frac{k}{2} \rfloor}^i = 2, \]
\begin{equation} \label{eq:credit2} 
\#_k ^i = 0 ,
\end{equation}
 and, by (\ref{eq:creditFH}),
\begin{equation} \label{eq:credLeaf1} 
cr(i,k) =  2 \lfloor \lg k \rfloor.
\end{equation}
One can verify by means of direct inspection that in this case the right-hand side of the equation (\ref{eq:credLeaf1}) also yields the credit for pulling down the parent of the node at index $ k $,
that is,
\begin{equation} \label{eq:credParent1} 
cr(i,\lfloor \frac{k}{2} \rfloor) =  cr(i,k) =2 \lfloor \lg k \rfloor.
\end{equation}
  If $ k > 1 $ is an index of a leaf with no sibling then then
\[ \#_{\lfloor \frac{k}{2} \rfloor}^i = 1, \]
 and, by (\ref{eq:creditFH}),
\begin{equation} \label{eq:credLeaf2} 
cr(i,k) =  2 \lfloor \lg k \rfloor - 1.
\end{equation}
As before, the right-hand side of the equation (\ref{eq:credLeaf2}) also yields the credit for pulling down the parent of the node at index $ k $ in this case, so
\begin{equation} \label{eq:credParent2} 
cr(i,\lfloor \frac{k}{2} \rfloor) = cr(i,k) = 2 \lfloor \lg k \rfloor - 1.
\end{equation}
Also,
\begin{equation} \label{eq:creditUB1} 
cr(1,1) = 0 .
\end{equation}
Since $ cr^{max} (i) $, defined by
\begin{equation} \label{eq:creditUB} 
cr^{max} (i) = \max \{ cr(i,k) \mid k \leq i \} , 
\end{equation}
is, by virtue of the definition of $ cr(i,k) $ at the beginning of this Section, equal to the maximum number $ C_{\tt RemoveMax()} ^{\tt max} (i+1) $ of comparisons of keys that the $ {\tt RemoveMax}() $ may perform on any heap on $ i+1 $ 
nodes\footnote{In particular, the heap produced by a pull down applied to a heap on $ i $ nodes.}, application of the equality
(\ref{eq:RemMaxmaxcredit}) page \pageref{eq:RemMaxmaxcredit} yields 
\begin{equation} \label{eq:creditUBform} 
cr^{max} (i) = \lfloor \lg i \rfloor +  \lfloor \lg( i-1)  \rfloor . 
\end{equation}

%

\medskip

The \textit{loss of credit} $ \lambda (i,k) $ relative to the said maximum $cr^{max} (i)$ for a move $ k $ in a heap on $ i $ nodes is defined as:
\begin{equation} \label{eq:deflamb}
\lambda (i,k) = cr^{max} (i) - cr(i,k).
\end{equation}
If $ \lambda (i,k) = 0 $ then the move $ k $ and the pull down $ H[k] $ that yielded the credit $ cr(i,k)  $  are called \textit{lossless}; otherwise, they are called \textit{lossy}. 

\medskip

The following Lemma characterizes lossless moves.

\begin{losslm} \label{lem:losslessM}
Let $ H $ be a heap of $ N $ nodes and let $ k $ be a valid move (that is, one that satisfies the inequality 
(\ref{eq:Remconstrain1}) page~\pageref{eq:Remconstrain1}). Move $ k $ is lossless if, and only if, one or more of the following conditions are true:

\begin{enumerate}
     \renewcommand\labelenumi{\theenumi}
     \renewcommand{\theenumi}{(\roman{enumi})}

\item \label{item:losslessM1} $ k = 2^{\lfloor \lg N \rfloor}$ , or 

\item \label{item:losslessM2} $ k =  2^{\lfloor \lg N \rfloor-1} $, or

\item  \label{item:losslessM3} $ k $ is a sibling node\footnote{Formally, $2 \lfloor  \frac{k}{2} \rfloor  < N$.}
 in the last level \footnote{Formally, $2 ^{\lfloor \lg N \rfloor} \leq k$.}
of $ H $, or

\item \label{item:losslessM4} $ k $ is the parent of a sibling node\footnote{Formally, $ 2 k  < N$. }
 of the last level \footnote{Formally, $2 ^{\lfloor \lg N \rfloor} \leq 2 k $. }
of $ H $.

\end{enumerate}

\end{losslm}
{\bf Proof} First, I will prove the \textit{if} part of the Lemma.
\medskip
\\
\ref{item:losslessM3}
Since $ k $ belongs to the last level of $ H $, $2 ^{\lfloor \lg N \rfloor} \leq k$. Thus, by equality (\ref{eq:credLeaf1}) page~\pageref{eq:credLeaf1}, 
\[ cr(N,k) \geq  2 \lfloor \lg 2 ^{\lfloor \lg N \rfloor} \rfloor = 2 \lfloor \lg N \rfloor \geq \]
[by equality (\ref{eq:creditUBform})]
\[ \geq cr^{max} (N) . \]
Thus,
\[ cr(N,k) = cr^{max} (N) . \]
Hence, by the equality (\ref{eq:deflamb}),
\[  \lambda (N,k) = 0 .\]
\medskip
\\
\ref{item:losslessM4}
has a similar proof except that it begins with the equality (\ref{eq:credParent1}) page~\pageref{eq:credParent1} rather than with (\ref{eq:credLeaf1}).
\medskip
\\
\ref{item:losslessM1}
Since $ k = 2^{\lfloor \lg N \rfloor}$, $ k $ belongs to the last level of $ H $, therefore is a leaf in that level. If $ k $ has a sibling then the case \ref{item:losslessM3} applies, which completes the proof of this case. If $ k $ has no sibling then its parent
$ \lfloor \frac{k}{2} \rfloor $ has one child only, that is
\[N =  2\lfloor \frac{k}{2} \rfloor = 2\lfloor \frac{2^{\lfloor \lg N \rfloor}}{2} \rfloor
= 2\lfloor 2^{\lfloor \lg N \rfloor-1} \rfloor = 2\times 2^{\lfloor \lg N \rfloor-1} =
2^{\lfloor \lg N \rfloor}.\]
Thus
\[N =2^{\lfloor \lg N \rfloor}.\]
Hence,
\[\lfloor \lg (N-1) \rfloor = \lfloor \lg (2^{\lfloor \lg N \rfloor}-1) \rfloor =\]
[by equality (\ref{eq:basic2}) page~\pageref{eq:basic2}]
\[=\lfloor \lg (2^{\lfloor \lg N \rfloor}) \rfloor - 1 =  \lfloor \lfloor \lg N \rfloor \rfloor - 1
= \lfloor \lg N \rfloor  - 1 ,\]
that is,
\begin{equation} \label{eq:losslessM100}
\lfloor \lg (N-1) \rfloor = \lfloor \lg N \rfloor  - 1 .
\end{equation}
By the equality (\ref{eq:credLeaf2}) page~\pageref{eq:credLeaf2},
\[ cr(i,k) =  2 \lfloor \lg 2^{\lfloor \lg N \rfloor} \rfloor - 1 =
2 \lfloor  \lfloor \lg N \rfloor \rfloor - 1  = 2  \lfloor \lg N \rfloor  - 1 =\]
[by equality (\ref{eq:losslessM100})]
\[ = \lfloor \lg N \rfloor + \lfloor \lg (N-1) \rfloor =  \]
[by equality (\ref{eq:creditUBform}) page~\pageref{eq:creditUBform}]
\[ = cr^{max} (i) .\]
Thus,
\[ cr(N,k) = cr^{max} (N) . \]
Hence, by the equality (\ref{eq:deflamb}),
\[  \lambda (N,k) = 0 .\]
\medskip
\\
\ref{item:losslessM4} follows from \ref{item:losslessM1} by virtue of equality (\ref{eq:credParent1}), in the case $ k $ has no sibling, or equality (\ref{eq:credParent1}), otherwise.
 This completes the proof of the \textit{if} part of the Lemma.
  
 \medskip
 
 If $ k $ is neither a leaf nor the parent of a leaf, which means that $ 4k \leq N $, then, by the equality (\ref{eq:no_of_childrenFH}) page~\pageref{eq:no_of_childrenFH},
\begin{equation} \label{eq:losslessM200}
\#_{ 2k }^N > 0.
\end{equation}
By virtue of equality (\ref{eq:creditFH})
page~\pageref{eq:creditFH},
\begin{equation} \label{eq:losslessM300}
cr(N, k) =  2  \lfloor \lg  k  \rfloor - 2   + \#_{\lfloor \frac{k}{2} \rfloor}^N + \#_k ^N
\end{equation}
and
\[  cr(N, 2k) =  2  \lfloor \lg 2 k  \rfloor - 2  + \#_{\lfloor k \rfloor}^N + \#_{2k} ^N = \]
\[ =  2  (\lfloor \lg k \rfloor +1) - 2 + \#_{ k }^N + \#_{2k} ^N = \]
\[ =  2  \lfloor \lg k \rfloor   + \#_{ k }^N + \#_{2k} ^N  .\]
Thus
\begin{equation} \label{eq:losslessM400}
cr(N, 2k) =  2  \lfloor \lg k \rfloor   + \#_{ k }^N + \#_{2k} ^N  .
\end{equation}
Subtracting (\ref{eq:losslessM300}) from (\ref{eq:losslessM400}), we get
\[cr(N, 2k)-cr(N, k) = 2 + \#_{ 2k }^N - \#_{\lfloor \frac{k}{2} \rfloor}^N \geq \#_{ 2k }^N. \]
Hence, by (\ref{eq:losslessM200}),
\[cr(N, 2k)> cr(N, k)\]
and, therefore,
\[\lambda(N,k)>0, \]
thus making $ k $ a lossy move.
\medskip
\\
If $ k $ is a leaf but not in the last level of $ H $ then 
$ \lfloor \lg  k  \rfloor = \lfloor \lg  N  \rfloor - 1 $, so, by virtue of equality (\ref{eq:creditFH})
page~\pageref{eq:creditFH},
\[  cr(N, k) =  2  (\lfloor \lg  N  \rfloor -1) - 2   + \#_{\lfloor \frac{k}{2} \rfloor}^N  
=  2 \lfloor \lg  N  \rfloor - 4   + \#_{\lfloor \frac{k}{2} \rfloor}^N  \leq \]
\[ \leq 2 \lfloor \lg  N  \rfloor - 2
< \lfloor \lg  N  \rfloor + \lfloor \lg  (N-1)  \rfloor = cr^{max}(N). \]
Thus
\[cr(N, 2k)> cr(N, k)\]
and, therefore,
\[\lambda(N,k)>0, \]
thus making $ k $ a lossy move.
\medskip
\\
If $ k $ is a parent thereof then,  by virtue of equality (\ref{eq:credParent1}), in the case $ k $ has no sibling, or equality (\ref{eq:credParent1}), otherwise,
\[cr(N, k) = cr(N, 2k)\]
and the same conclusion as for the previous case follows.
\medskip
\\
Since there are no other cases, this completes the prof of the \textit{only if} part, which completes the proof of the Lemma.
\hspace*{\fill} $\Box$

\medskip

One can immediately conclude from the above Credit Loss Characterization Lemma~\ref{lem:losslessM}~\ref{item:losslessM1} that for every $ i\geq 2 $, there is a heap $ H $ of $ i $ nodes
   and an index $ k $ such that pull down $ H[k] $ yields the maximum credit 
 $ cr(i,k) $ $ = $ $ cr^{max} (i) $\footnote{For instance, pull down 1 in a heap of $ i $ nodes that satisfies the equality 
 $ H[2^{\lfloor \lg i \rfloor}] = 1 $ yields the maximal credit $ cr^{max} (i) $.}, turning the inequality (\ref{eq:creditUB}) into equality for such $ i $, $ H $, and $ k $, and, therefore, making that inequality \textit{tight} for every $ i \geq 2 $.

\medskip

Function $ \lambda $ will allow me for easy evaluation of payoffs for some strategies.
\medskip

Let's consider for example a sequence of $ N+1 $ alternating pull downs $ \langle 1, 2,1, 2, ... \rangle$
 applied consecutively to a complete heap on $ N $ nodes. If $ N =1  $ then both moves are valid and lossless. For $ N \geq 2 $, since nodes 1 and 2 are the smallest nodes in any heap on N nodes, all these moves are valid as well, that is, the inequality (\ref{eq:Remconstrain1}) page~\pageref{eq:Remconstrain1} is satisfied for every $ N+1 \leq n \leq 2N+1 $ and $ i = H^{-1}_n[1], H^{-1}_n[2] $. It turns out that in the latter case, all these moves are also lossless, except, perhaps, for the second move that may lose 
 1~credit to $cr^{max} (n)$, as one can conclude from the following Lemma.

\begin{pqlm} \label{lem:1212}
Let $ H $ be a complete heap on $ N = 2^{\lceil \lg N \rceil}-1 $ nodes, let $ p, q $, with $ p < q \leq H [\frac{N+1}{2}]  $, be its leaves, and
let $ {\mathsf s} = \langle \mathsf{s}_i \mid 1 \leq i \leq M \rangle $, where $ 2 \leq M \leq N+1 $, be a sequence of $ M $ pull downs of $p$ and $q$,  applied consecutively to $ H $, that is 
given by this regular expression
\begin{equation} \label{eq:1212regexpr} 
{
\bf p
q}({\bf 
p
q} + {\bf 
q
p}) ^{*} 
+
{\bf p
q}({\bf 
p
q} + {\bf 
q
p}) ^{*} 
({\bf 
p
} + {\bf 
q
}).
\end{equation}

\begin{enumerate}
     \renewcommand\labelenumi{\theenumi}
     \renewcommand{\theenumi}{(\roman{enumi})}

\item \label{lem:1212i} 
All moves of $ {\mathsf s} $, except, perhaps, for move $ \mathsf{s}_2 $, are lossless.
\item \label{lem:1212ii} 
If  $ H [\frac{N+1}{2}] = q$ then move $ \mathsf{s}_2 $ is lossless.
\item \label{lem:1212iii} 
If  $ H [\frac{N+1}{2}] > q$ then move $ \mathsf{s}_2 $ 
loses $ 1 $ credit relative to the maximum 
$ 2\lfloor \lg N \rfloor$ for that move.
\end{enumerate} 
\end{pqlm}
{\bf Proof}.
 Let $ {\mathsf H} $ $ = $ $ \mathscr{G}_{H} ( {\mathsf s}) $ 
 \footnote{Function $  \mathscr{G}_{H} $  was defined by (\ref{eq:defseqheaps}) page \pageref{eq:defseqheaps}.}
be the sequence of heaps
produced by pull downs $ {\mathsf s} $ applied to $ H $. 

\medskip

Move $\mathsf{s}_1$ (the first move in the described sequence) pulls down $ p $, which resides in the last level $ \lfloor \lg N \rfloor $ of $H= {\mathsf H}_1 $ as a leaf with a sibling, down to index $ N+1 $ in $ {\mathsf H}_{2} $. 
By the Credit Loss Characterization  Lemma~\ref{lem:losslessM}~\ref{item:losslessM3}, 
move  $\mathsf{s}_1$ is lossless, with
\begin{equation} \label{eq:lampar1}
\lambda(N, \mathsf{s}_1) = 0.
\end{equation} 
Move $\mathsf{s}_{2}$ pulls down $ q $, which either resides in $ {\mathsf H}_{2} $ at index $ i = \frac{N+1}{2} $  as the parent of $ p $ (that has been pulled down to index $ N+1 $ by move $ \mathsf{s}_1 $) or as a leaf with a sibling.
In the latter case, it scores the same credit as move $ \mathsf{s}_1 $ did, that is,
\[ cr(N+1,\mathsf{s}_{2}) =  2\lfloor \lg N \rfloor   = \]
[since  $ N = 2^{\lfloor \lg N \rfloor + 1}-1 $ and, therefore, $ \lfloor \lg N \rfloor = \lfloor \lg (N+1) \rfloor -1 $]
\[ \lfloor \lg (N+1) \rfloor + \lfloor \lg N \rfloor -1.\]
Thus,
\begin{equation} \label{eq:parcr_i+1}
cr(N+1,\mathsf{s}_{2}) = \lfloor \lg (N+1) \rfloor + \lfloor \lg N \rfloor -1 .
\end{equation}
This, by (\ref{eq:creditUBform}) and  by inequality (\ref{eq:creditUB}) on page~\pageref{eq:creditUB}, yields one less than the maximum 
$ \lfloor \lg (N+1) \rfloor + \lfloor \lg N \rfloor $ 
of credit $ cr(N+1,k) $ for any $ k \leq  N+1$, thus making move $\mathsf{s}_2$ lossy, with
\begin{equation} \label{eq:lampar2}
\lambda (N+1, \mathsf{s}_2) = 1.
\end{equation}
This completes the proof of case~\ref{lem:1212iii}.

\medskip

In the former case ($ q $ resides in $ {\mathsf H}_{2} $ at index $ i = \frac{N+1}{2} $
$ = $  $ 2^{\lceil \lg N \rceil-1} $ as the parent of its only child $ p $), 
move $\mathsf{s}_2$ pulls down $ q $.  
By the Credit Loss Characterization  Lemma~\ref{lem:losslessM}~\ref{item:losslessM2}, 
move  $\mathsf{s}_2$ is lossless, with
\begin{equation} \nonumber 
\lambda(N, \mathsf{s}_1) = 0.
\end{equation} 
This completes the proof of case~\ref{lem:1212ii}.

\medskip
 
Moves $s _{2j-1} $ and $s _{2j} $, 
where $ 2 \leq j \leq  \frac{N+1}{2}  $ pull down $ p $ and $ q $, which reside in level 
$ \lfloor \lg (N+1) \rfloor $ of heaps $ {\mathsf H}_{2j-1} $ and 
 $ {\mathsf H}_{2j} $ as leafs with siblings.
 By the Credit Loss Characterization  Lemma~\ref{lem:losslessM}~\ref{item:losslessM3}, 
each of those moves is lossless, with
\begin{equation} \nonumber 
\lambda(N, \mathsf{s}_{2j-1}) = \lambda(N, \mathsf{s}_{2j}) = 0.
\end{equation} 
This, together with (\ref{eq:lampar1}), completes the proof of case~\ref{lem:1212i}.
\hspace*{\fill} $\Box$

\medskip

I will use special cases of sequences $ 
{\bf p
q}({\bf 
p
q} + {\bf 
q
p}) ^{*} $,
namely, 
$ p = 1, q = 2 $, in establishing some important constrains on optimal strategies, and later in Section~\ref{ProofWin}, 
 $ p = 1, q = 4 $, in design of an optimal strategy for the game of pull downs.

\section{Strategies and their payoffs} \label{strapay}

A \textit{strategy} ${\mathsf s}$ is an infinite sequence $\langle \mathsf{s}_ {n+1} \mid n \in \omega \rangle$ of valid moves consecutively applied to the 1-element heap $ {\mathsf H}_1 $.
Given a strategy ${\mathsf s} $,  I will call the sum of the the credits, defined in Section~\ref{sec:cremov} page~\pageref{sec:cremov}, for all moves of ${\mathsf s}$ between $ n $-th and $ m $-th move\footnote{Racall
that $ ith $ move of any strategy produces an $ i+1 $-element heap out of an $ i $-element heap}, inclusively, where $ 2 \leq n \leq m $, the  the \textit{payoff} $ P _{\mathsf s} (n,m) $. It is given by this formula:
\begin{equation} \label{eq:payoff} 
P _{\mathsf s} (n,m) = \sum _{i=n} ^m cr(i,\mathsf{s}_ i).
\end{equation}

\medskip

The following lemma shows that the payoff for the game, defined on page~\pageref{def:playerPayoff},  that the Player plays with a strategy $ \mathsf{s} $ is equal to the payoff \linebreak $ P _{\mathsf s} (2,N-1) $ for the strategy $ \mathsf{s} $.


\medskip

\begin{racrlm} \label{lem:RemAllcredit}
	Let $ H $ be a heap of $ N \geq 2 $ nodes and $\mathsf{s}$ be its creative sequence.
	\begin{equation} \label{eq:RemAll10}
	C_{{\tt RemoveAll()}} (H) = \sum _{i = 2} ^{N-1} cr(i, \mathsf{s}_i).
	\end{equation}
\end{racrlm} 
{\bf Proof} by induction on $ N $. For $ N = 2 $, both sides of (\ref{eq:RemAll10}) are 0, which observation completes the basis step.

\medskip

Let (\ref{eq:RemAll10}) be true for some $ N \geq 2 $, $ H $ be a heap of $ N+1 $ nodes, $\mathsf{s}$ be its creative sequence, $ H^{\prime} $ be the heap produced by execution of $ H.{\tt RemoveAll()} $, and 
$\mathsf{s}^{\prime}$ be the creative sequence for $ H^{\prime} $. By the above definition of $ H^{\prime} $,
\begin{equation} \label{eq:RemAll20}
C_{{\tt RemoveAll()}} (H) = C_{{\tt RemoveAll()}} (H^{\prime}) + C_{{\tt RemoveMax()}} (H).
\end{equation}
By the definition of creative sequence at the end of Section~\ref{sec:pull} on page~\pageref{def:creative}, \linebreak $\mathsf{s} _{1, N}$ $ = $ $\mathsf{s}^{\prime}$.
Since $ H^{\prime} $ has $ N $ nodes, by the inductive hypothesis we get
\begin{equation} \label{eq:RemAll30}
C_{{\tt RemoveAll()}} (H^{\prime}) = \sum _{i = 2} ^{N-1} cr(i, \mathsf{s}^{\prime}_i). 
\end{equation}
Substituting (\ref{eq:RemAll30}) to (\ref{eq:RemAll30}), we obtain
\[ C_{{\tt RemoveAll()}} (H) = \sum _{i = 2} ^{N-1} cr(i, \mathsf{s}^{\prime}_i) + C_{{\tt RemoveMax()}} (H)  = \]
[by the definition of credit at the beginning of Section~\ref{sec:cremov} page~\pageref{sec:cremov}]
\[  = \sum _{i = 2} ^{N-1} cr(i, \mathsf{s}^{\prime}_i) + cr(N, \mathsf{s}_N)  = \]
\[  = \sum _{i = 2} ^{N} cr(i, \mathsf{s}^{\prime}_i) ,\]
which yields (\ref{eq:RemAll10}), thus completing the inductive step.
\hspace*{\fill} $\Box$

\medskip

If $ m = 2^{\lfloor \lg m \rfloor+1}-2 $ and $ n = 2^{\lfloor \lg m \rfloor}-1 $ then all moves between $ n $-th and $ m $-th move, inclusively, construct the level $ \lfloor \lg m \rfloor $ of the output heap. In such a case,
both the argument of the $ n $-th move and the result of the $ m $-th move of $ {\mathsf s} $ are consecutive\footnote{The difference between their depths is equal to 1.} complete heaps; I will call the payoff
$ P _{\mathsf s} (n,m) $, given by the equality (\ref{eq:payoff}), a \textit{level payoff} and denote it by $ P^{lev} _{\mathsf s} (\lfloor \lg m \rfloor) $. Formally,
\begin{equation} \label{eq:levelpayoff} 
P^{lev} _{\mathsf s} (K)  = \sum _{i=2^K-1} ^{2^{K+1}-2} cr(i,\mathsf{s}_ i).
\end{equation}
Informally, it is the sum of credits of moves $ {\mathsf s}_{2^K-1,2^{K+1}-2} $ of $ {\mathsf s} $ that added the next level $ K $ to the complete heap $\mathscr{T} ({\mathsf s} _{1,2^K-2})$ \footnote{Function $\mathscr{F}$ has been defined on page \pageref{def:createdT}.} on $ 2^K-1 $ nodes created by the first $ 2^K-2 $ moves $ {\mathsf s} _{1,2^K-2} $ of $ {\mathsf s} $ consecutively applied to the 1-element heap $ H_1 $.

\medskip 

Similarly, I will call the sum of the the losses for all moves of ${\mathsf s}$ between $ n $-th and $ m $-th move, inclusively, where $ 2 \leq n \leq m $, the  the \textit{accumulated loss} $ \Lambda_{\mathsf s} (n,m) $. It is given by this formula:
\begin{equation} \label{eq:defL} 
\Lambda_{\mathsf s} (n,m) = \sum _{i=n} ^m \lambda _{\mathsf s} (i),
\end{equation}
where
\begin{equation} \label{eq:lamb=lambs} 
\lambda _{\mathsf s} (i) = \lambda  (i, \mathsf{s}_ i) ,
\end{equation} 
with abbreviated notation
\begin{equation} \label{eq:defLabbr} 
\Lambda_{\mathsf s} (m) = \Lambda_{\mathsf s} (2,m)
\end{equation}
that yields
\begin{equation} \label{eq:defLabbr2} 
\Lambda_{\mathsf s} (n, m) = \Lambda_{\mathsf s} (m) - \Lambda_{\mathsf s} (n-1) .
\end{equation}
If $ \Lambda_{\mathsf s} (n,m) = 0 $ then I call the sequences $ \langle k_i \mid n \leq i \leq m \rangle $ of moves and $ \langle {\mathsf H}_i[k_i] \mid n \leq i \leq m \rangle $ of pull downs of $ \mathsf s $ \textit{lossless}; otherwise, I call those sequences \textit{lossy}.

\medskip

The \textit{level loss} at level
$ K \geq 1 $ is given by this formula:
\begin{equation} \label{eq:levelloss} 
\Lambda^{lev} _{\mathsf s} (K)  = \Lambda_{\mathsf s} (2^K-1,2^{K+1}-2) .
\end{equation}
If $ \Lambda^{lev} _{\mathsf s} (K)  = 0 $ then I call the level $ K $ \textit{lossless}; otherwise, I call it \textit{lossy}.

\medskip

It follows directly from the definition (\ref{eq:levelloss}) that for any natural number $ D $:

\begin{equation} \label{eq:sumlevellossComplNew} 
\sum _{K=1} ^{D} \Lambda^{lev} _{\mathsf s} (K)  = \Lambda_{\mathsf s} (2,2^{D+1}-2)
= \Lambda_{\mathsf s} (2^{D+1}-2) .
\end{equation}

\medskip

If $ M $ is the size of the largest complete heap of no more than $ m $ nodes and $ D $ is the depth of that heap then
\[  \Lambda _{\mathsf s} (2,m-1) =  \Lambda _{\mathsf s} (2,M-1) +  \Lambda _{\mathsf s} (M,m-1) =
\]
[by (\ref{eq:sumlevellossComplNew}), taking into account that $ M = 2^{\lfloor \lg M \rfloor+1}-1 $]
\[ =  \sum _{K=1} ^{D}  \Lambda^{lev} _{\mathsf s} (K) + \Lambda _{\mathsf s} (M,m-1) =
\]
[by virtue of equalities (\ref{eq:completeCharactNodes}) and (\ref{eq:completeCharactDepth})
page~\pageref{eq:completeCharactDepth}]
\[ =  \sum _{K=1} ^{\lfloor \lg (m+1) \rfloor-1}  \Lambda^{lev} _{\mathsf s} (K) +
\Lambda _{\mathsf s} (2^{\lfloor \lg (m+1) \rfloor}-1,m-1) .
\]
Thus, for any $ m \geq 3 $
\begin{equation} \label{eq:sumlevelloss} 
\Lambda _{\mathsf s} (2,m-1) = \sum _{K=1} ^{\lfloor \lg (m+1) \rfloor-1}\Lambda^{lev} _{\mathsf s} (K)
+  
\Lambda _{\mathsf s} (2^{\lfloor \lg (m+1) \rfloor}-1,m-1).
\end{equation}
If, moreover, $ m = 2^{\lfloor \lg m \rfloor+1}-1 $ then $ \lfloor \lg (m+1) \rfloor-1 $ $ = $ $ \lfloor \lg m \rfloor $ and $ 2^{\lfloor \lg (m+1) \rfloor}-1 $ $ = $ $ m $, so that $ \Lambda _{\mathsf s} (2^{\lfloor \lg (m+1) \rfloor}-1,m-1) $ $ = 0 $, and (\ref{eq:sumlevelloss}) reduces to 
\begin{equation} \label{eq:sumlevellossCompl} 
\Lambda _{\mathsf s} (2,m-1) = \sum _{K=1} ^{\lfloor \lg m \rfloor}\Lambda^{lev} _{\mathsf s} (K).
\end{equation}

\medskip

If $ \tilde{H} $ is a residue of heap $ H $ then the payoff earned by the creative sequence $ \mathscr{S}_{\tilde{H}} (H)$ \footnote{Function $ \mathscr{S}  $ defined on page~\pageref{def:creative}.} is given by: 
\begin{equation} \label{eq:PHH0} 
P_{\tilde{H} \! H } = \sum _{i=n} ^{m-1} cr(i,\mathscr{S}_{\tilde{H}} (H)_i) ,
\end{equation}
where $ n $ and $ m $ are the sizes of heaps $ \tilde{H} $ and $ H $ respectively. If $ \tilde{H} $ is the 1-element heap then I will use $ P_{H } $ as an abbreviation of  $ P_{\tilde{H} \! H } $. By 
the equality (\ref{eq:RemAll10}) in the {\tt RemoveAll} Cost Lemma~\ref{lem:RemAllcredit} page \pageref{lem:RemAllcredit},
\begin{equation} \label{eq:PHH1}
P_{\tilde{H} \! H } 
= C_{\tt RemoveAll()} (H) - C_{\tt RemoveAll()} (\tilde{H}),
\end{equation}
 where $C_{\tt RemoveAll()} (F)$
is the number of comparisons of keys performed by the execution of $ F.{\tt RemoveAll}() $ on heap $ F $.
Thus,
\begin{equation} \label{eq:PHH2} 
P_{\tilde{H} \! H } 
= P_{\mathsf s} (n,m-1) ,
\end{equation}
where $ \mathsf s $ is any strategy whose $ n-1 $st move produces heap  $ \tilde{H} $ and $ m-1 $-st move produces heap $ H $.
\medskip

Under the same assumptions as above, the accumulated loss (of credits) for the creative sequence $ \mathscr{S}_{\tilde{H}} (H)$
is given by:
\begin{equation} \label{eq:LambdaHH0} 
\Lambda_{\tilde{H} \! H } = \sum _{i=n} ^{m-1} \lambda(i,\mathscr{S}_{\tilde{H}} (H)_i) .
\end{equation}
If $ \tilde{H} $ is the 1-element heap then I will use $ \Lambda_{H } $ as an abbreviation of  $ \Lambda_{\tilde{H} \! H } $. 
on heap $ F $.
Thus,
by (\ref{eq:defL}),
it satisfies this equality:
\begin{equation} \label{eq:lambdaHH2} 
\Lambda_{\tilde{H} \! H } 
= \Lambda_{\mathsf s} (n,m-1) ,
\end{equation}
where $ \mathsf s $ is any strategy whose $ n-1 $st move produces heap  $ \tilde{H} $ and $ m-1 $-st move produces heap $ H $.
\medskip

I will call the payoff $ P ^{\,U\!B} (n,m) $ yielded by the sum of credits that match the maximum
$cr^{max} (i)$ given by  (\ref{eq:creditUBform}) page~\pageref{eq:creditUBform} for all moves of between $ n $ and $ m $, inclusively, where $ 2 \leq n \leq m $, the \textit{upper-bound payoff}. It is given by this formula:
\begin{equation} \label{eq:ubpayoff} 
P ^{\,U\!B} (n,m) = \sum _{i=n} ^m cr ^{max}(i) = \footnote{By vitrue of (\ref{eq:creditUBform}) page \pageref{eq:creditUBform}.}
\sum _{i=n} ^m (\lfloor \lg i \rfloor +  \lfloor \lg( i-1)  \rfloor). \; 
\footnote{It is known that for $ n = 2 $, the right-hand side of (\ref{eq:ubpayoffn2m-1}) reduces to a closed-form formula $ (2m-1)\lfloor \lg (m-1) \rfloor
- 2 ^{\lfloor \lg (m-1) \rfloor +2}+4$; for instance, proof of Theorem 8.1 in \cite{suc:elem} provides a derivation of it. Thus for any $ n > 2 $, the right-hand side of (\ref{eq:ubpayoff}) reduces to a closed-form formula $ (2m+1)\lfloor \lg m \rfloor
- 2 ^{\lfloor \lg m \rfloor +2} - (2n-1)\lfloor \lg (n-1) \rfloor
+ 2 ^{\lfloor \lg (n-1) \rfloor +2} $.}
\end{equation}

\medskip

For any heap $ H $ on $ m $ nodes and its residue $ \tilde{H} $ on $ n < m $ nodes we have:

\[   P ^{\,U\!B} (n,m-1) - \Lambda_{\tilde{H} \! H }  = \]
[by (\ref{eq:ubpayoff}) and (\ref{eq:LambdaHH0})]
\[ = \sum _{i=n} ^m cr ^{max}(i) -  \sum _{i=n} ^{m-1} \lambda(i,\mathscr{S}_{\tilde{H}} (H)_i) =  \]
[by (\ref{eq:deflamb})]
\[ = \sum _{i=n} ^m cr ^{max}(i) -  \sum _{i=n} ^{m-1} ( cr ^{max}(i) -cr(i,\mathscr{S}_{\tilde{H}} (H)_i)) =  
\sum _{i=n} ^{m-1} cr(i,\mathscr{S}_{\tilde{H}} (H)_i) = \]
[by (\ref{eq:PHH0})]
\[ =  P_{\tilde{H} \! H } .\]
Thus
\begin{equation} \label{eq:PHH=PUB-LambHH}
 P_{\tilde{H} \! H} =  P ^{\,U\!B} (n,m-1) - \Lambda_{\tilde{H} \! H } .
\end{equation}
\medskip

As a special case of (\ref{eq:ubpayoff}), we have:

\begin{equation} \label{eq:ubpayoffn2m-1}
P ^{\,U\!B} (2,m-1) =  \sum _{i=2} ^{m-1} (\lfloor \lg i \rfloor +  \lfloor \lg( i-1)  \rfloor) = 2 \sum _{i=2} ^{m-1} \lfloor \lg i \rfloor - \lfloor \lg (m-1) \rfloor,
\end{equation}

which, by virtue of  equality (20) in \cite{suc:elem}, yields:

\begin{equation} \label{eq:PUBremAll}
P ^{\,U\!B} (2,m-1) \geq C_{\tt RemoveAll()} ^{\tt max} (m).
\end{equation}
Thus $ P ^{\,U\!B} (2,m-1) $ is and upper bound for $ C_{\tt RemoveAll()} ^{\tt max} (m) $. 
 
\medskip

The definitions (\ref{eq:deflamb}), (\ref{eq:payoff}), (\ref{eq:defL}) and (\ref{eq:ubpayoff}), yield for every $ 2 \leq n \leq m $:

\begin{equation} \label{eq:payoffbound} 
P _{\mathsf s} (n,m) + \Lambda_{\mathsf s} (n,m) = P ^{\,U\!B} (n,m).
\end{equation}

\medskip

In particular, by virtue of (\ref{eq:creditUBform}) and  the inequality (\ref{eq:creditUB}) page~\pageref{eq:creditUB}, we have for every $ 2 \leq n \leq m $:
\begin{equation} \label{eq:payoffbound2} 
P _{\mathsf s} (n,m) \leq P ^{\,U\!B} (n,m).
\end{equation}

Equation (\ref{eq:payoffbound}) will allow me for easy derivation of the value of payoff $ P _{\mathsf s} (n,m) $ from its upper bound $ P ^{\,U\!B} (n,m) $, given by a known closed-form formula, once I have computed the loss of credit
$ \Lambda_{\mathsf s} (n,m) $.

\medskip

Thus the Player draws at random an integer $ N \geq 2$, choses any strategy $ {\mathsf w} $, and plays the first $ N-1 $ moves of it. By the definition of Player's payoff on page~\pageref{def:playerPayoff}, his total payoff 
is equal to the number of comparisons that ${\mathsf H}_{N}. {\tt RemoveAll()} $ will perform while run on the heap $ {\mathsf H}_{N} = \mathscr{T} ({\mathsf w}_{1,N-1}) $ \footnote{\label{foo:funTdef} The function $ \mathscr{T} $ has been defined on page~\pageref{def:createdT}.} produced by the $ N-1 $st move of the strategy $ {\mathsf w} $. This, by virtue of the definition (\ref{eq:payoff}) page~\pageref{eq:payoff} of $ P_{\mathsf w}(N-1) $ and the $ {\tt RemoveAll} $ Cost Lemma~\ref{lem:RemAllcredit} page~\pageref{lem:RemAllcredit} is equal to 
\begin{equation} \label{eq:payofftot} 
P_{\mathsf w}(N-1) = P_{\mathsf w}(2, N-1) =  \sum _{i=2} ^{N-1} cr(i,w_i).
\end{equation} 
His total loss of credit, defined as
\begin{equation} \label{eq:lossftot} 
\Lambda_{\mathsf w}(N-1) = \Lambda_{\mathsf w}(2, N-1) =  \sum _{i=2} ^{N-1} \lambda(i,w_i),
\end{equation}
determines whether the strategy ${\mathsf w}$ he choose is \textit{optimal}, according to the following Lemma.

\medskip

\begin{lem} \label{lem:winmin} 
	A strategy ${\mathsf s} $ is optimal for  $ N \geq 2$ if, and only if, for every strategy $ {\mathsf u} $,
		\begin{equation} \label{eq:winmin}
		\sum _{i=2} ^{N-1} \lambda _{\mathsf s} (i) \leq 
		 \sum _{i=2} ^{N-1} \lambda _{\mathsf u} (i).
		\end{equation} 
\end{lem}
{\bf Proof}. ${\mathsf s} $ is optimal for  $ N \geq 2$ if, and only if,  for every strategy $ {\mathsf u} $,
\[  \sum _{i=2} ^{N-1} cr(i,s_i) \geq  \sum _{i=2} ^{N-1} cr(i,u_i) ,\]
or
\[  \sum _{i=2} ^{N-1} (cr^{max}(i) - cr(i,s_i)) \geq  \sum _{i=2} ^{N-1} (cr^{max}(i) - cr(i,u_i)) ,\]
or, by (\ref{eq:deflamb}) on page~\pageref{eq:deflamb}, (\ref{eq:winmin}) holds.
\hspace*{\fill} $\Box$

\begin{figure}
	\centering
	\includegraphics[scale=.10]{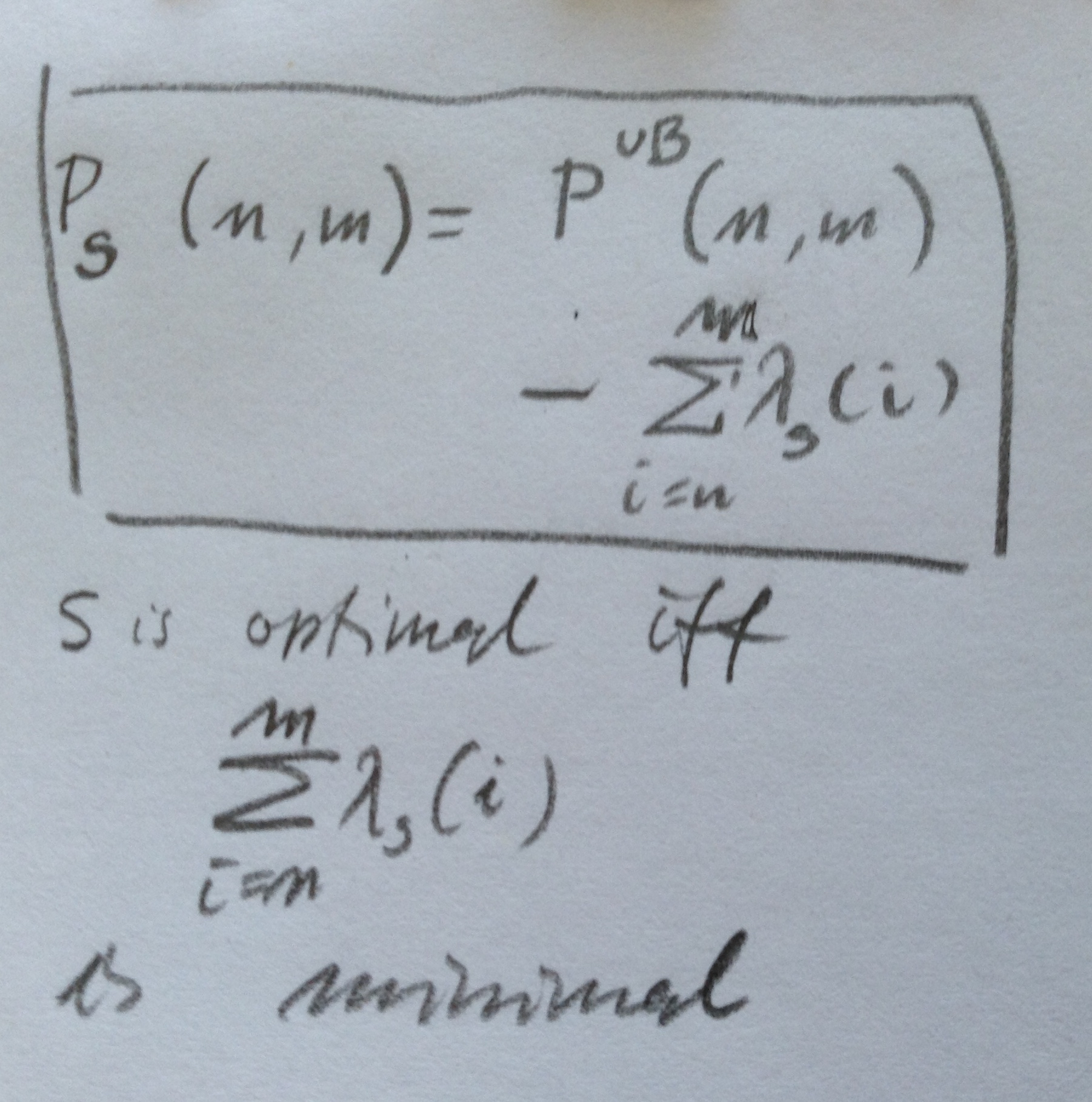}
	\caption{\label{page:minimal_loss} A draft of the definition of optimal strategy from  an earlier manuscript.}
\end{figure}

\medskip

The Player wins if the payoff $ P_{\mathsf w}(N-1) $ for the strategy $ {\mathsf w} $ he choose is optimal, that is, cannot be beaten by any strategy for the same $ N $.

\medskip

\begin{theorem} \label{thm:win} 
A strategy ${\mathsf w} $ is optimal for  $ N \geq 2$ if, and only if, its first $ N-1 $ moves produce a heap $ {\mathsf H}_{N} = \mathscr{T} ({\mathsf s}_{1,N-1}) $ $^{\ref{foo:funTdef}} $ 
 that forces ${\mathsf H}_{N}. {\tt RemoveAll()} $ to perform the worst-case number of comparisons of keys.
\end{theorem}
{\bf Proof}. Suppose that ${\mathsf w} $ is an optimal strategy for  $ N$ but the heap $ {\mathsf H}_{N} $ is not a worst-case heap for $ {\tt RemoveAll} $. In such a case, Player $ B $ can pick up an $ N $-element worst-case heap $ G_{N} $ for $ {\tt RemoveAll} $, run $G_{N}. {\tt RemoveAll()} $ on it while recording a sequence $ \langle \kappa _{i} \mid 1 \leq i < N \rangle $ of patches that were used to fill the vacancies left by the removed maximal elements, and then play any strategy $ \mathsf{v} $ whose first $ N $ moves are given by the reversed sequence
\linebreak  $ \langle \kappa _{N-i} \mid 1 \leq i < N \rangle $. Since the payoff for the Player $ A $, by the definition  (\ref{eq:creditFH}) of $ cr(i,k) $ page~\ref{eq:creditFH}, is equal to the number of comparisons performed by
\linebreak ${\mathsf H}_{N}. {\tt RemoveAll()} $, and the payoff for the Player $ B $ is,  equal to the number of comparisons performed by $G_{N}. {\tt RemoveAll()} $, the payoff for the Player $ B $ is larger than the payoff of the Player $ A $, and so the Player $ B $ wins, contrary to the assumption that the Player $ A $ ${\mathsf w} $ had an optimal strategy for  $ N$.
\hspace*{\fill} $\Box$

\medskip

The player could try to apply a brute force and devise a greedy strategy that maximizes credits for its all moves. 
This could be done relatively easily with a program that attempts to generate  a greedy substrategy for any given size $ N $ of a worst-case heap. Unfortunately, as I illustrate in \ref{Hereditary} page~\pageref{Hereditary}, any such greedy substrategy must fail to produce any worst-case heap of more than 22 nodes.

\medskip

In the next Section, I will use a bit subtler approach to generation of worst cases for $ {\tt RemoveAll} $.

\section{Strategies $\mathsf{par}$ and $\mathsf{win}$, and the lower bounds they establish}
\label{StraParWin}

\begin{figure}[h] 
\begin{center}
\includegraphics[scale=.75]{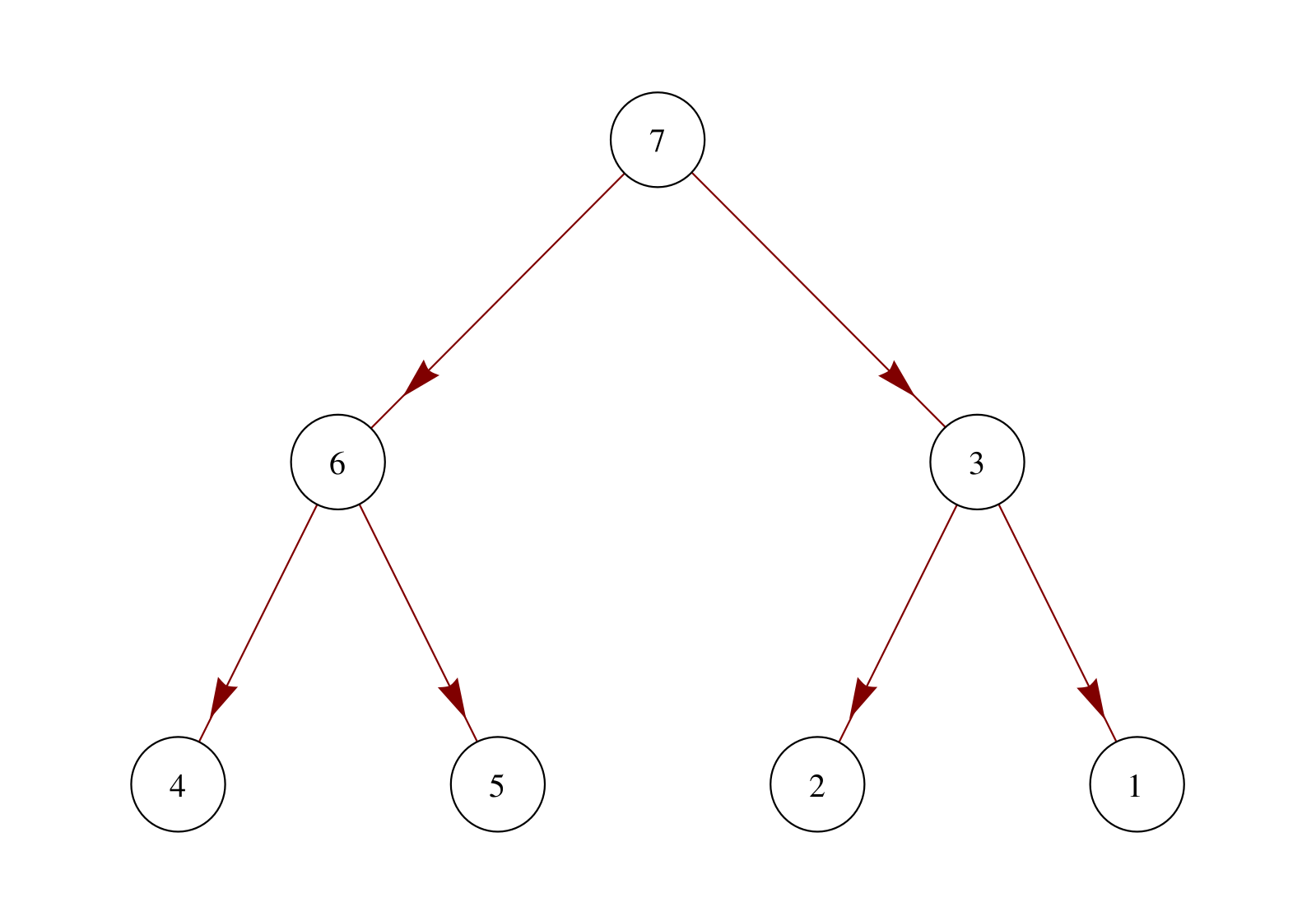} 
\end{center}
  \caption{A worst-case heap $ {\mathsf H}_7 $ of 7 nodes for $ {\tt RemoveAll} $. \label{fig:worst7}}
\end{figure}

\begin{figure}[h] 
\begin{center}
\includegraphics[scale=1]{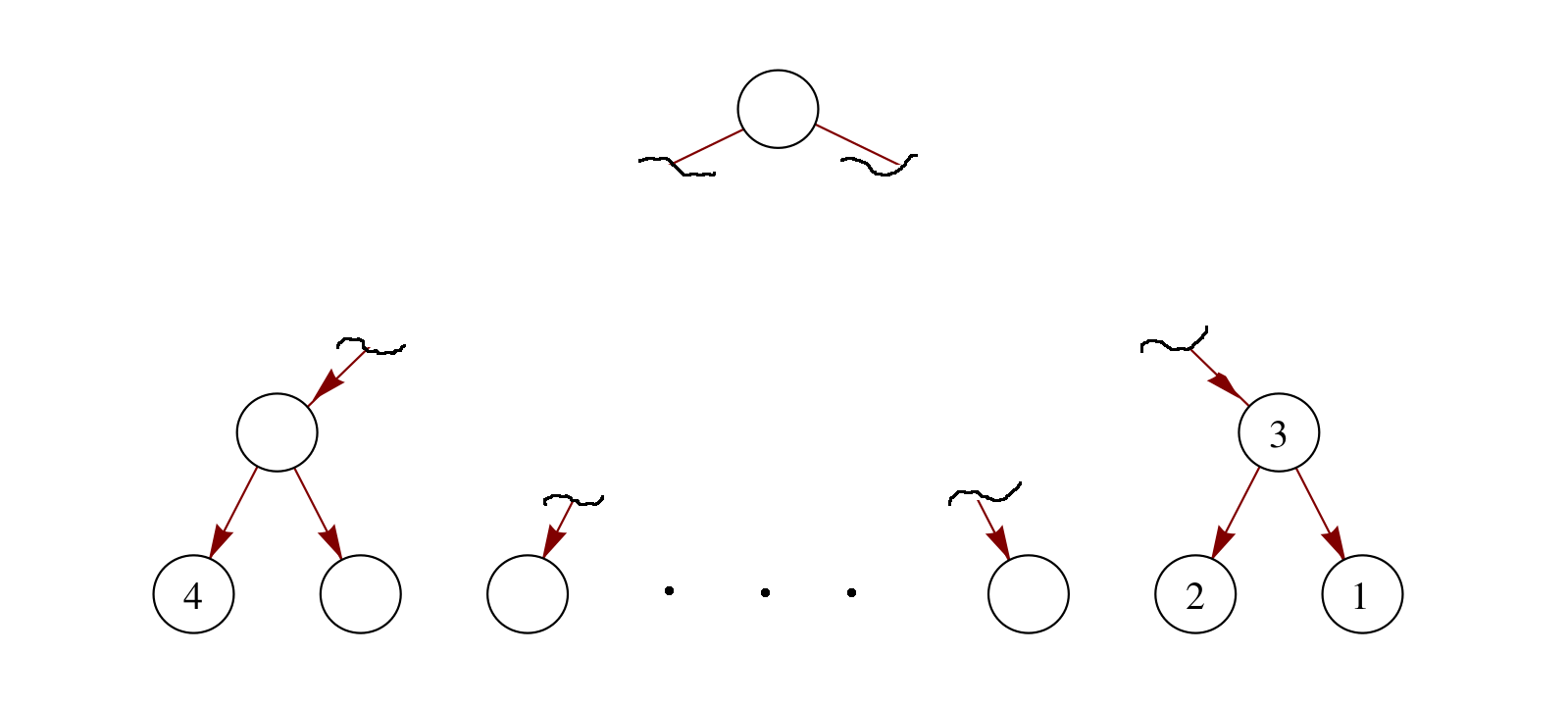} 
\end{center}
  \caption{The invariant fragment of the heap $ {\mathsf H}_{i} $ on $m = 2^{\lfloor \lg i \rfloor + 1} - 1 $ nodes, where $ i \geq 7 $, for the strategy  $ \mathsf{par} $, with locations of nodes 1, 2, 3, and 4 shown. \label{fig:worst7inv}}
\end{figure}

Given an integer $ N \geq 2 $, I will construct an optimal strategy $ \mathsf{win}(N) $ for that $ N $. For $  N \leq 7  $, one can take any worst-case heap $ H $ of $ 7 $ nodes\footnote{Beginning with a worst-case heap on $ 8 $ nodes and using heaps of $ N =  2^{\lceil \lg N \rceil}$ nodes as benchmarks, instead, would be, perhaps, more rational but it would also make the their illustration on Figures~\ref{fig:worst7} and \ref{fig:worst7inv} less neat.} (for example, the heap $ [ 7 , 6 , 3 , 4, 5 , 2, 1 ] $ visualized on Figure~\ref{fig:worst7}) and extract its creative sequence $ \mathscr{S} (H)$ of
the first $ 6 $ pull downs (for example, the creative sequence $ \langle 1, 1, 1, 1, 2, 1 \rangle  $ for the heap $ [ 7 , 6 , 3 , 4, 5 , 2, 1 ] $, given by the first 6 moves $ \langle 1, 2, 3, 4, 4, 5 \rangle $ of the sequence shown in \ref{sec:ex12} page \pageref{sec:ex12}) of a strategy from it; we will see in a moment that $ \mathscr{S} (H) ^{\frown} \mathsf{s} $ (e.g., $ \langle 1, 1, 1, 1, 2, 1 \rangle ^{\frown} \mathsf{s} $), where $ \mathsf{s} $ is an infinite\footnote{Just to make the entire sequence infinite.} sequence of pull downs, for instance, $ \langle 1 , 1 , 1 ,  ...  \rangle $, is an optimal strategy for every $ 2 \leq N \leq 7 $. Some straightforward experimentations with different sequences of pull downs allow one to find optimal strategies for $ N \leq 12 $ (for example, $ \langle 1, 1, 1, 1, 2, 1, 1, 4, 1, 4, 1 \rangle ^{\frown} \mathsf{s}  $ is such an optimal strategy the first 11 moves of which produce a 12-node worst-case heap visualized on Figures~\ref{fig:ExHeap}, \ref{fig:ArrayHeap}, and \ref{fig:mkworstout} pages \pageref{fig:ExHeap}
and \pageref{fig:mkworstout}; see \ref{sec:ex12}, page \pageref{sec:ex12} for details of construction of such a heap). 

\medskip

For $ N > 12 $,  such a construction of is easy if the sought-after worst-case heap is complete, that is, if $ N = 2 ^{\lceil \lg N \rceil} -1 $; the mentioned above sequence $ \mathscr{S} (H) $ followed by any infinite
sequence of pull downs that makes their concatenation $ \mathsf{s} $'s level loss $ \Lambda^{lev} _{\mathsf s} (K) =1$ for every $ K \geq 3 $ (for instance,  $ \langle 1, 1, 1, 1, 2, 1 \rangle  $ followed by the infinite sequence $ \langle 1 , 2 , 1 , 2, ... , 1, 2,... \rangle $ of alternating pull downs of 1 and 2) will do.
However, showing that such a construction results in an optimal strategy for any $ N = 2 ^{\lceil \lg N \rceil} -1 $ is nothing but a routine exercise due to the number of cases to consider in order to carry on the proof of it\footnote{See \cite{kru:heap} for an example of such a proof.}.

\medskip

 If $ N \neq 2 ^{\lceil \lg N \rceil} -1 $ then the construction of an optimal strategy, although  still relatively simple\footnote{FIX 
 THIS A combination of the strategies $ \langle 1 , 2 , 1 , 2, ... , 1, 2 \rangle ^{\frown} \mathsf{s}  $ and $ \langle 1, 1, 1, 1, 2, 1, 1, 4, 1, 4, 1 \rangle ^{\frown} \mathsf{s}  $}, is, perhaps, a bit harder to discover; its detailed proof\footnote{At least the one I know.} is definitely convoluted.

\medskip

\begin{figure}[h] 
\begin{center}
\includegraphics[scale=.3]{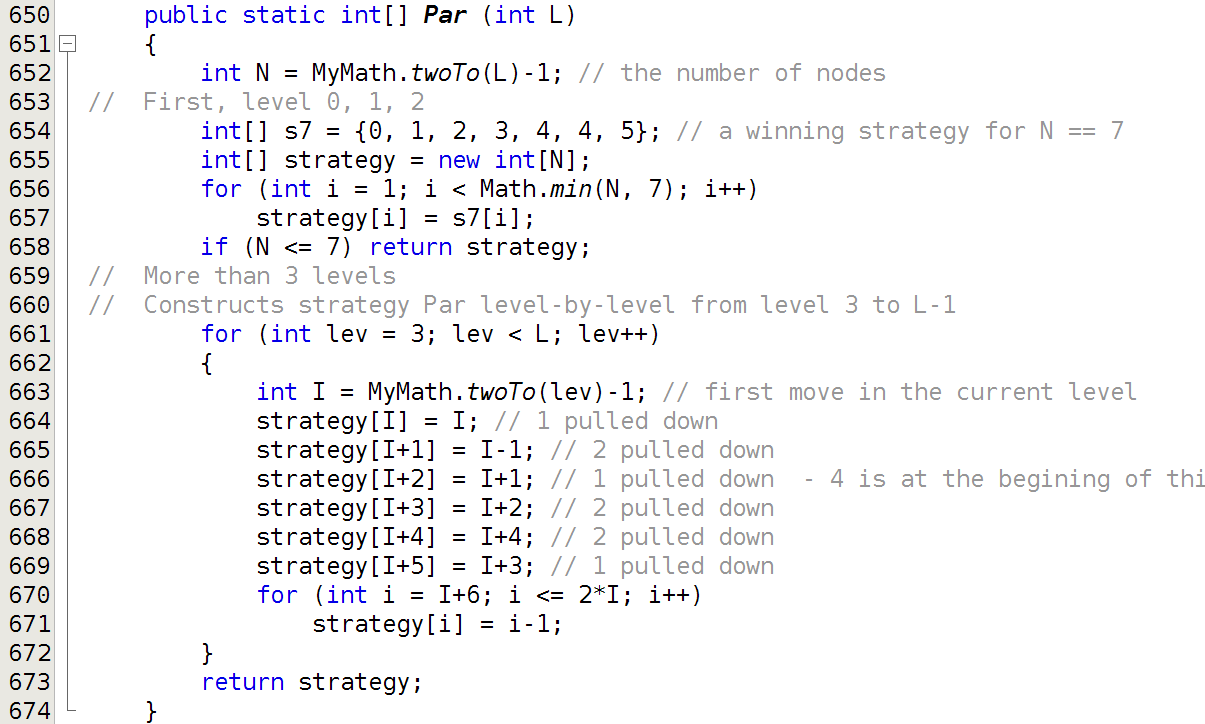} 
\end{center}
\caption{A Java code that implements strategy  $ \mathsf{par} $. Method $ {\tt Par (L)} $ returns the first $ L $ levels of moves of  $ \mathsf{par} $. The static method $ {\tt twoTo}(n) $ in class $ {\tt MyMath} $ computes $2 ^{ n}  $. The sequence of moves $ \langle 1, 2, 3, 4, 4, 5 \rangle $ in line 454 produces the heap $ {\mathsf H}_7 $ of Figure~\ref{fig:worst7} created by the sequence
$ \langle 1, 1, 1, 1, 2, 1 \rangle  $ of pull downs;  it comes from the equality (\ref{eq:defK16kV16})
page~\pageref{eq:defK16kV16}. Instruction in line 663 is a bit wasteful and could be replaced with 
$ {\tt I = lev;} $ 
followed by 
$ {\tt I = 2*I+1;} $ 
before the line 672,
but it comes handy as is while evaluating the outcomes of  method $ {\tt Par (L)} $. \label{fig:Par} }
\end{figure}

\label{defPar}
I begin with constructing the base strategy  $ \mathsf{par} $, whose Java code is shown on Figure~\ref{fig:Par}, that is an optimal strategy for every $ N > 2 ^{\lceil \lg N \rceil} - 4 $ (in particular, for $ N = 2 ^{\lceil \lg N \rceil} -1 $), but not for any other $ N $.

\medskip

The 7th heap $ {\mathsf H}_7 $ in the game $\mathscr{G} ({\mathsf par}) = \mathsf{H} $ carried on with strategy $ \mathsf{par} $ is visualized on Figure~\ref{fig:worst7}. As I have indicated, the creative sequence for $ {\mathsf H}_7 $ is $ \langle 1, 1, 1, 1, 2, 1 \rangle  $. Thus, by (\ref{eq:defK16kV16})
page~\pageref{eq:defK16kV16}, the first six moves ${\mathsf par}_{1,6}$ of $ {\mathsf par} $ are
$ \langle 1, 2, 3, 4, 4, 5 \rangle  $.

\medskip

One can easily verify (a program or an argument will do) that application of operation ${\mathsf H}_7 . {\tt RemoveAll}() $ to $ {\mathsf H}_7 $ performs 14 comparisons of keys, which number happens to be equal to the known\footnote{See, for instance, \cite{suc:elem} for a proof of the formula for any heap on $ N \geq 2$ (not just on $ N=7 $) nodes.} upper bound
\[ C_{\tt RemoveAll()} ^{\tt max} (7) =  2\sum _{i=2} ^{7-1} \lfloor \lg i \rfloor - \lfloor \lg (7-1) \rfloor  \]
on the number of comparisons of keys that the $ {\tt RemoveAll} $ performs on any heap of 7 nodes. Thus 14 is the worst-case number of comparisons of keys that the $ {\tt RemoveAll} $ performs on any heap of 7 nodes, and, therefore, $ {\mathsf H}_7 $ is a worst-case heap for $ {\tt RemoveAll} $.

\medskip

The heaps $ {\mathsf H}_6 $ through $ {\mathsf H}_1 $ of $ \mathsf{H} $ are defined as the \textit{residua} of successive applications of $ {\tt RemoveMax}() $ to $ {\mathsf H}_7 $.  The creative sequences of those \textit{residua} are the beginning subsequences of the creative sequence for $ {\mathsf H}_7 $.
Because
\begin{equation} \label{eq:par1-6}
C_{\tt RemoveAll()} ^{\tt max} (7) = \sum _{i=3} ^{7} C_{\tt RemoveMax()} ^{\tt max} (i) ,
\end{equation}
where $ C_{\tt RemoveMax()} ^{\tt max} (i) $ is the maximum number of comparisons of keys that the $ {\tt RemoveMax} $ performs on any heap of $ i $ nodes,
those creative sequences are lossless. As a result
 all the \textit{residua} of $ {\mathsf H}_7 $, that is, the heaps $ {\mathsf H}_6 $ through $ {\mathsf H}_1 $, are automatically worst-case heaps for $ {\tt RemoveMax} $ and, therefore, for $ {\tt RemoveAll} $. 

\medskip

After the 7th move, the strategy  $ \mathsf{par} $ is defined inductively level-by-level. For any $ I  $ with, 

\begin{equation} \label{eq:iEQ2tofloorlg}
7 \leq I = 2^{\lfloor \lg I \rfloor + 1} - 1 =\footnote{Since I > 1.} \;
 2^{\lceil \lg I \rceil} - 1 = 2^{\tt lev} -1,
\end{equation} 
it takes the heap $ {\mathsf H}_{I} $ on $ I $ nodes, whose scheme is visualized on Figure~\ref{fig:worst7inv} (for instance, heap $  {\mathsf H}_7  $ of Figure~\ref{fig:worst7} falls under that scheme), that it has constructed so far,  and adds to it the next level $ {\tt lev} $ by performing on $ {\mathsf H}_{I} $ a sequence
$  \mathsf{par} _{I,  2I}$ 
of $ I + 1 $ consecutive moves such that
the resulting heap $ {\mathsf H}_{2I + 1} $ on $ 2I + 1 $ nodes falls under the scheme of Figure~\ref{fig:worst7inv} and
all these moves are lossless except for one  move $\mathsf{par}  _{I+1} $ (the second move in the described sequence) that scores a credit that is one less than the maximal score.


\medskip

More specifically, move $\mathsf{par} _I$ (the first move in the described sequence), implemented by the statement at line 664 in the method $ {\tt Par (L)} $ shown on Figure~\ref{fig:Par}, pulls down 1 that resides at index $ I $ of heap  $ {\mathsf H}_{I} $.
That move, among other effects, brings node 3 to the last index $I$ in the level $ \lfloor \lg I \rfloor $ of $ {\mathsf H}_{I+1} $ and makes node 1 the child of node 4 at index $ \frac{I+1}{2} $. Nodes 2 and 4 maintain their indicies $ I - 1 $ and $ \frac{I+1}{2} $, respectively, in heap $ {\mathsf H}_{I+1} $.
\medskip

Move $\mathsf{par} _{I+1}$, implemented by the statement at line 665 in the method $ {\tt Par (L)} $ shown on Figure~\ref{fig:Par}, pulls down 2 that resides at index $ I-1 $ of heap  $ {\mathsf H}_{I+1} $. Nodes 3 and 4 are not affected because they reside at the same level $ \lfloor \lg I \rfloor $ as 2 did in $ {\mathsf H}_{I+1} $. 
\medskip
 
Move $\mathsf{par}  _{I+2} $, implemented by the statement at line 666 in the method $ {\tt Par (L)} $ shown on Figure~\ref{fig:Par}, pulls down 1 that resides at index $ I+1 $ of heap  $ {\mathsf H}_{I+2} $. That move, among other effects, brings its parent 4 to the first index 
$ I+1 $ of the last level $ \lfloor \lg I \rfloor +1 $ of of $ {\mathsf H}_{I+3} $. Nodes 2 and 3 are not affected.

\medskip

 After that, the remainder moves $ \langle \mathsf{par} _{n} \mid I+3 \leq n \leq 2I  \rangle $, implemented by the statements at lines 667 through 672 in the method $ {\tt Par (L)} $ shown on Figure~\ref{fig:Par}, keep pulling down 1 and 2 that reside as leaves in heaps $ {\mathsf H}_{I+3},..., {\mathsf H}_{2I}$, making sure that they always pull down a node with a sibling, and that the two last moves are pulling down 2 and then 1. Nodes 3 and 4 are not affected by any of those moves, except that node 3 becomes the parent of node 2 in ${\mathsf H}_{2I}$ and the parent of nodes 1 and 2 in ${\mathsf H}_{2I+1}$.

\medskip

Obviously, all the moves are valid because no other node of any heap is smaller than 2 and 1.

\medskip

The resulting heap $ {\mathsf H}_{2I+1} $ falls under the scheme of Figure~\ref{fig:worst7inv}; the first node 
$ {\mathsf H}_{2I+1}[I+1] $ in the last level of $ {\mathsf H}_{2I+1} $ is 4, the last two nodes $ {\mathsf H}_{2I+1}[2I] $ and 
$ {\mathsf H}_{2I+1}[2I+1] $ in the last level of $ {\mathsf H}_{2I+1} $ are 2 and 1 in that order, and the parent 
$ {\mathsf H}_{2I+1}[I]$ of 1 and 2 in $ {\mathsf H}_{2I+1} $ is 3. 
Moreover, by the pq~Lemma~\ref{lem:1212}~\ref{lem:1212i} and \ref{lem:1212iii} page \pageref{lem:1212} substituting 1 for $ p $ and 2 for $ q  $,
taking into account that $  {\mathsf H}_{I}[\frac{I+1}{2}] = 4 > 2 $,
 all the moves were lossless, except for the
$I+1= 2^{\lfloor \lg (I+1) \rfloor} $th move $\mathsf{par}  _{I+1} $ (the second move $\mathsf{s}  _{2} $ in the sequence of  the pq~Lemma~\ref{lem:1212}) that scored a credit one less than the maximal credit for the $ I+1 $st move.

\medskip

This way we proved the following two theorems.

\medskip

\begin{theorem} \label{thm:parinvariant} 
Every heap $ H $ on $ N = 2^{\lceil \lg N \rceil} -1 $ nodes, where $ N \geq 7 $, produced by strategy $ \mathsf{par} $ satisfies the invariant visualized on Figure~\ref{fig:worst7inv}.
\end{theorem}
{\bf Proof} follows from the above discussion.
\hspace*{\fill} $\Box$

\medskip

\begin{theorem} \label{thm:parcredits} 
For every $ i \geq 1 $,  
\begin{equation} \label{eq:parcredits100}
\lambda_{\mathsf{par}}(i)  = 
\left\{ \begin{array}{ll}
1 \mbox{ if } \; 8  \leq  i = 2 ^{\lfloor \lg i \rfloor}   \\ \\
0 \mbox{ otherwise}.
\end{array} \right.
\end{equation}
\end{theorem}
{\bf Proof}. If $ i \leq 7 $ then $ \lambda_{\mathsf{par}}(i)  = 0 $ because the first 6 moves of strategy $ \mathsf{par} $ are lossless. If $ i \geq 7 $ then, by (\ref{eq:par1-6}), $ \lambda_{\mathsf{par}}(i)  = 0 $ for all $ i $ except $I+1= 2^{\lfloor \lg (I+1) \rfloor} $ in the above discussion, that is, $ i = 2^{\lfloor \lg i \rfloor}$, for which $ \lambda_{\mathsf{par}}(i)  = 1 $.
\hspace*{\fill} $\Box$
\medskip

\begin{figure}[h] 
\begin{center}
\includegraphics[scale=1]{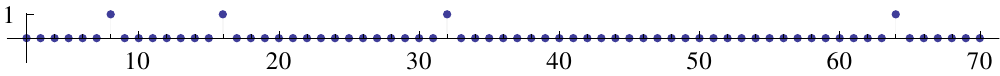}  
\end{center}
\caption{A discrete graph of function $ \lambda_{\mathsf{par}}(i)  $; the characteristic function of the set of powers of 2 that are greater or equal 8. \label{fig:lambdapar} }
\end{figure}

Figure~\ref{fig:lambdapar} shows a graph of $ \lambda_{\mathsf{par}}(i)  $.

\medskip

\begin{figure}[h] 
\begin{center}
\includegraphics[scale=.5]{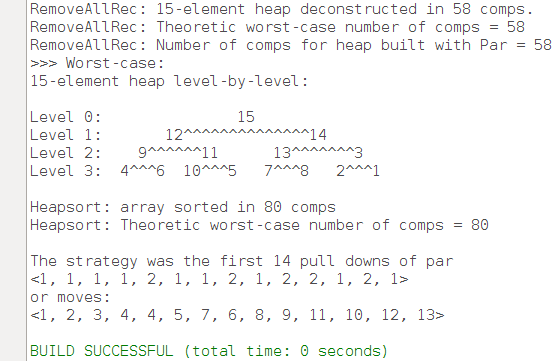}  
\end{center}
\caption{A complete heap $ {\mathsf H} _{15} $ produced by the first $ 14 $ pull downs of the strategy $ \mathsf{par} $ (an output of my Java program). The pull downs and the moves are shown at the bottom of the output. \label{fig:par15out}}
\end{figure}

A Java code that implements strategy  $ \mathsf{par} $ is shown on Figure~\ref{fig:Par}, and an example of a heap produced by strategy $ \mathsf{par} $ is shown on Figure~\ref{fig:par15out}. 
The subsequence $ \mathsf{par} _{1,14}   $ of the first 14 moves of $ \mathsf{par} $ that produced it is:
\[ \langle 1, 2, 3, 4, 4, 5, 7, 4, 8, 9, 10, 12, 12, 13 \rangle , \]
and the sequence $ \langle {\mathsf H}_n [\mathsf{par} _n] \mid 1 \leq n \leq 14  \rangle $ of corresponding pull downs  is:
\[ \langle 1, 1, 1, 1, 2, 1, 1, 2, 1, 2, 2, 1, 2, 1 \rangle . \]
The sequence ${\mathsf H}_{1,14} = \mathscr{G}_{{\mathsf H}_1} ( {\mathsf par}_{1,14} ) $ of 14 heaps created by these moves is shown in the \ref{sec:ex12} page~\pageref{sec:ex12}.

\medskip


Let for every $ i \geq 1 $, the \textit{delayed loss} function $ \lambda^* $ be defined by:
\begin{equation} \label{eq:parStarDef}
\lambda^*(i)  = 
\left\{ \begin{array}{ll}
1 \mbox{ if } \;   i = 2 ^{\lceil \lg i \rceil} -4  \\ \\
0 \mbox{ otherwise}.
\end{array} \right.
\end{equation}

\medskip

I will show in Section~\ref{ProofWin} that function $\sum_{i=2}^{n} \lambda^*(i) $ establishes a lower bound on accumulated loss of credit for moves $ 1 $ through $ n $ for any strategy. The following Lemma provides a closed-form formula for that lower bound.

\begin{lem} \label{lem:Lambda*closed}
	For every $ n \geq 1 $,
	\begin{equation} \label{eq:Lambda*closed}
	\sum_{i=2}^{n} \lambda^*(i) = \max \{ {\lfloor \lg (n+4)  \rfloor} , 3 \} -3.
	\end{equation}
\end{lem}
{\bf Proof}. By virtue of definition (\ref{eq:parStarDef}), $ \sum_{i=2}^{n} \lambda^*(i) $ is equal to the
number of $ i $'s that are less than or equal $ n $, with $12 \leq i = 2^{\lceil \lg i  \rceil} -4$,
or $16 \leq i+4 = 2^{\lceil \lg i  \rceil}$. Since $ 2^x $ is a $1-1$ function, that number is the same as the number of different values of $ \lceil \lg i  \rceil $ between 4 and ${ \lg (n+4) }$, which is equal to ${\lfloor \lg (n+4)  \rfloor} - 3$ for ${\lfloor \lg (n+4)  \rfloor} \geq 3$ (same as $ n \geq 4 $) or $ 0 $ otherwise. 
\medskip

Hence,
\[ \sum_{i=2}^{n} \lambda^*(i) = \max \{ {\lfloor \lg (n+4)  \rfloor} - 3, 0 \} = \max \{ {\lfloor \lg (n+4)  \rfloor} , 3 \} -3, \]
which yields (\ref{eq:Lambda*closed}).
\hspace*{\fill} $\Box$

\medskip

\begin{figure}[h] 
	\begin{center}
		\includegraphics[scale=1]{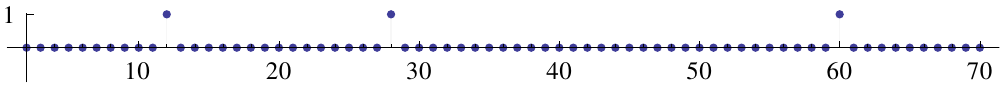} 
		\includegraphics[scale=1]{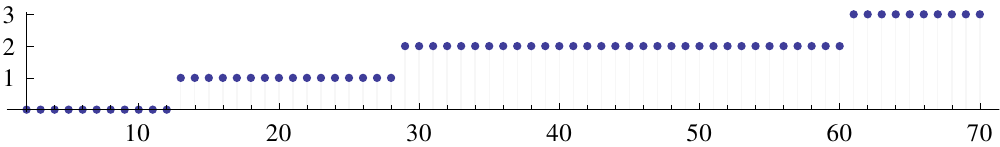} 
	\end{center}
	\caption{Discrete graphs of the delayed loss function $  \lambda^*(N) $ (upper graph) and its sum 
		$ \sum_{i=2}^{N-1} \lambda^*(i) $ (lower graph), for $ 2 \leq N \leq 70 $. \label{fig:lambdaStar} }
\end{figure}

Graphs of functions $ \lambda^*(N) $ and 	$ \sum_{i=2}^{N-1} \lambda^*(i) $ are visualized on Figure~\ref{fig:lambdaStar}. 

\medskip

\begin{lem} \label{lem:parstar}
For every $ K \geq 1 $, the following equality holds: 
\begin{equation} \label{eq:parStar1}
\Lambda^{level}_{\mathsf{par}}(K)  = 
\sum_{i=2^K-1}^{2^{K+1}-2} \lambda^*(i).
\end{equation}
\end{lem}
{\bf Proof}.  The least solution of the equation 
\begin{equation} \label{eq:parStarEq}
i = 2 ^{\lceil \lg i \rceil} -4 
\end{equation}
that occurs in (\ref{eq:parStarDef})
is $ i=12 $, so $ \lambda^*(i) = 0 $ for $ i < 7 $. Thus
$ \sum_{i=2^K-1}^{2^{K+1}-2} \lambda^*(i) = 0 $ for $ K =  1, 2 $.
Since, by (\ref{eq:parcredits100}), $ \lambda_{\mathsf{par}}(i) = 0 $ for $ i < 7 $, also
$ \Lambda^{level}_{\mathsf{par}}(K)  $ $ = $ [by the equation (\ref{eq:levelloss}) page \pageref{eq:levelloss}] $ \sum_{i=2^K-1}^{2^{K+1}-2} \lambda_{\mathsf{par}}(i) = 0 $ for $ K =  1, 2 $.
Hence (\ref{eq:parStar1}) holds for $ K =  1, 2 $.

\medskip

For each $ K \geq 3 $, the equation ({\ref{eq:parStarEq}}) has exactly one solution that satisfies
$ 2^K-1 \leq i \leq 2^{K+1}-2$, namely, $ i = 2 ^{K+1} -4$, and so does the equation $ i = 2 ^{\lfloor \lg i \rfloor} $ that occurs in (\ref{eq:parcredits100}), namely,  $ i = 2 ^{K} $. Thus both   $ \sum_{i=2^K-1}^{2^{K+1}-2} \lambda_{\mathsf{par}}(i)$, that by (\ref{eq:levelloss}) is equal to $\Lambda^{level}_{\mathsf{par}}(K) $,
and $  \sum_{i=2^K-1}^{2^{K+1}-2} \lambda^*(i) $ are equal 1. Hence, by  
(\ref{eq:levelloss}), the equation (\ref{eq:parStar1}) holds for each $ K \geq 3 $.
\hspace*{\fill} $\Box$

\medskip

\begin{lem} \label{lem:sumparstar}
For every $ D \geq 1 $, the following equality holds:
\begin{equation} \label{eq:sumparStar1}
\Lambda_{\mathsf{par}}(2^{D+1}-2)   = \sum_{i=2}^{2^{D+1}-2} \lambda^*(i).
\end{equation}
\end{lem}
{\bf Proof}. By the equality~(\ref{eq:sumlevellossComplNew}) page~\pageref{eq:sumlevellossComplNew},
\[\Lambda_{\mathsf{par}}(2^{D+1}-2) = \sum_{K=1}^D\Lambda^{level}_{\mathsf{par}}(K)  =\]
[by the equation~(\ref{eq:parStar1}) in Lemma~\ref{lem:parstar}]
\[   = \sum _{K=1}^D \sum_{i=2^K-1}^{2^{K+1}-2} \lambda^*(i)
=   \sum_{i=1}^{2^{D+1}-2} \lambda^*(i) = \]
[since $ \lambda^*(1) = 0 $] 
\[=   \sum_{i=2}^{2^{D+1}-2} \lambda^*(i) . \]
Hence, (\ref{eq:sumparStar1}) holds.
\hspace*{\fill} $\Box$

\begin{figure}[h] 
\begin{center}
\includegraphics[scale=1]{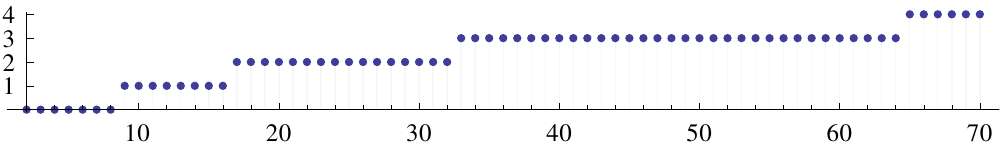} 
\end{center}
\caption{A discrete graph of function $ \Lambda_{\mathsf{par}}(N-1) $ $ = $ $ \sum _{i=1}^{N-1} \lambda_{\mathsf{par}}(i) $, where $  \lambda_{\mathsf{par}}(i) $ is the function visualized on Figure~\ref{fig:lambdapar} page~\pageref{fig:lambdapar}. Clearly, for $ N \geq 5 $, $ \Lambda_{\mathsf{par}}(N-1) $ $ = $ $ \lceil \lg N  \rceil - 3 $ $ = $ $ {\lfloor \lg (N-1)  \rfloor} - 2 $. \label{fig:sumlambdapar} }
\end{figure}

\medskip

Figure~\ref{fig:sumlambdapar} visualizes graph of function $ \Lambda_{\mathsf{par}}(N-1) $.

\medskip

Not surprisingly, it coincides with function $ \Lambda^*_{\mathsf{par}}(N-1) $ for some $ N $, as the following Lemma states.

\medskip

\begin{lem} \label{lem:sumparstar2}
	For every $ N > 2^{\lceil \lg N \rceil} - 4 $, the following equality holds:
	\begin{equation} \label{eq:sumparStar2}
	\Lambda_{\mathsf{par}}(N-1)   = \sum_{i=2}^{N-1} \lambda^*(i).
	\footnote{One can also show, using the fact that $ {\mathsf{par}} $ loses to another another strategy for every $ N \leq 2^{\lceil \lg N \rceil} - 4 $, that the equation (\ref{eq:sumparStar2}) holds only if $ N > 2^{\lceil \lg N \rceil} - 4 $, but I do not need that fact in the proofs presented in this paper.}
	\end{equation}
\end{lem}
{\bf Proof}. If $ N = 2^{\lfloor \lg N \rfloor}  $ then the equation (\ref{eq:sumparStar2}) follows from (\ref{eq:sumparStar2}) since 
\[  \lambda_{\mathsf{par}}(2^{\lfloor \lg N \rfloor}-1) = 0 = \lambda^*(2^{\lfloor \lg N \rfloor}-1).  \]
If $  2^{\lfloor \lg N \rfloor} < N $ and $ 2^{\lceil \lg N \rceil} - 4 < N $ then 
\begin{equation} \label{eq:sumparstar2.1}
\Lambda_{\mathsf{par}}(2^{\lfloor \lg N \rfloor}, N-1) = 1,
\end{equation}
and
\begin{equation} \label{eq:sumparstar2.2}
\sum_{i=2^{\lfloor \lg N \rfloor}}^{N-1} \lambda^*(i) = 1.
\end{equation}
Thus
\[ \Lambda_{\mathsf{par}}(N-1) = \Lambda_{\mathsf{par}}(2^{\lfloor \lg N \rfloor}-1) +  \Lambda_{\mathsf{par}}(2^{\lfloor \lg N \rfloor}, N-1) = \]
[by (\ref{eq:sumparstar2.1})]
\[ = \Lambda_{\mathsf{par}}(2^{\lfloor \lg N \rfloor}-1) + 1 =   \]
[by (\ref{eq:sumparstar2.2}) and the already proven case of (\ref{eq:sumparStar2}) for  $ N = 2^{\lfloor \lg N \rfloor}  $]
\[ =  \sum_{i=2}^{2^{\lfloor \lg N \rfloor - 1}} \lambda^*(i)   + \sum_{i=2^{\lfloor \lg N \rfloor}}^{N-1} \lambda^*(i)  =  \sum_{i=2}^{N-1} \lambda^*(i) .\]
Hence, (\ref{eq:sumparStar2}) holds.
\hspace*{\fill} $\Box$

\medskip

The about facts about strategy $ {\mathsf{par}} $ is all I need to derive and prove the main results of this paper. Below is its more definitive characteristics that I quote here for an illustration. 

\medskip

The payoff $ P_{\mathsf{par}}(N-1) $ after $ N -1$ moves of the strategy $ \mathsf{par} $ for $ N \geq 2 $ is given by this formula:
\begin{equation} 
\label{eq:parcredits} 
2 (N - 1) \lfloor \lg (N - 1)  \rfloor - 
 2^{\lfloor \lg (N - 1)  \rfloor +2} + \min (\lfloor \lg (N - 1)  \rfloor, 2) + 4. 
\end{equation}

\begin{figure}[h] 
\begin{center}
\includegraphics[scale=1]{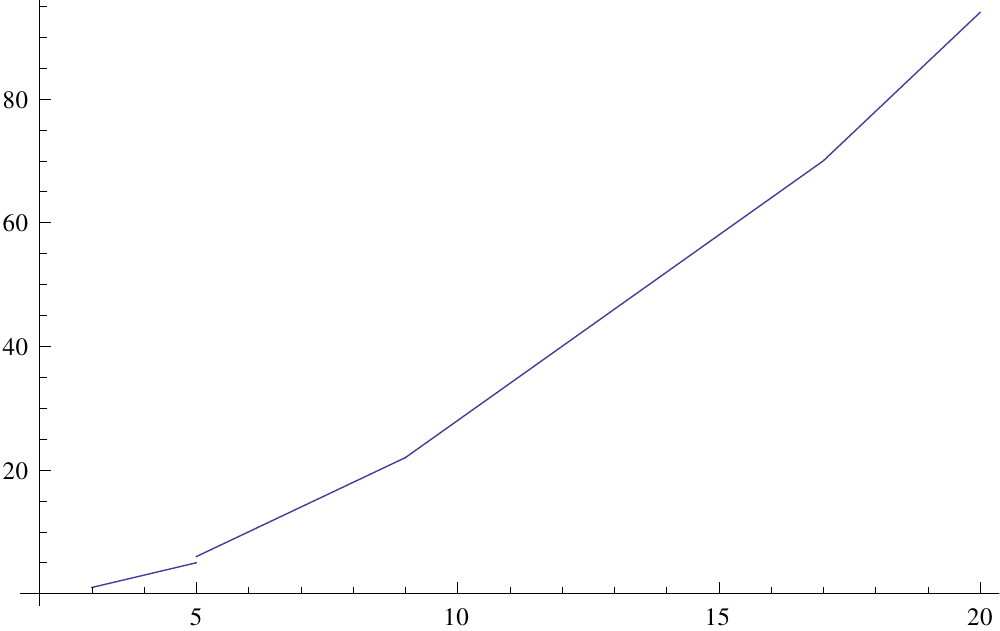} 
\end{center}
\caption{A graph of function $ P_{\mathsf{par}}(N-1) $ of payoff for strategy $ \mathsf{par} $. Also, a lower bound on the number of comparisons of keys performed in the worst-case  by $ {\tt RemoveAll} $ on an $ N $-element heap. \label{fig:baseLB2}}
\end{figure}

\medskip

Figure~\ref{fig:baseLB2} shows a graph of the total payoff function $ P_{\mathsf{par}} $ for $ \mathsf{par} $.
Of course, $ P_{\mathsf{par}} $, given by (\ref{eq:parcredits}), is a lower bound for the number of comparisons of keys that $ {\tt RemoveAll} $ performs in the worst case on an $ N $-element heap.

\medskip

\begin{figure}[h] 
\begin{center} 
\includegraphics[scale=.3]{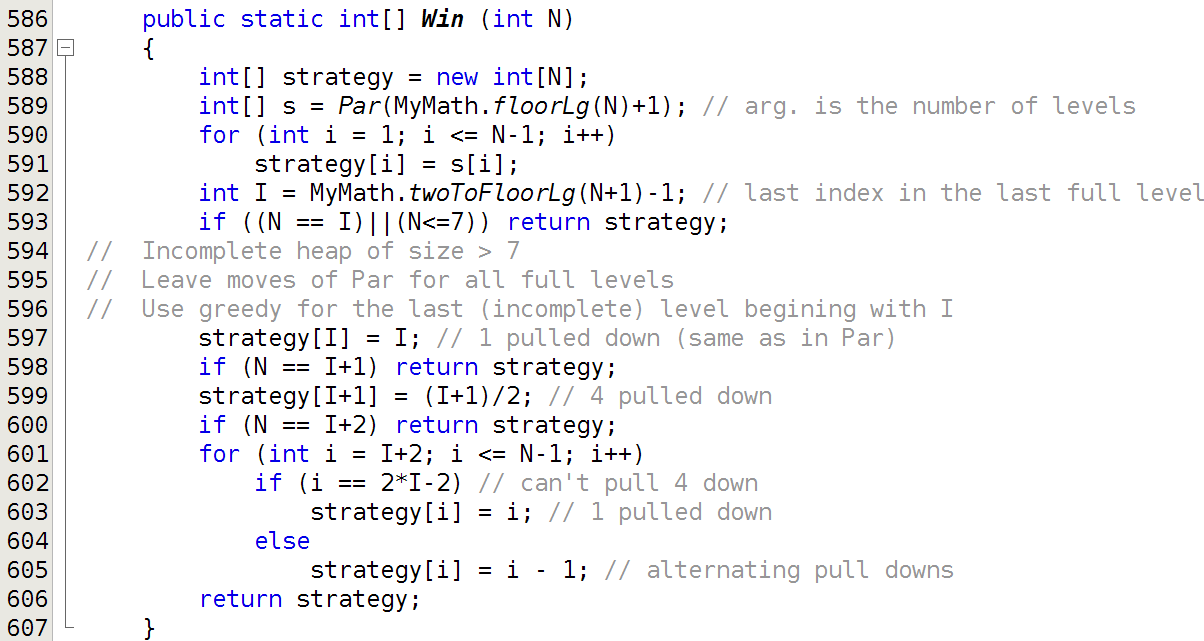} 
\end{center}
\caption{
	A Java code that implements strategy  $ \mathsf{win}(N) $. Method $ {\tt Win(N) }$ returns the first $ N-1 $ moves of $ \mathsf{win}(N) $.  The static method $ {\tt floorLg(N)} $ in class $ {\tt MyMath} $ computes ${\lfloor \lg N \rfloor}  $. 
The static method $ {\tt twoToFloorLg(N+1)} $ in class $ {\tt MyMath} $ computes $2^{\lfloor \lg (N+1) \rfloor}  $.  
Computation of the moves in the last level (lines 597 through 605) overrides the respective moves of the strategy   $ \mathsf{par} $ (returned by the call to method $ {\tt Par}$ in line 589 and copied to array $ {\tt strategy} $ in  $ {\tt for} $-loop in lines 590 and 591) in order to simplify handling of the special case of the method $ {\tt Win(N) }$ for $ N\leq 7 $.
\label{fig:Win} }
\end{figure}

\medskip

The strategy $ \mathsf{par} $ loses 1 credit per level, beginning with level 3, and in the very second move of each such level. Thus switching to a greedy strategy in the last level of the produced heap (moves $ 2^{\lfloor \lg N \rfloor} $ through $ N-1 $, provided that $ N > 2^{\lfloor \lg N \rfloor} $) may help one to avoid the loss of credit in the last level and, therefore, result in a set of strategies that are better than  $ \mathsf{par} $ for infinitely many $ N $.

\medskip

It turns out that, given $ N \geq 8 $, a strategy  $ \mathsf{win}(N) $ that is a combination $ \mathsf{par+gre} $ of the strategy $ \mathsf{par} $, played for all levels from 0 to ${\lfloor \lg (N+1) \rfloor} -1 $, and a greedy\footnote{One that does not admit any delayed gratification, that is, takes a loss only if no lossless move is valid at the moment.} strategy $ \mathsf{gre} $ for the level ${\lfloor \lg (N+1) \rfloor} $ if $ N > 2^{\lfloor \lg (N+1) \rfloor} -1 $, is an optimal strategy for $ N $, that is, it produces a worst-case heap in its first $ N-1 $ moves. 
I am going to show that for every $ N \geq 9 $, the above arrangement allows $ \mathsf{win}(N) $ to postpone the loss at the move $  2^{\lfloor \lg N  \rfloor}  $, that the strategy $ \mathsf{par} $ would have to incur, up until the move  $  2^{\lceil \lg N  \rceil}  -4$ of $ \mathsf{win}(N) $ if $  2^{\lceil \lg N  \rceil}  -4 < N$, or avoiding that loss altogether, otherwise. 
In the latter case, for all $ N \leq 2^{\lceil \lg N  \rceil}  -4 $, strategy $ \mathsf{win}(N) $ will beat $ \mathsf{par} $ by 1 credit after completion of their first $ N-1 $ moves. This fact and the formula (\ref{eq:parcredits})  for $ P_{\mathsf{par}}(N-1) $ will yield the formula for payoff $ P_{\mathsf{win}(N)}(N-1) $ for strategy $ \mathsf{win}(N) $.

\medskip

\label{defWin}
 I define the greedy strategy $ \mathsf{gre} $ that in the case of 
\begin{equation} \label{eq:Ngre} 
 N > 2^{\lfloor \lg (N+1) \rfloor} -1 \geq 7
\end{equation} 
 is played in the last level ${\lfloor \lg N \rfloor} $ $ = $ [by (\ref{eq:Ngre})] ${\lfloor \lg (N+1) \rfloor} $ of the constructed heap as follows:
\begin{enumerate}
     \renewcommand\labelenumi{\theenumi}
     \renewcommand{\theenumi}{(\roman{enumi})}

\item \label{step1}  It pulls down 1 (the last element of the heap of $ 2^{\lfloor \lg (N+1) \rfloor} -1 $ nodes that is shown on Figure~\ref{fig:worst7inv}).
\item \label{step2}  It pulls down 4 (the new parent of 1).
\item It repeats steps \ref{step1} and \ref{step2} (in that order) until \ref{step2} is invalid.
\item It pulls down 1.
\item It pulls down 1 and 2.
\item After that, it keeps executing pull downs of 1 indefinitely (just to make $ \mathsf{win}(N) $ an infinite sequence)
\end{enumerate}

\medskip

More specifically, $\mathsf{gre}$ is defined inductively level-by-level, from level 3 up. For any
$ N $ that satisfies inequality (\ref{eq:Ngre}),
it takes the largest complete heap $ H $ of no more than $ N $ nodes, which by (\ref{eq:completeCharactNodes}), 
 happens to have 
\begin{equation} \label{eq:defI} 
I  =  2^{\lfloor \lg (N+1) \rfloor} -1
\end{equation}
nodes, that was created by the first $ I-1 $ moves of $\mathsf{par}$ and, therefore, by virtue of Theorem~\ref{thm:parinvariant}, falls under the scheme visualized on Figure~\ref{fig:worst7inv}, and performs on it a sequence
$  \mathsf{win}(N) _{I,  N-1}$ 
of $ N-I $ consecutive moves as follows.

\medskip

The move $\mathsf{win}(N)_I$ (the first move $\mathsf{gre}_1$ in the described sequence), implemented by the statement at line 597, with $ I $ computed by the statement at line 592, in the method $ {\tt Win (N)} $ shown on Figure~\ref{fig:Win}, pulls down 1.
That move, among other effects, brings node 3 to the last position $i$ in the level $ \lfloor \lg i \rfloor $ of $ {\mathsf H}_{i+1} $ and makes 1 the only child of 4.

\medskip

The move $\mathsf{win}(N)_{I+1}$ (or $\mathsf{gre}_2$), implemented by the statement at line 599 in the method $ {\tt Win (N)} $ shown on Figure~\ref{fig:Win}, pulls down 4.

\medskip

After that, the moves $ \langle \mathsf{win}(N)_{n} \mid I+2 \leq n \leq \min\{2I-3,N-1\}  \rangle $, implemented by the first part of the $ {\tt for} $-loop where $ i < 2I - 2 $ (the statements at lines 601 through 606) in the method $ {\tt Win (N)} $ shown on Figure~\ref{fig:Win}, keep pulling down 1 and 4 that reside as leaves in heaps $ {\mathsf H}_{I+2},..., {\mathsf H}_{\min\{2I-3,N-1\}}$, making sure that they always pull down a node with a sibling. All those are valid moves since 1 is the least node in any heap and the only two nodes 2 and 3 that are less than 4 reside at the last two indicies $ I-1 $ and $ I $ of level 
$ \lfloor \lg I \rfloor $ in heaps $ {\mathsf H}_{I+2},..., {\mathsf H}_{\min\{2I-3,N-1\}}$.

\medskip

An example of situation after move $ \mathsf{win}(N)_{2I - 3}  $ for $  N = 28 $ and $ I = 15 $ is shown on    Figure~\ref{fig:win12-28out} page 
\pageref{fig:win12-28out}.

\medskip

If $ {\tt Win(N) }$ is not done at this point, that is, if $ 2I - 2 < N $, then the move $\mathsf{win}(N)_{2I-2}$, implemented by the statement at line 603 in the method $ {\tt Win (N)} $ shown on Figure~\ref{fig:Win}, pulls down 1, which resides in $ {\mathsf H}_{2I-1} $ as the parent of 1 that is a leaf at index $ 2I-2 $ and has no sibling. So, it scores the same credit as move $\mathsf{win}(N)_{I+1}$ that pulled down 4 for the first time did, that is,
\[ cr(2I-2,\mathsf{win}(N)_{2I-2}) =  2 \lfloor \lg (I+1) \rfloor - 1   = \]
[by virtue of (\ref{eq:iEQ2tofloorlg})]
\[ \lfloor \lg (2I-2) \rfloor + \lfloor \lg (2I-3) \rfloor - 1.\]
This, by (\ref{eq:creditUBform}) and the inequality (\ref{eq:creditUB}) on page~\pageref{eq:creditUB}, yields one less than the maximum 
of credit $ cr(2I-2,k) $ for any $ k \leq  2I-2$, thus making move $\mathsf{win}(N)_{2I-2}$ lossy, with
\begin{equation} \label{eq:lampar2a}
\lambda_{\mathsf{win}(N)}(2I-2) = 1.
\end{equation}

\medskip

 After that, if $ {\tt Win(N) }$ is still not done, that is, if $ 2I - 1 < N $, then the 
 reminder move $  \mathsf{win}(N)_{2I-1} $, 
 implemented by the statement at line 605 in the method $ {\tt Win (N)} $ shown on Figure~\ref{fig:Win}, pulls down 
 2 that resides as a leaf with siblings in heap $ {\mathsf H}_{2I-1} $. 
By the Credit Loss Characterization  Lemma~\ref{lem:losslessM}~\ref{item:losslessM3} page~\pageref{item:losslessM3}, 
move  $  \mathsf{win}(N)_{2I-1}$ is lossless, with
\begin{equation} \nonumber 
\lambda(N, \mathsf{s}_1) = 0.
\end{equation}  
At this point, $ {\tt Win(N) }$, that computes only the first $ N-1 $ moves of $\mathsf{win}(N)$, terminates because $ 2I -2 \geq N $, so the next move after the  move $ 2I-3 $ would fall beyond that range.
 
Obviously, the last 
two moves are valid because  no other node of any heap is smaller than 1 and 2. Moreover, by  The pq~Lemma~\ref{lem:1212}~\ref{lem:1212ii} and \ref{lem:1212iii} page \pageref{lem:1212} substituting 1 for $ p $ and 4 for $ q  $, 
taking into account that $  {\mathsf H}_{I}[\frac{I+1}{2}] = 4 $,
all the moves up to and including the move $ \mathsf{win}(N)_{\min \{2I - 3, N-1\} } $ were lossless.
Thus, since $ 2I-3\geq N-1 $, all the moves between $ \mathsf{win}(N)_I $ and $ \mathsf{win}(N)_{N-1} $ were lossless, except for move $\mathsf{win}(N) _{2I-1} $  that scored a credit one less than the maximal credit for $ 2I-1 $st move.

\medskip

This way I proved the following theorem.

\medskip

\begin{theorem} \label{thm:winscore} 
For every $ i < N $, 
\begin{equation} \label{eq:wincredits100}
\lambda_{\mathsf{win}(N)}(i)  = 
\left\{ \begin{array}{ll}
\lambda_{\mathsf{par}}(i) \mbox{ if } \; N \leq 7 \mbox{ or }  N = 2 ^{\lceil \lg N \rceil} -1
\mbox{ or } \; i < 2 ^{\lfloor \lg (N+1) \rfloor} -1 \\ \\
1 \mbox{ if } \; 8 \leq N \mbox{ and } \; 2 ^{\lfloor \lg (N+1) \rfloor} -1 \leq  i =  2 ^{\lceil \lg i \rceil} -4   \\ \\
0 \mbox{ otherwise}.
\end{array} \right.
\end{equation}
\end{theorem}
\begin{figure}[h] 
\begin{center}
\includegraphics[scale=1]{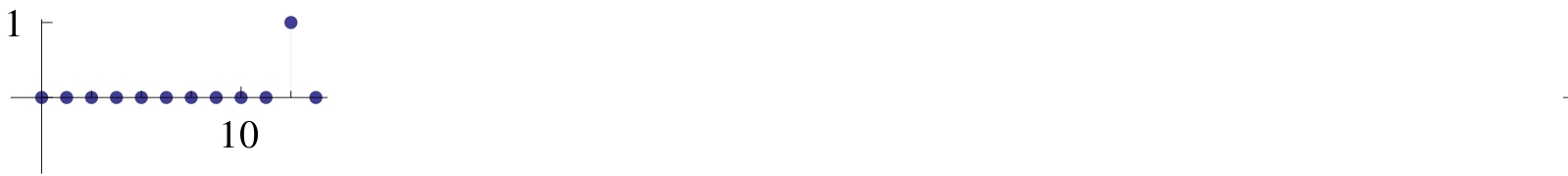} 
{\footnotesize \textit{N} = 14} 
\includegraphics[scale=1]{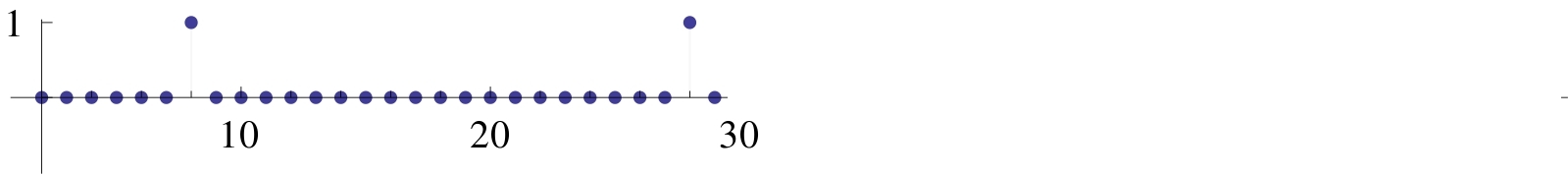} 
{\footnotesize \textit{N} = 30} 
\includegraphics[scale=1]{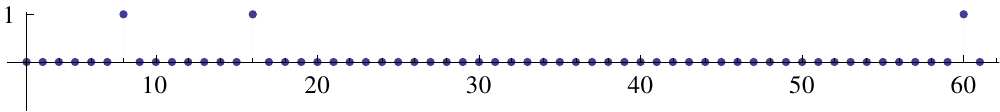} 
{\footnotesize \textit{N} = 62}
\end{center}
\caption{Discrete graphs of functions $ \lambda_{\mathsf{win}(N)}(i)  $ for $ N=14, 30 $, and $ 62 $ and
$ 1 \leq i \leq N-1 $. \label{fig:lambdawin} }
\end{figure}

{\bf Proof} Follows from the above discussion. In particular, the equality (\ref{eq:wincredits100}) can be easily extracted from the code of Java program shown on Figure~\ref{fig:Win} page~\pageref{fig:Win}.
\hspace*{\fill} $\Box$

\medskip

Since $ \lambda_{\mathsf{win}(12)}(i) = 0 $ for $ i \leq 11 $, $ {\mathsf{win}(12)} $ is an optimal strategy for $ 11 $. Thus the heaps
$ {\mathsf H}_2,..., {\mathsf H}_{12}$ created by the first 11 moves of $ {\mathsf{win}(12)} $  are all lossless and, therefore, worst-case heaps. Hence, all worst-case heaps of 12 nodes or less are lossless, too.

\medskip

I am going to prove in Section~\ref{ProofWin} that for every $ N \geq 2 $ (not just for $ N = 11 $), $ {\mathsf{win}(N)} $ is optimal for $ N-1 $.

\medskip

\begin{lem} \label{lem:sumwinstar}
Let for every $ i \geq 1 $, the \textit{delayed loss} function $ \lambda^* $ be defined by the equation
(\ref{eq:parStarDef}) in Lemma~\ref{lem:parstar} page \pageref{lem:parstar}.
For every $ N \geq 3 $, the following equality holds:
\begin{equation} \label{eq:sumwinStar1}
\Lambda_{\mathsf{win}(N)}(N-1)  = \sum_{i=2}^{N-1} \lambda^*(i).
\end{equation}
\end{lem}
{\bf Proof}. By (\ref{eq:parStarDef}) page~\pageref{eq:parStarDef},  for $ 2^{\lfloor \lg (N+1) \rfloor}-1 \leq i \leq  N-1$, $ \lambda^*(i) $ $ = $ $ 1 $ if, and only if, $ i =  2^{\lceil \lg i \rceil} - 4  $, the latter of which, by (\ref{eq:wincredits100}) page~\pageref{eq:wincredits100}, holds for $ 2^{\lfloor \lg (N+1) \rfloor}-1 \leq i \leq  N-1$ if, and only if,  $ \lambda(i)_{\mathsf{win}(N)}(i) $ $ = $ $ 1 $. Thus, 
\[ \lambda^*(i) = 0 = \lambda(i)_{\mathsf{win}(N)}(i) ,\]
 so that
\[ \Lambda_{\mathsf{win}(N)}(2^{\lfloor \lg (N+1) \rfloor}-1,N-1)  = \sum_{i=2^{\lfloor \lg (N+1) \rfloor}-1}^{N-1} \lambda^*(i) .\]
Therefore, since 
\[ \Lambda_{\mathsf{win}(N)}(N-1) = \Lambda_{\mathsf{win}(N)}(2^{\lfloor \lg (N+1) \rfloor}-2) + \Lambda_{\mathsf{win}(N)}(2^{\lfloor \lg (N+1) \rfloor}-1,N-1) ,\]
 in order to prove (\ref{eq:sumwinStar1}) it suffices to show that
\begin{equation} \label{eq:sumwinStar2}
\Lambda_{\mathsf{win}(N)}(2^{\lfloor \lg (N+1) \rfloor}-2)  = \sum_{i=2}^{2^{\lfloor \lg (N+1) \rfloor}-2} \lambda^*(i).
\end{equation}
Since by (\ref{eq:wincredits100}) page~\pageref{eq:wincredits100}, for all $ i \leq 2^{\lfloor \lg (N+1) \rfloor}-2$, $ \lambda_{\mathsf{par}}(i) $ $ = $ $ \lambda(i)_{\mathsf{win}(N)}(i) $, 
\[ \Lambda_{\mathsf{win}(N)}(2^{\lfloor \lg (N+1) \rfloor}-2) = \Lambda_{\mathsf{par}}(2^{\lfloor \lg (N+1) \rfloor}-2)  = \]
[by the equality (\ref{eq:sumparStar1}) in Lemma~\ref{lem:sumparstar} page~\pageref{lem:sumparstar}, substituting $ \lfloor \lg (N+1) \rfloor - 1 $ for $ D $]
 \[\sum_{i=2}^{2^{\lfloor \lg (N+1) \rfloor}-2} \lambda^*(i).\]
 This completes the proof.
\hspace*{\fill} $\Box$

\medskip

A Java code that implements the strategy $ \mathsf{win}(N) $ is shown on Figure~\ref{fig:Win}.
A worst-case heap on $ 28 $ nodes produced by the strategy $ \mathsf{win}(28) $ and the sequence of the first $ 27 $ pull downs of $ \mathsf{win}(28) $, which is the same as the sequence of the first $ 27 $ pull downs of $ \mathsf{win}(31) $, is shown on Figure~\ref{fig:win12-28out}. The sequence of 14 heaps created by the last 14 of these moves is shown in the \ref{sec:win31} page~\pageref{sec:win31}.

\medskip

\begin{figure}[h] 
\begin{center}
\includegraphics[scale=.5]{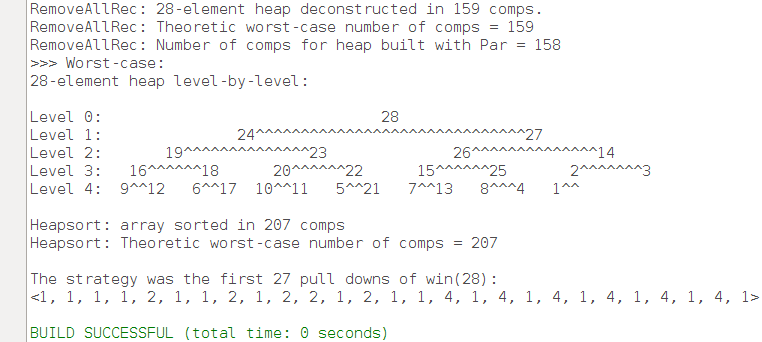}    
\end{center}
\caption{The heap $ H _{28} $ produced by the first 27 pull downs of the strategy $ \mathsf{win}(28) $ (an output of my Java program). The first $ 27 $ pull downs are at the bottom of the output. All pull downs in the last level (the last 13 pull downs) are lossless; however, there is no lossless pull down that the heap $ H _{28} $ would admit. \label{fig:win12-28out}}
\end{figure}

\medskip

Now, we are ready to compute the accumulated loss for each strategy $ \mathsf{win}(N) $.

\begin{lem} \label{lem:LambdawinIt}
	For every $ N \geq 2 $,
\begin{equation} \label{eq:LambdawinIt}
\Lambda_{\mathsf{win}(N)}(N-1) = \max \{ {\lfloor \lg (N+3)  \rfloor} , 3 \} -3.
\end{equation}
\end{lem}
{\bf Proof}. By the equality (\ref{eq:sumwinStar1}) in Lemma~\ref{lem:sumwinstar} page~\pageref{eq:sumwinStar1}, $ \Lambda_{\mathsf{win}(N)}(N-1) $ $ = $ $ \sum_{i=2}^{N-1} \lambda^*(i) $. 
Application of Lemma~\ref{lem:Lambda*closed} page \pageref{lem:Lambda*closed} completes the proof.
 \hspace*{\fill} $\Box$ 
 \medskip 
 
\begin{figure}[h] 
\begin{center}
\includegraphics[scale=1]{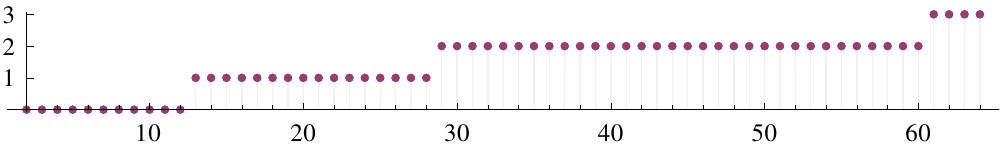} 
\end{center}
\caption{A discrete graph of function $ \Lambda_{\mathsf{win}(N)}(N-1) $ $ = $ $ \sum _{i=1}^{N-1} \lambda_{\mathsf{win}(N)}(i) $, where $  \lambda_{\mathsf{win}(N)}(i) $ is the function visualized on Figure~\ref{fig:lambdapar} page~\pageref{fig:lambdapar}. Clearly, for $ N \geq 5 $, $ \Lambda_{\mathsf{\mathsf{win}(N)}}(N-1) $ $ = $  $ {\lfloor \lg (N+3)  \rfloor} - 3 $. \label{fig:sumlambdawin} }
\end{figure}

 A graph of function $ \Lambda_{\mathsf{win}(N)}(N-1) $ is shown on Figure~\ref{fig:sumlambdawin}. 

\begin{2LBth} \label{thm:wincredits} 
The payoff $ P_{\mathsf{win}(N)}(N-1) $ after $ N -1$ moves of the strategy $ \mathsf{win}(N) $ for $ N \geq 2 $ is given by this formula:
\begin{equation} \label{eq:wincredits} 
2 (N - 1) \lfloor \lg (N - 1)  \rfloor - 
 2^{\lfloor \lg (N - 1)  \rfloor +2} + \min (\lfloor \lg (N - 1)  \rfloor, 2) + 4 + c, 
\end{equation}
where $ c $ is a binary function on the set of integers defined by:
\begin{equation} \label{eq:wincreditsC}
c = 
\left\{ \begin{array}{ll}
1 \mbox{ if } \; N \leq 2 ^{\lceil \lg N \rceil} - 4   \\ \\
0 \mbox{ otherwise}.
\end{array} \right.
\end{equation}
\end{2LBth}

\begin{figure}[h] 
\begin{center}
\includegraphics[scale=1]{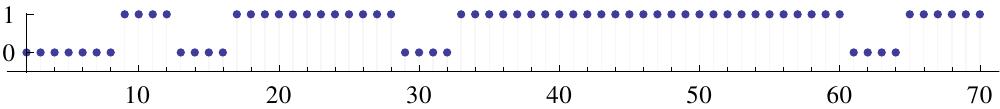} 
\end{center}
\caption{A discrete graph of function $ c $. \label{fig:c3} }
\end{figure}

Function $ c $ is visualized on Figure~\ref{fig:c3}. It raises from 0 to 1 when the strategy
$ \mathsf{par} $ performs a lossy move in the level $ \lfloor \lg N \rfloor $, and drops from 1 to 0  when the strategy
$ \mathsf{win}(N) $ performs a lossy move in that level.

\medskip

{\bf Proof}. Since, by (\ref{eq:payoffbound}) page~\pageref{eq:payoffbound}, $ P_{\mathsf{win}(N)}(N-1) $ $ = $ $ P ^{\,U\!B} (N-1) - \Lambda_{\mathsf{win}(N)} (N-1) $, by virtue of (\ref{eq:LambdawinIt}), it suffices to prove that  
for any $ N \geq 2 $, (\ref{eq:wincredits}) is equal to
\begin{equation} \label{eq:wincredits2000}
P ^{\,U\!B} (N-1) - \max \{ {\lfloor \lg (N+3)  \rfloor} , 3 \} +3 .
\end{equation}
 Once we pinpointed correct formulas, verification of that equality is easy.

\medskip

By Theorem 8.1 in \cite{suc:elem},
\[ P ^{\,U\!B} (N-1) =  (2N-1)\lfloor \lg (N-1) \rfloor
- 2 ^{\lfloor \lg (N-1) \rfloor +2} + 4 ,\]
thus (\ref{eq:wincredits2000}) is equal to
\begin{equation} \label{eq:wincredits2001}
(2N-1)\lfloor \lg (N-1) \rfloor
- 2 ^{\lfloor \lg (N-1) \rfloor +2} + 4 - \max \{ {\lfloor \lg (N+3)  \rfloor} , 3 \} +3.
\end{equation}
It suffices to show that (\ref{eq:wincredits}) is equal to (\ref{eq:wincredits2001}) for all $ N \geq 2 $.

\medskip

If $ N \leq 5 $ then $ \min (\lfloor \lg (N - 1)  \rfloor, 2)  $ 
$ = $ $ \lfloor \lg (N - 1)  \rfloor$, 
$ \max \{ {\lfloor \lg (N+3)  \rfloor} , 3 \} - 3 $ $ = $ $ 0 $, and $ c = 0 $ so that
(\ref{eq:wincredits}) reduces to:
\[2 (N - 1) \lfloor \lg (N - 1)  \rfloor - 
 2^{\lfloor \lg (N - 1)  \rfloor +2} + \lfloor \lg (N - 1)  \rfloor + 4 , 
=\]
\[(2N - 1) \lfloor \lg (N - 1)  \rfloor - 
 2^{\lfloor \lg (N - 1)  \rfloor +2}  + 4 , 
\]
and so does (\ref{eq:wincredits2001}). Thus (\ref{eq:wincredits}) and (\ref{eq:wincredits2001}) are equal 
in this case.

\medskip

If $ N \geq 6 $ then $ \min (\lfloor \lg (N - 1)  \rfloor, 2)  $ 
$ = $ $ 2$, and
$ \max \{ {\lfloor \lg (N+3)  \rfloor} , 3 \} - 3 $ $ = $ $ \lfloor \lg (N+3)  \rfloor -3 $, so that
(\ref{eq:wincredits}) reduces to:
\[2 (N - 1) \lfloor \lg (N - 1)  \rfloor - 
 2^{\lfloor \lg (N - 1)  \rfloor +2} + 6 + c =\]
\begin{equation} \label{eq:wincredits2005}
=(2N - 1) \lfloor \lg (N - 1)  \rfloor - 
 2^{\lfloor \lg (N - 1)  \rfloor +2} - \lfloor \lg (N - 1)  \rfloor+ 6 + c ,
\end{equation}
and (\ref{eq:wincredits2001}) reduces to
\begin{equation} \label{eq:wincredits2006}
 (2N-1)\lfloor \lg (N-1) \rfloor
- 2 ^{\lfloor \lg (N-1) \rfloor +2} - \lfloor \lg (N+3)  \rfloor +7  .
\end{equation}
Thus, in order to show that (\ref{eq:wincredits}) is equal to (\ref{eq:wincredits2001}) it suffices to
show that (\ref{eq:wincredits2005}) is equal to (\ref{eq:wincredits2006}), or, subtracting the common term
$ (2N-1)\lfloor \lg (N-1) \rfloor
- 2 ^{\lfloor \lg (N-1) \rfloor +2} + 6 $ from  (\ref{eq:wincredits2005}) and (\ref{eq:wincredits2006}), to show that
\begin{equation} \nonumber 
- \lfloor \lg (N - 1)  \rfloor + c = - \lfloor \lg (N+3)  \rfloor +1  .
\end{equation}
that is, 
\begin{equation} \nonumber 
\lfloor \lg (N+3)  \rfloor - \lfloor \lg (N - 1)  \rfloor  =  1 - c ,
\end{equation}
or, incorporating the definition (\ref{eq:wincreditsC}) of $ c $,
 \begin{equation} \label{eq:lg-lg=1-c}
\lfloor \lg (N+3)  \rfloor - \lfloor \lg (N-1) \rfloor  = 
\left\{ \begin{array}{ll}
0 \mbox{ if } \;  N \leq 2 ^{\lceil \lg N \rceil} -4   \\ \\
1 \mbox{ otherwise}.
\end{array} \right.
\end{equation}

 \medskip
 
 I will show that for every $ N \geq 3 $, (\ref{eq:lg-lg=1-c}) holds.
 
\medskip 
 
Indeed, if $N \leq 2 ^{\lceil \lg N \rceil} -4 $ then $  \lg (N+3)   $ $ \leq $
$  \lg (2 ^{\lceil \lg N \rceil} -1)   $ $ < $ 
$  \lg 2 ^{\lceil \lg N \rceil}    $ $ = $ 
 $  \lceil \lg N  \rceil  $. So, $ \lfloor \lg (N+3)  \rfloor $ $ < $
$ \lceil \lg N \rceil $, that is, $ \lfloor \lg (N+3)  \rfloor $ $ \leq $
$ \lceil \lg N \rceil -1 $
 $ = $ $ \lfloor \lg  (N-1)   \rfloor $. Thus
$ \lfloor \lg (N+3)  \rfloor $ $ \leq $   $ \lfloor \lg  (N-1)   \rfloor $. 
Hence, since 
$ \lfloor \lg  N   \rfloor $ is a non-decreasing function,
$ \lfloor \lg (N+3)  \rfloor $ $ = $  $ \lfloor \lg  (N-1)   \rfloor $ and
$ \lfloor \lg (N+3)  \rfloor - \lfloor \lg  (N-1)   \rfloor = 0 $, thus proving  (\ref{eq:lg-lg=1-c}) in this case.

\medskip

If, however, $N > 2 ^{\lceil \lg N \rceil} -4 $ then $N \geq 2 ^{\lceil \lg N \rceil} -3 $ and $ \lfloor \lg (N+3)  \rfloor $ $ \geq $
$ \lfloor \lg 2 ^{\lceil \lg N \rceil}  \rfloor $ $ = $ 
 $ \lfloor \lceil \lg N  \rceil  \rfloor $ $ = $ 
 $  \lceil \lg N  \rceil  $ $ = $ $ \lfloor \lg  (N-1)   \rfloor +1 $
 $ > $ $ \lfloor \lg  (N-1)   \rfloor  $. Thus
$ \lfloor \lg (N+3)  \rfloor $ $ > $   $ \lfloor \lg  (N-1)   \rfloor $. Hence,
$ \lfloor \lg (N+3)  \rfloor - \lfloor \lg  (N-1)   \rfloor \geq 1 $ and,
 since
for $ N \geq 3 $,  $ \lfloor \lg (N+3)  \rfloor - \lfloor \lg  (N-1)   \rfloor \leq 1 $,
$ \lfloor \lg (N+3)  \rfloor - \lfloor \lg  (N-1)   \rfloor = 1 $, thus proving  (\ref{eq:lg-lg=1-c}) in this case, too.

\medskip

Since there are no other cases, this completes the proof of  (\ref{eq:lg-lg=1-c}).

\medskip

Thus (\ref{eq:wincredits}) and (\ref{eq:wincredits2001}) are equal 
in this case, too. This completes the proof of the theorem.
\hspace*{\fill} $\Box$

\medskip

\begin{figure}[h] 
\begin{center}
\includegraphics[scale=1]{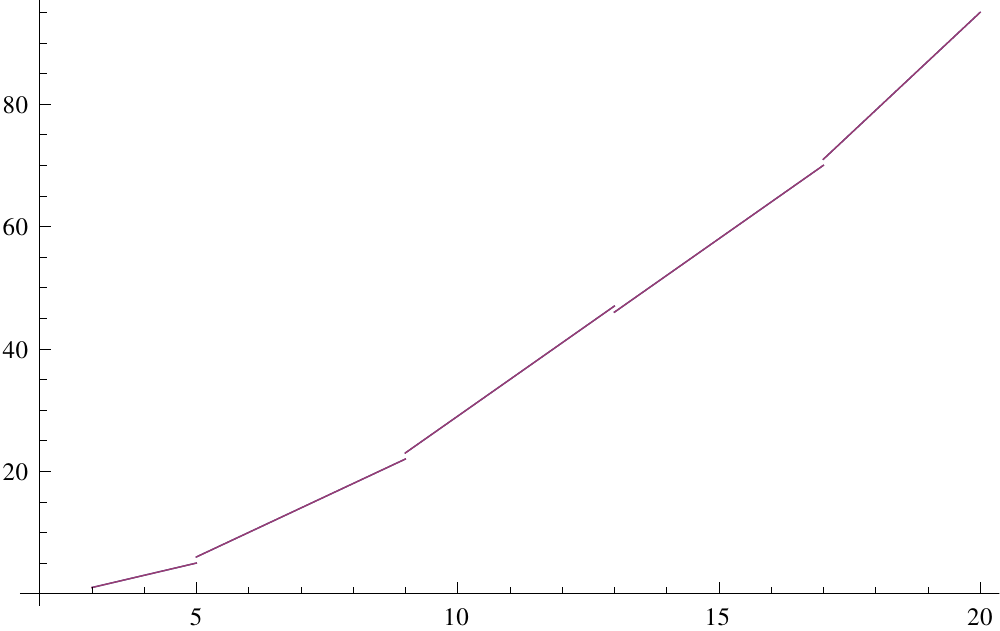} 
\end{center}
\caption{A graph of function $ P_{\mathsf{win}(N)}(N-1) $ of payoff for strategy $ \mathsf{win}(N) $. Also, a lower bound on the number of comparisons of keys performed in the worst-case  by $ {\tt RemoveAll} $ on an $ N $-element heap. \label{fig:cLB}}
\end{figure}

Figure~\ref{fig:cLB} shows a graph of the total payoff function $ P_{\mathsf{win}(N)} $ for $ \mathsf{win}(N) $.
Obviously, $ P_{\mathsf{win}(N)} $, given by (\ref{eq:wincredits}), is a lower bound on the number of comparisons of keys that $ {\tt RemoveAll} $ performs in the worst case on an $ N $-element heap. I will prove in the next section that it is an upper bound on that number, too.

\section{The proof of the winning} \label{ProofWin} 

Granted the optimality of $ \mathsf{par} $  for each $N=2^{\lfloor \lg N \rfloor} - 1$, all heaps produced by $ \mathsf{gry} $ are worst-case heaps as long as the corresponding pull downs are lossless. If no greedy strategy exists that would perform more consecutive lossless pull downs on any worst-case heap on $2^{\lfloor \lg N \rfloor} - 1$ nodes than $ \mathsf{gre} $ does on the worst-case heap $ {\mathsf H}_{2^{\lfloor \lg N \rfloor} - 1} $ then, taking into account that, by virtue of the equality (\ref{eq:wincredits100}) page \pageref{eq:wincredits100}, $ \mathsf{gre} $ loses no more than 1 credit relative to $ cr^{max} $ in the entire ${\lfloor \lg N \rfloor}$th level, all heaps produced by $ \mathsf{gre} $ are worst-case heaps, thus making the combination $ \mathsf{par} + \mathsf{gry} $ an optimal strategy for $ N $.

\medskip

So, the basic question that needs to be answered in order to decide the optimality of $ \mathsf{win}(N) $ for $ N $ is how many consecutive lossless pull downs a worst-case heap on  $ 2^{\lfloor \lg N \rfloor} - 1 $ nodes may admit. I am going to show that no more than $ 2^{\lfloor \lg N \rfloor} - 3 $, that is the same as $ \mathsf{win}(N) $ does admit on the heap $ {\mathsf H}_{2^{\lfloor \lg N \rfloor} - 1} $. From that the optimality of  $ \mathsf{win}(N) $ for $ N $ will follow.

\medskip

If $ N =  2^{\lfloor \lg N \rfloor}$ then pulling down 1 yields $2(\lfloor \lg N \rfloor - 1)$ credit, which is maximal, so it implements the desired lossless strategy for the level ${\lfloor \lg N \rfloor}$.

\medskip
  
If $ N =  2^{\lfloor \lg N \rfloor} + 1$ then pulling down 1 twice or pulling down 1 and then pulling down its new parent yields $2(\lfloor \lg N \rfloor - 1)$
$ + $ $2(\lfloor \lg N \rfloor - 1) - 1$ $ = $ $4\lfloor \lg N \rfloor - 3$ credit, which is maximal, so it implements the desired lossless strategy for the level ${\lfloor \lg N \rfloor}$.

\medskip

If $ N >  2^{\lfloor \lg N \rfloor} + 1$ then the first two pull downs in the last level ${\lfloor \lg N \rfloor}$ determine the maximum credit that the best greedy strategy in that level can collect.

\medskip

Let $ p $ and $ q $ be the first two nodes that ended up in the last level ${\lfloor \lg N \rfloor}$ as a result of the first two pull downs. (In the example of strategy $ \mathsf{gry} $ for $ N =  2^{\lfloor \lg N \rfloor} + 1$, it would be 1 and the first node in level $ {\lfloor \lg N \rfloor} - 1$ after the strategy $ \mathsf{par} $ was executed.)

\medskip

Consider heap $ H_{2^{\lfloor \lg N \rfloor} + 1} $ that was produced after these two pull downs. 
\begin{figure}[h] 
\begin{center}
\includegraphics[scale=.7]{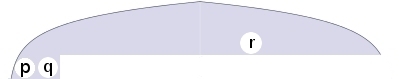}  
\end{center}
  \caption{Example heap $  H_{2^{\lfloor \lg N \rfloor} + 1} $ on $N = 2^{\lfloor \lg N \rfloor} + 1$ nodes with nodes $ p $, $ q $, and $ r $ shown. Nodes $ p $ and $ q $ are the only nodes in level $ \lfloor \lg N \rfloor $. Node $r$ (less than $ \max \{ p,q \} $) is in the level $ \lfloor \lg N \rfloor -1 $. The index of $ r $ is $ j $, and the indicies of  $ p $, $ q $ are $ N -1 $, $ N $, respectively. \label{fig:heapqr}}
\end{figure}
How far any strategy could proceed without a loss of credit relative to the upper-bound payoff $ P ^{\,U\!B} (2^{\lfloor \lg N \rfloor} + 1,m) $? It turns out that the index of the first node $ r $ in the level ${\lfloor \lg N \rfloor - 1}$ of heap $ H_{2^{\lfloor \lg N \rfloor} + 1} $ that is less than the maximum of $ p $ and $ q $ (see Figure~\ref{fig:heapqr}) 
puts the limit on the number of consecutive lossless pull downs that the strategy in question can make. This intuitively obvious fact has a surprisingly unobvious proof, quite a typical sample of the fine-grain complexity of this subject matter.

\begin{lem} \label{lem:parwin1} 
Let $ H $ be a heap on $N = 2^{\lfloor \lg N \rfloor} + 1 $ nodes, where $ N \geq 5$ ,
and let $ p $ and $ q $ be its last two nodes. If $ j $ is the index of a node $ r $ in the level ${\lfloor \lg N \rfloor - 1}$ of heap $ H $ with $ r < \max \{ p,q \} $ then heap $ H $ does not admit a sequence of more than 
$ 2 j - N$ consecutive lossless pull downs.
\end{lem}
{\bf Proof}. Since $ r < \max \{ p,q \} $, $ r $ is not a parent of $ p $ and $ q $, the only nodes in the last level $ \lfloor \lg N \rfloor $ of $ H $. So, $ r $ is not a parent of $ p $ or $ q $. Thus $ r $ is a leaf in $ H $, and so 
\begin{equation} \label{eq:parwin1-90} 
N < 2j .
\end{equation}
 Because $ \lfloor \lg j \rfloor $, the level number of $ r $, equals to ${\lfloor \lg N \rfloor - 1}$, 
\begin{equation} \label{eq:parwin1-100} 
\lfloor \lg (2j+1) \rfloor  = \lfloor \lg 2j \rfloor  =  \lfloor \lg N \rfloor .
\end{equation}
Let's assert that the lemma is false. Let $ \mathsf{v} $ be a sequence of $ 2j-N+1 $ consecutive lossless pull downs executed on $ H $.
By (\ref{eq:parwin1-90}), $ 2j-N+1 \geq 2$.  Let $ 2 \leq 2i \leq 2j-N+1 $, let $ H ^{\prime ^{2i}} $ be the heap on $ N+2i $ nodes produced by the first $ 2i $ pull downs $ \langle v _1,...,v_{2i} \rangle $ of $ \mathsf{v} $, and let $ S_{2i} $ be the set of all ancestors (proper and improper) of the nodes in the last level $ K = \lfloor \lg (N+2i) \rfloor$ of heap $ H ^{\prime ^{2i}} $, with convention $ H ^{\prime ^{0}} $ $ = $ $ H $. 
By virtue of (\ref{eq:parwin1-100}),
\begin{equation} \label{eq:LlgN}
K = \lfloor \lg N \rfloor.
\end{equation}
Since the number $ N+2i $ of nodes of $ H ^{\prime ^{2i}} $ is odd, the  number of nodes in the level $ K $ of $ H ^{\prime ^{2i}} $ is even. In particular, every element of level $ K $ in $ H ^{\prime ^{2i}} $ has a sibling in $ H ^{\prime ^{2i}} $, also an element of that level.

\medskip

First, I am going to show that a pair $ v_{2i-1}, v_{2i}$ of lossless pull downs may only pull down elements of  $ S_{2i-2} $. Let $ s $, 
$ t $ be the nodes pulled down by $ v_{2i-1}, v_{2i}$, respectively. For the pull down $ v_{2i-1}$ to be lossless, by the Credit Loss Characterization Lemma~\ref{lem:losslessM} page~\pageref{lem:losslessM}, taking into account that the last level of each heap in question has at least 2 nodes so that each of the conditions \ref{item:losslessM1} and \ref{item:losslessM2} of that lemma entails the disjunction of conditions \ref{item:losslessM3} and \ref{item:losslessM4}, $ s $  must be a sibling of a node in the level $ K $ of heap $ H ^{\prime ^{2i-2}} $ or a parent thereof. In particular, it must be an element of $ S_{2i-2} $.
For the pull down $ v_{2i}$ to be lossless, by the Credit Loss Characterization Lemma~\ref{lem:losslessM}, $ t $  must be a sibling of a node in the level $ K $ of heap $ H ^{\prime ^{2i-1}} $ or a parent thereof. 
The only such node that can possibly be not in $ S_{2i-2} $ is the (new) parent of $ s $ in $ H ^{\prime ^{2i-1}} $, which, however, has one child only ($ s $, that is) because $ H ^{\prime ^{2i-1}} $ has an even number of elements and so its last level has an odd number of nodes, so that the last element ($ s $, that is) of the last level of $ H ^{\prime ^{2i-1}} $ cannot have a sibling. Therefore, by , by the Credit Loss Characterization Lemma~\ref{lem:losslessM}, the (new) parent of $ s $ in $ H ^{\prime ^{2i-1}} $ cannot be pulled down or otherwise $v_{2i}$ would be lossy.

\medskip

Second, I prove by induction on $ i $ that for every $ i $ with $ 0 \leq 2i \leq 2j-N+1 $, the two smallest elements of
$ S_{2i}$ are $ \{ p,q \} $. 
Indeed, $ p $ and $ q $ are the only nodes in the last level of the heap $ H $ $ = $  $ H ^{\prime ^{\, 0}} $, so
$ S_0 = \{ p,q \} \cup A_0 $, where $ A_0 $ is the set of proper ancestors of $ \{ p,q \} $ in the heap $ H $. Since no element in $ A_0 $ is less than $ \max \{ p,q \}  $, the two smallest elements of
$ S_0$ are $ \{ p,q \} $. 
For $ i>0 $, we have $ S_{2i} = S_{2i-2} \cup A_{2i}$, where $ A_{2i} $ is the set of proper ancestors of the nodes in the last level of heap $ H ^{\prime ^{2i}} $, because moves $ v_{2i-1}, v_{2i}$  pulled down elements of  $ S_{2i-2} $. Since no element in $ A_{2i} $ is less than $ \max \{ p,q \}  $, the two smallest elements of
$ S_{2i}$ are the same as the two smallest elements of
$ S_{2i-2}$, that is, $ \{ p,q \} $.

\medskip

\begin{figure}[h] 
\begin{center}
\includegraphics[scale=.7]{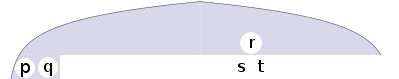}  
\end{center}
  \caption{Example heap $  H_{2^{\lfloor \lg N \rfloor} + 1} $ of Figure~\ref{fig:heapqr} page~\pageref{fig:heapqr} with $ r $'s future children $ s $ and $ t $ (not necessarily different from $ p $, $ q $) shown. The nodes $ s $ and $ t $ are pulled down by the last two moves $ v_{2j-N} $ and $ v_{2j-N+1} $ the of sequence $ \mathsf{v} $. Their new indicies (after the said pull downs) are $ 2j $ and $ 2j+1 $, respectively. If the first $ 2j-N $ pull downs of $ \mathsf{v} $ are lossless then $ 2j + 1-N $th pull down $ v_{2j + 1-N} $ of $ t $ is lossy. \label{fig:heapqrst}}
\end{figure}Because $ r < \max \{ p,q \} $,
$ r \not \in S_{2m} $ for any $ 0 \leq 2m \leq 2j-N+1  $. Therefore, none of the moves of the sequence $ \mathsf{v} $ pulls down $ r $, which stays put during the execution of $ \mathsf{v} $. Thus
 the last two moves $ v_{2j-N} $ and $ v_{2j-N+1} $ the of sequence $ \mathsf{v} $ result in attaching the (new) children $ s $, $ t $ to node $ r $ (see Figure~\ref{fig:heapqrst}). They must both pull down elements from the set $ S_{2j-N+1} $. Since there is at most one element of $ S_{2j-N+1} $ that is less than $ r $ and qualifies as its child, such a case is clearly impossible. This contradicts the assertion that the lemma is false.
\hspace*{\fill} $\Box$


\medskip

The following technical lemma uses function $ \Lambda_{\tilde{X} \! X } $ of accumulated loss of credit given by the equation (\ref{eq:LambdaHH0}) page \pageref{eq:LambdaHH0}. It derives properties of some heaps $ H $ and their \textit{residua} $ \tilde{H} $ from properties of virtually unrelated heaps $ G $ and their \textit{residua} $ \tilde{G} $ of the same sizes as $ H $ and  $ \tilde{H} $, respectively. Although the facts spelled out by this lemma may seem inconsequential and intuitively obvious, and have  elementarily-algebraic and straightforward proofs, they have allowed me to cut the lengths\footnote{By about a factor of four or so.} of some of the lengthiest proofs of some fundamental results in this Section.

\begin{diaglem} \label{lem:diag} 
Let $ H $ and $ G $ be heaps on $ N > 2 $ nodes, and $ \tilde{H} $ and $ \tilde{G} $ be their respective residua on $2 \leq M < N $ nodes.
If $ H $ is a worst-case heap then the following are true:
\begin{enumerate}
     \renewcommand\labelenumi{\theenumi}
     \renewcommand{\theenumi}{(\roman{enumi})}
\item \label{item:diag1} 
If 
$ \tilde{G} = \tilde{H}$
then $ \Lambda_{\tilde{H} \! ,  H} \leq \Lambda_{\tilde{G} \! ,  G} $.
\item \label{item:diag2} 
If $ \tilde{G} $ is a worst-case heap and  $ \Lambda_{\tilde{H} \! ,  H} \geq \Lambda_{\tilde{G} \! ,  G} $ then $ \tilde{H} $ and $ G $ are a worst-case heaps and $ \Lambda_{\tilde{H} \! ,  H} = \Lambda_{\tilde{G} \! ,  G} $.
\item \label{item:diag4} 
If $ \tilde{G} $ is a worst-case heap but $ \tilde{H} $ is not, and $ \Lambda_{\tilde{G} \! ,  G} \leq 1$ then $ G $ is a worst-case heap.
\item \label{item:diag5} 
If $ \tilde{G} $ is a worst-case heap but $ G $ is not, and $ \Lambda_{\tilde{G} \! ,  G} \leq 1$ then $ \tilde{H} $ is a worst-case heap and $ \Lambda_{\tilde{H} \! ,  H} = 0$.
\end{enumerate}
\end{diaglem}
{\bf Proof}. 
\begin{figure}[h] 
\begin{center}
\includegraphics[scale=1.2]{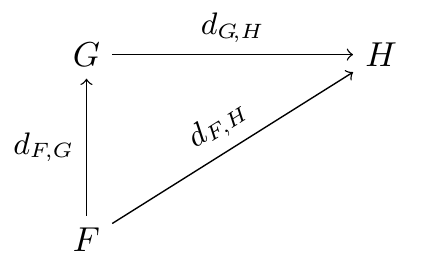} 
\end{center}
\caption{An example of a commuting diagram; $ d_{F \! ,  G} $ $ +$  $ d_{G \! ,  H} $
$ = $ $ d_{F \! ,  H} $.  \label{fig:diag100}}
\end{figure}
Consider a differential diagram (a weighted directed graph) $ D $, an example of which is shown on Figure~\ref{fig:diag100},
 whose nodes $ H $, $ G $, ..., are heaps and weights of edges
$ (H,G) $ are differences
\begin{equation} \label{eq:defdGH}
d_{G \! , H} = C_{\tt RemoveAll()} (H) - C_{\tt RemoveAll()} (G), 
\end{equation}
 where $C_{\tt RemoveAll()} (X)$
is the number of comparisons of keys performed by the execution of $ X.{\tt RemoveAll}() $ on heap $ X $. Obviously, any differential diagram $ D $ commutes, that is, any two paths in $ D $ with the same source and destination have equal weights.

\begin{figure}[h] 
\begin{center}
\includegraphics[scale=1.2]{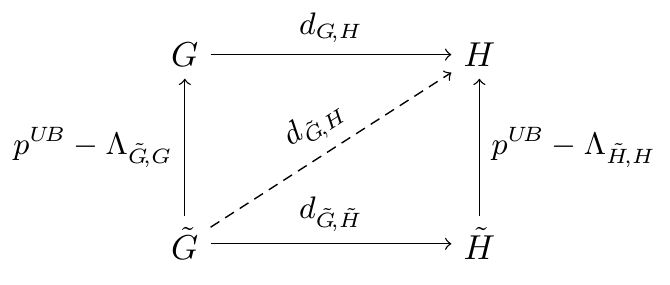} 
\end{center}
\caption{Another commuting diagram.  \label{fig:diag101}}
\end{figure}

In particular, the diagram with heaps $ G$, $ H $ on $ N $ nodes each, 
and their respective residua $ \tilde{G} $ and $ \tilde{H} $ on $ M < N $ nodes each, visualized on Figure~\ref{fig:diag101}, commutes.

\medskip

Here $ p^{U\! B} $ $ = $ $ P^{U\!B} ( M,N-1) $ is the upper bound on credit for the sequence of pull downs that produces an $ N $-element heap from and $ M $-element heap,  and
$ \Lambda_{\tilde{G} \! ,  G} $ and $ \Lambda_{\tilde{H} \! ,  H} $
are accumulated losses of credits given by the equation (\ref{eq:LambdaHH0}) page \pageref{eq:LambdaHH0}, so that 
\[ p^{U\! B} -  \Lambda_{\tilde{G} \! ,  G} =   \] 
[by (\ref{eq:PHH=PUB-LambHH}) page \pageref{eq:PHH=PUB-LambHH}]
\[ = P _{\tilde{G} \! ,  G} = \]
[by (\ref{eq:defdGH}) and (\ref{eq:PHH1}) page \pageref{eq:PHH1}]
\[ =  d _{\tilde{G} \! ,  G}  ,\]
and
\[ p^{U\! B} -  \Lambda_{\tilde{H} \! ,  H} =   d _{\tilde{H} \! ,  H}  .\]

\medskip

Since the diagram on Figure~\ref{fig:diag101} commutes, we have:
\[ p^{U\! B} - \Lambda_{\tilde{G} \! ,  G} + d_{G \! ,  H} = d_{\tilde{G} \! ,  H} = d_{\tilde{G} \! ,  \tilde{H}} + p^{U\! B} - \Lambda_{\tilde{H} \! ,  H},\]
which proves 
\begin{equation} \label{eq:diagmain1} 
d_{G \! ,  H} +  \Lambda_{\tilde{H} \! ,  H}  = d_{\tilde{G} \! ,  \tilde{H}}  + \Lambda_{\tilde{G} \! ,  G} .
\end{equation}

\medskip

Let's assume that $ H $ is a worst-case heap. In such a case, 
\begin{equation} \label{eq:diag201} 
d_{G \! ,  H} \geq 0.
\end{equation}
If $ \tilde{G} = \tilde{H} $ then $ d_{\tilde{G} \! ,  \tilde{H}} = 0$ and,  by (\ref{eq:diagmain1}),
\[\Lambda_{\tilde{H} \! ,  H} = d_{\tilde{G} \! ,  \tilde{H}}+ \Lambda_{\tilde{G} \! ,  G}  - d_{G \! ,  H}  =  \Lambda_{\tilde{G} \! ,  G} - d_{G \! ,  H}
\leq \Lambda_{\tilde{G} \! ,  G},\]
that is
\begin{equation} \label{eq:diag202} 
\Lambda_{\tilde{H} \! ,  H} \leq \Lambda_{\tilde{G} \! ,  G}.
\end{equation}
This proves \ref{item:diag1}.

\medskip

If 
\begin{equation} \label{eq:diag203} 
\Lambda_{\tilde{H} \! ,  H} \geq \Lambda_{\tilde{G} \! ,  G}.
\end{equation}
then by (\ref{eq:diagmain1})
\[d_{\tilde{G} \! ,  \tilde{H}} =    d_{G \! ,  H} + \Lambda_{\tilde{H} \! ,  H} - \Lambda_{\tilde{G} \! ,  G}  \geq d_{G \! ,  H} \geq \]
[by (\ref{eq:diag201})]
\[ \geq 0,\]
that is,
\begin{equation}  \nonumber
d_{\tilde{G} \! ,  \tilde{H}}   \geq 0.
\end{equation}
Thus if $ \tilde{G} $ is a worst-case heap then $ \tilde{H} $ is a worst-case heap, too.
Hence,
\begin{equation} \label{eq:diag204} 
d_{\tilde{G} \! ,  \tilde{H}}   = 0.
\end{equation}
By \ref{item:diag1}, inequality (\ref{eq:diag202}), and inequality (\ref{eq:diag203})
\begin{equation} \label{eq:diag205} 
\Lambda_{\tilde{H} \! ,  H} = \Lambda_{\tilde{G} \! ,  G}.
\end{equation}
By virtue of (\ref{eq:diagmain1}) and (\ref{eq:diag204}) and (\ref{eq:diag205}),
\[d_{G \! ,  H}   = 0.\]
Therefore, $ G $ is a worst-case heap.
This and (\ref{eq:diag205}) prove \ref{item:diag2}.

\medskip

If $ \tilde{G} $ is a worst-case heap but $ \tilde{H} $ is not then
\begin{equation} \label{eq:diag206} 
d_{\tilde{G} \! ,  \tilde{H}}   \leq -1.
\end{equation}
Therefore, if $ \Lambda_{\tilde{G} \! ,  G} \leq 1 $ then (\ref{eq:diagmain1}) yields:
\[d_{G \! ,  H} =   d_{\tilde{G} \! ,  \tilde{H}}  + \Lambda_{\tilde{G} \! ,  G} - \Lambda_{\tilde{H} \! ,  H} \leq -1+1 - 
\Lambda_{\tilde{H} \! ,  H} \leq 0.\] 
Since $ H $ is a worst-case heap, so is $ G $.
This proves \ref{item:diag4}.

\medskip

If $ \tilde{G} $ is a worst-case heap but $ G $ is not, and $ \Lambda_{\tilde{G} \! ,  G} \leq 1$ then, by \ref{item:diag4}, $ \tilde{H} $ is a worst-case heap, and so
\begin{equation} \label{eg:diag901}
d_{\tilde{G} \! ,  \tilde{H}} = 0.
\end{equation}
Also, since $ H $ is a worst-case heap,
\begin{equation} \label{eq:diag900} 
d_{G \! ,  H} > 0.
\end{equation}
From (\ref{eg:diag901}) and  (\ref{eq:diagmain1}) I infer
\begin{equation} \label{eq:diag902}  \nonumber
d_{G \! ,  H} +  \Lambda_{\tilde{H} \! ,  H}  =   \Lambda_{\tilde{G} \! ,  G} ,
\end{equation}
that is,
\begin{equation} \label{eq:diag903}  \nonumber
 \Lambda_{\tilde{H} \! ,  H}  =   \Lambda_{\tilde{G} \! ,  G} - d_{G \! ,  H} <
\end{equation}
[by (\ref{eq:diag900})]
\begin{equation} \label{eq:diag90g} \nonumber
< \Lambda_{\tilde{G} \! ,  G}   \leq 1.
\end{equation}
Hence,
\[  \Lambda_{\tilde{H} \! ,  H} < 1 ,\]
that is,
\[  \Lambda_{\tilde{H} \! ,  H} =0 .\]
This proves \ref{item:diag5}.
\hspace*{\fill} $\Box$ 

\medskip

The following lemma imposes rather rigid limits on how far one optimal strategy can fall behind another. It is a consequence of the existence of strategies that are lossless up until level 3 and, like the substrategy $ \langle 1,2,... \rangle $  of alternating pull downs of 1 an 2, lose only one credit per level relative to their respective upper bounds $ P^{U\!B} ( 2^K-1,2^{K+1}-2) $ 
\footnote{One can directly compute $ P^{U\!B} ( 2^K-1,2^{K+1}-2)  $ for any $ K \geq 1 $ as 
$ \sum _{i=2^K-1} ^{2^{K+1}-2} cr ^{U \! B}(i) $ $ = $ $ \sum _{i=2^K-1} ^{2^{K+1}-2} 
2 \lfloor \log (i+1) \rfloor - 3 $ $ = $ $ \sum _{i=2^K-1} ^{2^{K+1}-2} 
2 K - 3 $ $ = $ $ K 2^{K+1} - 3 $. Intuitively speaking, there are $ 2^K $ pull downs between $ 2^K-1 $ and $ 2^{K+1}-2 $, each collecting maximal credit of $2K$, except for the first one that loses 2 credits relative to $2K$ and for the second one that that loses 1.}
 for each level $ K \geq3 $, with which optimal strategies must compete in order to win.

\begin{complem} \label{lem:2loss} 
For every worst-case heap $ H $  
 on $N \geq 2$ nodes,
\begin{enumerate}
     \renewcommand\labelenumi{\theenumi}
     \renewcommand{\theenumi}{(\roman{enumi})}
\item  \label{item:2loss1} 
the last level of $ H $ can lose no more than 1 credit, and
\item \label{item:2loss2}
if the last level of $ H $ is lossy than heap $ H $'s complete residue $ \tilde{H} $ on \linebreak $ M  $ $ = $ $ 2^{\lfloor \lg N \rfloor} - 1  $ nodes is a worst-case heap.
\end{enumerate}
\end{complem}
{\bf Proof}.
Let $ H $ be a 
worst-case heap on $N \geq 2 $ nodes,  $ \tilde{H} $ be the $ H $'s complete residue on $M = 2^{\lfloor \lg N \rfloor} - 1 $ nodes, and  $ \Lambda_{\tilde{H} \! ,  H} $ be the loss of credit, relative to the upper bound $  P^{U\!B} ( M, N-1) $, incurred by the sequence of $ N -M $ consecutive pull downs that produce $ H $ out of $ \tilde{H} $. Since 
$ N \leq  2^{\lfloor \lg N \rfloor+1} - 1 $,
we have:
\[ N-M \leq  2^{\lfloor \lg N \rfloor+1} - 1 -  2^{\lfloor \lg N \rfloor} + 1 =  2^{\lfloor \lg N \rfloor}, \]
that is
\begin{equation} \label{eq:2loss200}
N - M \leq M+1.
\end{equation}
Consider a sequence  $ \langle 1,2,... \rangle$ of $ N-M $ alternating pull downs of $ 1 $s and $ 2 $s consecutively applied to a heap $ \tilde{G} $ on $ M $ nodes.
Let $ G $ be the heap on $ N $ nodes that is a result of that application and 
$ \Lambda_{\tilde{G} \! ,  G} $ be the loss of credit, relative to the upper bound $  P^{U\!B} ( M, N-1) $, incurred by that sequence.
By virtue of  the pq~Lemma~\ref{lem:1212}~\ref{lem:1212i} 
page~\pageref{lem:1212i}, applicable because of the inequality (\ref{eq:2loss200}), 
\begin{equation} \label{eq:2loss209} 
\Lambda_{\tilde{G} \! ,  G } \leq 1.
\end{equation}
\medskip

\ref{item:2loss1} Let $ \tilde{G} $ be equal to $ \tilde{H} $. Application of Diagram
 Lemma~\ref{lem:diag}~\ref{item:diag1} yields
\[ \Lambda_{\tilde{H} \! ,  H} \leq \Lambda_{\tilde{G} \! ,  G } ,\]
that is, by virtue of (\ref{eq:2loss209}),
\[\Lambda_{\tilde{H} \! ,  H} \leq 1.\]
This completes the proof of \ref{item:2loss1}.

\medskip

\ref{item:2loss2} If the last level of $ H $ is lossy then
\[\Lambda_{\tilde{H} \! ,  H} \geq 1,\]
that is, by virtue of (\ref{eq:2loss209}),
\begin{equation} \label{eq:2loss210} 
\Lambda_{\tilde{G} \! ,  G } \leq \Lambda_{\tilde{H} \! ,  H} .
\end{equation}

Let $ \tilde{G} $ be a worst-case heap. By virtue of Diagram
 Lemma~\ref{lem:diag}~\ref{item:diag2}, applicable because of (\ref{eq:2loss210}) $\tilde{H}$ is a worst-case heap.
This completes the proof of \ref{item:2loss2}.
\hspace*{\fill} $\Box$ 
 
\medskip

\begin{lem} \label{lem:lossless}
Heap $ H $ on $ N =  2^{\lfloor \lg N \rfloor} $ nodes, where $ N \geq 2 $, admits lossless sequence of of $ N - 1 $ consecutive pull downs if, and only if, its last node $ H [N] $ is $ 1 $ and its parent $ H [\frac{N}{2}] $ is $ 2 $.
\end{lem}
{\bf Proof}. ($ \Leftarrow $) The sequence of $ N -1$ consecutive alternating pull downs \linebreak $ \langle 1, 2, 1, 2, ... ,1,2,1 \rangle $ is an example of the said lossless sequence.

\medskip

($ \Rightarrow $) If $ 1 $ is not in the level $ \lg N $ of heap $ H $ then it has to be pulled down at certain point before completion of any sequence of $ N-1 $ consecutive pull downs. Because $ 1 $ has no children, that pulling down cannot be lossless except  
 in a heap on $ K =  2^{\lceil \lg K \rceil} - 1 $ nodes, which is not the case here. So, $ H [N] = 1 $. The only lossless pull downs that can be applied to $ H $ in this case is pulling down $ 1 $ or its parent $ H [\frac{N}{2}] $. Each of those will move the parent of $ 1 $ from level $ \lg N -1$ down to level $ \lg N $. If the parent $ H [\frac{N}{2}] $ of $ 1 $ in heap $ H $ is not $ 2 $ then $ 2 $ resides somewhere else as a leaf in the  level $ \lg N -1 $ of heap $ H $ after the index $\lfloor \frac{N}{2} \rfloor$, and will terminate any lossless sequence of pull downs right after $ 1 $ became the left child of $ 2 $. So the said parent $ H [\frac{N}{2}] $ must be $ 2 $ in order for the said sequence of of $ N-1 $ consecutive lossless  pull downs to exist. 
\hspace*{\fill} $\Box$

\begin{lem} \label{lem:lossless12}
If heap $ H $ on $ N =  2^{\lfloor \lg N \rfloor} $ nodes, where $ N \geq 4 $, admits sequence of $ N-1 $ consecutive lossless pull downs then in any such sequence $ 1 $ has to be pulled down at least once and $ 2 $ has to be either pulled down at least twice or demoted at least once and pulled down at least once.
\end{lem}
{\bf Proof}. By Lemma \ref{lem:lossless}, the last node $ H [N] $ is $ 1 $ and its parent $ H [\frac{N}{2}] $ is $ 2 $. 
The only lossless pull down that heap $ H $ admits is pulling down $ 1 $ or its parent $ 2 $, any of which creates a heap $ H^\prime $ on $ N +1 $ nodes whose last two nodes (the only two nodes in the last level $ k $ of heap $ H $) are $ 1 $ and $ 2 $. The parent $ a $ of $ 2 $ in heap $ H^\prime $ is the same as the parent of $ 2 $ in heap $ H $ as it got demoted with the demotion of $ 2 $ caused by the pull down of $ 1 $ or with the pull down of $ 2 $. Since $ 1 $ and $ 2 $ are siblings in heap $ H^\prime $, $ a $ is also the parent of $ 1 $ in heap $ H^\prime $. The right sibling of $ 2 $ in heap $ H $ is not $ 1 $ because $ 1 $ is the left child of $ 2 $ in heap $ H $. So, the right sibling of $ 2 $ in heap $ H $ is greater than $ 2 $ and, therefore, the parent $ a $ of $ 2 $ in heap $ H $ is greater than $ 3 $. Thus $ 3 $ is not the parent of $ 1 $ or $ 2 $ in heap $ H^\prime $, so $ 3 $ must be a leaf in heap $ H^\prime $ and as such it must belong to the level $\lfloor \lg N \rfloor -1$ in that heap. As a leaf in heap  $ H^\prime $, it cannot be pulled down without a loss. So it will remain in the level $\lfloor \lg N \rfloor -1$, acquiring, eventually $ 1 $ and $ 2 $ as its children since no other node can be a child of $ 3 $ and all nodes in the level $\lfloor \lg N \rfloor -1$ of any heap on  $ 2^{\lfloor \lg N \rfloor +1} -1 $ nodes produced from $ H $ by any sequence of $ N-1 $ pull downs has two children each.
So, in order for the sequence of $ N - 1 $ to be lossless, both $ 1 $ and $ 2 $ must be pulled down or otherwise they won't become the children of $ 3 $.

\medskip

Whatever was the case in the sequence described above, $ 1 $ is pulled down at least once (while making it a child of $ 3 $), and $ 2 $ is either demoted once (if $ 1 $ is pulled in the first move) and pulled down once (while making it a child of $ 3 $) or pulled down twice (in the first move and while making it a child of $ 3 $).
\hspace*{\fill} $\Box$

\begin{lem} \label{lem:losslessNot123}
If heap $ H $ on $ N =  2^{\lfloor \lg N \rfloor} $ nodes, where $ N \geq 4 $, admits lossless sequence of of $ N-1 $ consecutive pull downs then any such lossless sequence produces a complete heap $ H^\prime $ on $ 2^{\lfloor \lg N \rfloor +1}-1 $ nodes whose first node of its last level is greater than or equal to $ 4 $.
\end{lem}
{\bf Proof}. By Lemma \ref{lem:lossless}, $ H[N]=1 $ and $ H[\frac{N}{2}]=2 $. Therefore,  neither $ 1 $ nor $ 2 $ are  children of node $ 3 $ in heap $ H $ ($ 1 $ is the child of $ 2 $ in $ H $ and sibling of $ 2 $ in $ H $ is greater than $ 2 $) nor its descendants.

\medskip

By Lemma \ref{lem:lossless12}, node $ 1 $ cannot maintain its index $ N $ because it is pulled down from that index. Node $ 2 $ can only be moved at index $ N $ as a result of its demotion. It cannot maintain its index $ N $ because is pulled down from that index. So the first node of its last level of heap $ H^\prime $ is neither $ 1 $ nor $ 2 $.

\medskip

Node $ 3 $ gets never moved at index $ N $, because it is not an ancestor of nodes $ 1 $ and $ 2 $. So, it cannot be there in heap $ H^\prime $. (An analysis of the part of the proof of Lemma \ref{lem:lossless12} regarding position of $ 3 $ in heap $ H $ could yield a more refined argument.)

\medskip

So, the said first node in the last level of  heap $ H^\prime $ is neither $ 1 $, $ 2 $, nor $ 3 $. Hence, it is greater than or equal to $ 4 $.
\hspace*{\fill} $\Box$

\begin{lem} \label{lem:losslessNot123+}
If heap $ H $ on $ N =  2^{\lceil \lg N \rceil}-1 $ nodes, where $ N \geq 3 $, admits a sequence of $ N+1 $ consecutive lossless pull downs then any such lossless sequence produces a complete heap $ H^\prime $ on $ 2^{\lceil \lg N \rceil +1}-1 $ nodes whose first node of its last level is greater than or equal to $ 4 $.
\end{lem}
{\bf Proof}. 
If $ H $ admits a sequence of $ N+1 $ consecutive lossless pull downs then the heap $ G $ that is produced by the first of those pull down admits $ N $ consecutive lossless pull downs. Since the size of $ G $ is $ M $  $= $  $ 2^{\lceil \lg N \rceil} $ $ = $ $ 2^{\lceil \lg M \rceil} $ and $ N  = M-1 $, Lemma~\ref{lem:losslessNot123} applies, a consequence of which yields the thesis of this lemma. 
\hspace*{\fill} $\Box$ 
  
\medskip
 
The following seemingly inconsequential Lemma is critical for my proof that  $ \mathsf{win}(N) $ is an optimal strategy for $ N $. 
  
\begin{charlem} \label{lem:parwin4}
The first node of the last level of  any complete worst-case heap on more than three nodes is
greater or equal to $4$.
\end{charlem}
{\bf Proof} Let $ H $ be a complete worst-case heap on $ 2^{k} - 1 $ nodes, where $ k \geq 3 $. I will prove by induction on $ k $ that the first element  $ H [2^{k-1}] $ of the last level $ k - 1 $ of $ H $ satisfies this inequality:

\begin{equation} \label{item:parwin4}
H [2^{k-1}] \geq 4.
\end{equation}

\medskip

There are only four worst-case heaps $ H $ on $ 7 $ nodes, and for each of them $ H [4] = 4 $ or $ H [4] = 5 $. (One way of showing it is to inspect the results of all possible 6-elements sequences of lossless pull downs.)
Thus the lemma's thesis (\ref{item:parwin4}) is true for $ k = 3 $. 

\medskip

For the inductive step, let's assert that
 $ k $ is the smallest number greater than $ 3 $ for which the lemma's thesis (\ref{item:parwin4}) is false.
Let $ \mathsf{v} $ be a strategy that produces $ H $ in its $ 2^{k} - 2 $nd move $ v_{2^{k} - 2} $, and $ \mathsf{H} $ $ = $
$ \langle H_i \mid i \in \omega ^+ \rangle$  be the sequence of heaps produced by $ \mathsf{v} $. We have
$ H =  H_{2^{k} - 1}$. By the above assertion, we have:
\begin{equation} \label{eg:1parwin4}
H_{2^{k} - 1}[2^{k-1}] \leq 3.
\end{equation}

\medskip

By Lemma~\ref{lem:losslessNot123+}, if the last level $ k-1 $ of heap $ H_{2^{k} - 1}$ were lossless, that is, if all pull downs $ v_{2^{k-1} - 1} $ through $ v_{2^{k} - 2} $ earned maximal credits, then the first element $ H_{2^{k} - 1}[2^{k-1}]$ of level $ k-1 $ would have to be at least $ 4 $, which, by (\ref{eg:1parwin4}), is not the case. So, the last level $ k-1 $ of heap $ H_{2^{k} - 1}$ is lossy and, by Competition
Lemma \ref{lem:2loss}~\ref{item:2loss2}, the previous complete heap $ H_{2^{k-1} - 1}$ must be a worst-case heap. By the induction hypothesis, the first node $ H_{2^{k-1} - 1}[2^{k-2}]$ of the last level $ k-2 $ of heap $ H_{2^{k-1} - 1}$ must satisfy this inequality:
\begin{equation} \label{eg:2parwin4}
H_{2^{k-1} - 1}[2^{k-2}] \geq 4.
\end{equation}

\medskip

Because of that, the only scenario under which the inequality (\ref{eg:1parwin4}) is satisfied
 is that  the node  $ H_{2^{k} - 1}[2^{k-1}]$ of heap $ H_{2^{k} - 1}$ was pulled down there by the move $v_{2^{k-1} -1 }$ executed on heap $ H_{2^{k-1} - 1}$ and never pulled down 
 thereafter. The heap $ H_{2^{k-1}} $ (not to be confused with heap $ H_{2^{k}-1} $) produced by move $v_{2^{k-1} -1 }$ has only one node $ p = H_{2^{k-1}}  [2^{k-1}] $ in its last level $ k-1 $, which by (\ref{eg:1parwin4}), satisfies $ p \leq 3 $. Its parent, $ q =  H_{2^{k-1}}  [2^{k-2}] $, satisfies $ q \geq 4 $ by virtue of (\ref{eg:2parwin4}).  

\medskip

\begin{figure}[h] 
\begin{center}
\includegraphics[scale=.7]{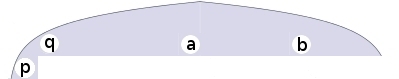}  
\end{center}
  \caption{Example heap $  H_{2^{k-1}} $ on $ 2^{k-1} $ with nodes $ a $, $ b $, $ p $, and $ q $ shown. Node $ p $ is the only node in level $ k-1 $. Nodes $a$, $b$, and $q$ (the parent of $p$) are in the level $ k-2 $. \label{fig:heapabpq}}
\end{figure}

\begin{figure}[h] 
\begin{center}
\includegraphics[scale=.7]{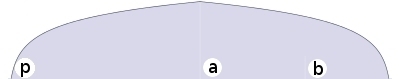}  
\end{center}
  \caption{Heap $  H_{2^{k}-1} $ on $ 2^{k}-1 $ nodes produced by the sequence of pull downs $ \langle v_{2^{k-1} }, ... , v_{2^{k} -2} \rangle$ from example heap $  H_{2^{k-1}} $ of Figure~\ref{fig:heapabpq}. Nodes $a$, $b$, and $p$ are in the level $ k-1 $. \label{fig:heapabp}}
\end{figure}

Let $ a $ and $ b $ be the elements of the set $ \{ 1,2,3 \} \setminus \{ p \} $. (For algebraists, one can put $ a = p \!\!\! \mod 3 +1 $ and $ b = (p+1) \!\!\! \mod 3 +1 $.) An example of such an arrangement 
is shown on Figure~\ref{fig:heapabpq}.
Neither $ a $ nor $ b $ is the parent $ q $ of $ p $ in heap $  H_{2^{k-1}} $ since $ q \geq 4 $. 
So,  none of them can have more than one child, which means that they must belong to level $ k-2 $ of heap $  H_{2^{k-1}} $ since all nodes above that level have two children each. 
Because the only element of level $ k-1 $ of heap $  H_{2^{k-1}} $ is $ p $, and neither $ a $ nor $ b $ is the parent of $ p $, 
both nodes $ a $ and $ b $ are leaves in the level $ k-2 $ of heap $ H_{2^{k-1}} $ and are less than any other node of heap $ H $ that is pulled down in moves $ \langle v_{2^{k-1}}, ... , v_{2^{k}-2} \rangle$ (recall that node $ p $ is not to be moved from its index $ 2^{k-1} $). 
Therefore, neither of them can acquire two children. Since all nodes in level $ k-2 $ of heap $  H_{2^{k}-1} $ have two children each, they both must be pulled down, eventually, from the level $ k-2 $ to the level $ k-1 $ by some moves in $ \langle v_{2^{k-1}}, ... , v_{2^{k}-2} \rangle$, as shown on Figure~\ref{fig:heapabp}, 
each of those pull downs, by virtue of (\ref{eq:credLeaf1}) and (\ref{eq:credLeaf2}) page~\pageref{eq:credLeaf2}, earning at most $2 (k-1) - 1$ credit\footnote{Each of them will earn only $2 (k-2)$ credit if the larger of the two gets pulled down before the smaller of the two.}, that is,
each causing a loss of credit to the maximum of credit $2 (k-1)$. Thus the sequence of pull downs  $ \langle v_{2^{k-1} }, ... , v_{2^{k} -2} \rangle$ must lose more than one credit and, therefore, the sequence of pull downs   $ \langle v_{2^{k-1}-1}, ... , v_{2^{k} -2} \rangle$ must lose more than one credit. Hence, by Competition
 Lemma \ref{lem:2loss}~\ref{item:2loss1} and \ref{item:2loss2}, the resulting heap $ H $ $ = $ $ H_{2^{k} - 1}$ is not a worst-case heap, contrary to the assertion made at the beginning of this proof.
\hspace*{\fill} $\Box$

\medskip

At this point I am ready to conclude a fundamental fact that comes handy while proving the optimality of strategies $ \mathsf{par} $ and $ \mathsf{win} $.

\medskip

\begin{theorem} \label{thm:parwin}
 Let $ H $ be a complete worst-case heap on $N = 2^{\lceil \lg N \rceil} - 1$ nodes, where $ N \geq 7 $. 
 $ H $ does not admit a sequence of more than $N - 2$ consecutive lossless pull downs.\end{theorem}
{\bf Proof}. Let $2 \leq K = \lfloor \lg N \rfloor $ and $ H $ be a complete worst-case heap on $N = 2^{K+1} - 1$ nodes, where $ K \geq 2 $. By the Worst-case Heap Characterization 
 Lemma~\ref{lem:parwin4}, the first element $p= H[2^{K}] $ of the last level $ K $ of $ H $ is $ 4 $ or larger.
Let $ \mathsf{v} = \langle v_i \mid 1 \leq i \leq n \rangle $ be a sequence of consecutive lossless pull downs that heap $ H $ can admit, and let $ n $ be the length of $ \mathsf{v} $. Since the first two pull downs on a complete heap can clearly be lossless, I may assume without a loss of generality that $ n \geq 2 $.  Let for any $ 0 \leq i \leq n $, $ H ^{\prime ^{i}} $ be the heap on $ N+i $ nodes produced by the first $ i $ pull downs $ \langle v _1,...,v_{i} \rangle $ of $ \mathsf{v} $, with convention $ H ^{\prime ^{0}} $ $ = $ $ H $.   In particular, $ H ^{ \prime} $ $ = $ $ H ^{ \prime ^1} $ is the heap produced by pull down $ v_1 $ executed on $ H $, and
 $ H ^{ \prime \prime} $ $ = $ $ H ^{ \prime ^2} $ is the heap produced by pull down $ v_2 $ executed on $ H ^{\prime} $.

\medskip

Move $ v_1 $ can pull down, and losslessly so, any node of the last level $ K $ of heap $ H $. In particular, it can pull down $ p $, in which case $ p $ will become one of the two nodes of the last level $ K+1 $ of heap $ H ^{ \prime \prime} $. If $ v_1 $ pulls down a node $ q $ different than $ p $ then the second move $ v_2 $, in order to be lossless, must either pull down the parent $ p $ of $ q $ in heap $ H ^{ \prime} $, or pull down $ q $, which will case a demotion of its parent $ p $ in heap $ H ^{ \prime} $. In either case, $ p $ will end up in the level $ K+1 $ of heap $ H ^{ \prime \prime} $. So, let $ p $ and $ q $ be the nodes of level $ K+1 $ of heap $ H ^{ \prime \prime} $. Since $ p \geq 4 $, we have $\max \{p,q\} \geq 4 $. 
Let's see where the nodes $ 1, 2 $ and $ 3 $ can reside in heap 
$ H ^{ \prime \prime} $.

\medskip

Each element of level $ K-1 $ of heap 
$ H ^{ \prime \prime} $ and above that level has two children, so neither node $ 1 $ nor node $ 2 $ can be there since none of them can have two children. So, nodes $ 1 $ and $ 2 $ reside somewhere in levels $ K $ and $ K+1 $ of heap 
$ H ^{ \prime \prime} $.

\medskip

If node $ 3 $ does reside above level $ K $ of heap 
$ H ^{ \prime \prime} $ then $ 3 $ must have two children. These can only be $ 1 $ and $ 2 $, and if so then $ 1 $ and $ 2 $ must be siblings and, therefore, must not reside in different levels of heap 
$ H ^{ \prime \prime} $. Level $ K+1 $ of heap 
$ H ^{ \prime \prime} $ consists of two nodes $ p $ and $ q $, one of which, $ p $, is larger than or equal to $ 4 $, so both $ 1 $ and $ 2 $ reside in level $ K $ of heap 
$ H ^{ \prime \prime} $ in such a case.

\medskip

If node $ 3 $ does not reside above level $ K $ of heap 
$ H ^{ \prime \prime} $ then it resides in level $ K $ or $ K+1 $ of 
$ H ^{ \prime \prime} $. In this case, all three nodes, $ 1 $, $ 2 $, and $ 3 $ reside in levels $ K $ and $ K+1 $ of heap 
$ H ^{ \prime \prime} $. Level $ K+1 $ of heap 
$ H ^{ \prime \prime} $ consists of two nodes $ p $ and $ q $, one of which, $ p $, is larger than or equal to $ 4 $, so at least two of the nodes $ 1 $, $ 2 $ and $ 3 $ reside in level $ K $ heap 
$ H ^{ \prime \prime} $ in such a case.

\medskip

Whatever the case, at least two of the nodes $ 1 $, $ 2 $ and $ 3 $ reside in level $ K $ of the heap 
$ H ^{ \prime \prime} $. Let $ j $ be the minimum of the indices of those  nodes $ 1 $, $ 2 $ and $ 3 $ that reside in level $ K $ of the heap 
$ H ^{ \prime \prime} $.  Because the maximum index in level $ K $ of $ H ^{ \prime \prime} $ is $ N $ and there are at least two such nodes, $ j \leq N -1 $.

\medskip

We have: $ H ^{ \prime \prime}[j] \leq 3 $. In particular, $ H ^{ \prime \prime}[j] < \max \{p,q\} $. Heap 
$ H ^{ \prime \prime} $ has $ N+2 $ nodes so, by Lemma~\ref{lem:parwin1}, it does not admit a sequence of more than $ 2 j - (N + 2)$ consecutive lossless pull downs. Now, 
$ 2 j - (N + 2)$ $ \leq $ $ 2N-2 - (N + 2)$ $ = $ $ N - 4 $. Therefore, the sequence $ \mathsf{v} $ of consecutive pull downs has the length $ n \leq  $ $ N - 4  + 2$ $ = $ $ N-2 $. Thus heap $ H $ does not admit a sequence of more than  $ N-2 $ consecutive pull downs.
\hspace*{\fill} $\Box$

\medskip

It may seem that the preceding technical lemmas, particularly, \linebreak Lemma~\ref{thm:parwin}, actually nailed the proof that  $ \mathsf{win}(N) $ is an optimal strategy for $ N $, but such an idea is a bit deceitful as it is tacitly based on an unpronounced assertion that only a quasi-greedy strategy whose $ 2^{\lfloor \lg N \rfloor} -2$nd move produces a worst-case heap , which $ \mathsf{win}(N) $ is an instance of, can be an optimal one. Unfortunately, there exist other strategies that are optimal for $ N $, for instance, strategies that lose to $ \mathsf{win}(N) $ in $ 2^{\lfloor \lg N \rfloor} -2$nd move but are lossless afterwards so that, if $ N \geq 2^{\lceil \lg N \rceil} -4 $, they can catch on with $ \mathsf{win}(N) $ at the end of the last level $ \lfloor \lg N \rfloor $ of the worst-case heaps that they produce. Here is an example of such an optimal strategy.

\medskip

\begin{example} \label{ex:opt} 
The heap $ G_4 = [4, 3, 2, 1] $ is a worst-case heap and is not a residue of any complete worst-case heap. However, it can be pulled down onto a worst-case heap of arbitrary large size via, for instance, $ G_7 = [ 7, 5,  6,  3,  4,  1,  2] $ that, of course, is not a worst-case heap. The following very simple strategy $ \mathsf{z}(N) $, optimal for 
$ N = I+1 $ and $ I+2 $, where $ I = 2^{\lceil \lg I \rceil} -4  $,\footnote{$ \mathsf{z}(N) $ is not optimal for any other $ N \geq 7 $.} may be constructed from $ G_7 $. 
Beginning with move 7, play a version of the strategy $ \mathsf{win}(N) $ on $ G_7 $, that is,
in any level beyond level 2, except for the last level if it is not full, play a version of the strategy $ \mathsf{par} $\footnote{Just keep pulling down 1 and 2 indefinitely.} that would assure that 3 resides at the first index of the last level of any complete heap $ G_M $ created this way, and 1 and 2 reside at the last two indices of that level. If the last level of $ G_N $ is not full then play the strategy $ \mathsf{gre} $ in that level. No complete heap produced with $ \mathsf{par} $ beginning from its 7th move on heap $ G_7 $ is a worst-case heap, simply because $ G_7 $ is not and $ \mathsf{par} $ loses 1 credit in every level beyond level 2. However, the restriction imposed by the Theorem~\ref{thm:parwin} does not apply, and, as it turns out, all complete heaps $ G_{2^K-1} $, where $ 3 \leq K \leq \lg N $, produced by
$ \mathsf{z}(N) $, admit more than $ 2^K-3 $  lossless pull downs each\footnote{$ 2^K-1 $, to be exact; 1 and 3 can be pulled down repeatedly without a loss of credit until 1 becomes the first child of 2 and 3 cannot be pulled down.}. This allows $ \mathsf{z}(N) $ to catch on with $ \mathsf{win}(N) $ in move $ m = 2^{\lceil \lg N \rceil} -4  $ and recover the 1 lost credit from its move 6 (that created heap $ G_7 $). For example the following heaps created by  $ \mathsf{z}(14) $ in its 12th and 13th moves are worst-case\footnote{But not hereditary worst-case heaps.} heaps: $G_{13} = [13,12,7,9,11,6,2,5,8,4,10,1,3] $ and $ G_{14} = [14,12,13,9,11,7,2,5,8,4,10,6,3,1] $; their only worst-case residua are heap $ G_6 = [6,5,2,3,4,1] $ and the residua of $ G_6 $.\footnote{Similar restriction applies to all heaps $ G_N $ on $N \geq 7$  nodes produced by  $ \mathsf{z}(N) $; in particular, no $ G_N $ has more than 7 worst-case residua.}
\end{example}

\medskip

Below I prove, by neat induction, an optimality criterion for strategy $ \mathsf{par} $, from which I am going to conclude the optimality of $ \mathsf{win}(N) $ for any given $ N $.

\begin{1oth} \label{thm:Spar}
	Strategy $ \mathsf{par} $ is optimal for $ N $ if, and only if, 
	\begin{equation} \label{eq:Spar}
	N > 2^{\lceil \lg N \rceil} -4 .
	\end{equation}
\end{1oth}
{\bf Proof}. The \textit{only if} part follows from the fact that strategy $ \mathsf{win}(N) $ beats $ \mathsf{par} $ at $ N-1 $ \footnote{In first $ N-1 $ moves, that is.} for all $ N \leq 2^{\lceil \lg N \rceil} -4 $.

\medskip

 I will prove the \textit{if} part the by induction on $ \lceil \lg N \rceil $.

 \medskip
 
 Since, by Theorem~\ref{thm:parcredits} page~\pageref{thm:parcredits}, the first 7 moves of $ \mathsf{par} $ are lossless,  $ \mathsf{par} $ is an optimal strategy for $ 2 \leq N \leq 8 $, that is for $ 1 \leq  \lceil \lg N \rceil \leq 3 $.

\medskip

Let $  \lceil \lg N \rceil \geq 3 $. The inductive hypothesis implies that $ \mathsf{par} $ is an optimal strategy for $ N = 2^{ \lceil \lg N \rceil} -1 $
\footnote{Because for $  \lceil \lg N \rceil \geq 2 $, $ \lceil \lg (2^{ \lceil \lg N \rceil} -1) \rceil $ $ = $ $ \lceil \lg N \rceil  $ and  $ 2^{ \lceil \lg N \rceil} -1 $ $ > $ $ 2^{\lceil \lg (2^{ \lceil \lg N \rceil} -1) \rceil} -4 $.}. Let $ M = 2^{ \lceil \lg N \rceil+1} -3 $.
 If the heap $ G $ created by $ \mathsf{par} $'s first $ M - 1 $ moves is a worst-case heap then, since by Theorem~\ref{thm:parcredits} page~\pageref{thm:parcredits} all the moves $ K < 2^{ \lceil \lg K \rceil} $ of $ \mathsf{par} $
are lossless, in particular, moves  $ M , M+1 $, and $ M+2 $ are, the heaps $ G^{\prime}, G^{\prime\prime} $, and $ G^{\prime\prime\prime} $ created by $ \mathsf{par} $'s first $ M , M+1 $, and $ M+2 $ moves (respectively) are worst-case heaps, too. In such a case, $ \mathsf{par} $ is an optimal strategy for all $ K $ with $ K > 2^{ \lceil \lg K \rceil} -4  $ as long as $ \lceil \lg K \rceil = \lceil \lg N \rceil + 1 $
\footnote{Because for $  \lceil \lg N \rceil \geq 2 $,
	$ \lceil \lg (M+i) \rceil  $ 
	$ = $ $ \lceil \lg N \rceil + 1 $ and  $ M+i $ $ > $ $ 2^{\lceil \lg (M+i) \rceil} -4 $ are true if, and only if, $ i = -1, 0,1,2 $.}, which observation completes the inductive step. So, all I have to prove at this point is that $ G $  is a worst-case heap.

\medskip

 Let us suppose to the contrary that $ G $ is not a worst-case heap. Let $ \tilde{G} $ be  $ G $'s residue on $ N $ nodes. Obviously, $ \tilde{G} $ is created by the first $ N-1 $ moves of $ \mathsf{par} $.  Also, 
\begin{equation} \label{eq:Spar10}
\Lambda_{\tilde{G} \, G } = 1.
\end{equation}

\medskip

Since $ \mathsf{par} $ is optimal for $ N $, $ \tilde{G} $ is a worst-case heap. Let $ H $ be a worst-case heap on $ M $ nodes and $ \tilde{H} $ be its residue on $ N $ nodes. By the Diagram Lemma~\ref{lem:diag} \ref{item:diag5} page \pageref{item:diag5}, $ \tilde{H} $ is a worst-case heap and
\begin{equation} \label{eq:Spar20}
\Lambda_{\tilde{H} \! H } = 0.
\end{equation}
Thus $ \tilde{H} $ is a complete worst-case heap on $ N \geq 7 $ nodes that admits
\[  M-N = 2^{ \lceil \lg N \rceil+1} -3 - 2^{ \lceil \lg N \rceil} +1 =2^{ \lceil \lg N \rceil} - 2 = N-1\]
consecutive lossless pull downs, contradicting Theorem~\ref{thm:parwin}.  Therefore, $ G $ is a worst-case heap.
\hspace*{\fill} $\Box$

\medskip

Since $ \mathsf{win}(N) $ coincides with $ \mathsf{par} $ for $ N > 2^{\lceil \lg N \rceil} -4 $ and beats $ \mathsf{par} $ at $ N-1 $ for $ N > 2^{\lceil \lg N \rceil} -4 $, the fact that $ \mathsf{par} $ never loses more than 1 credit to an optimal strategy allows me to easily conclude an optimality of $ \mathsf{win}(N) $.

\begin{2oth} \label{cor:win(N)}
	For every $ N \geq 2 $, $ \mathsf{win}(N) $ is an optimal strategy for $ N $.
\end{2oth}
{\bf Proof}. The first $ N-1 $ moves of  $ \mathsf{win}(N) $ coincide with the first $ N-1 $ moves of  $ \mathsf{par} $ if $ N > 2^{\lceil \lg N \rceil} -4 $, thus for every $ N > 2^{\lceil \lg N \rceil} -4 $, 
$ \mathsf{win}(N) $ is an optimal strategy for $ N $. 

\medskip

Let 
\begin{equation} \label{eq:corwinN10}
N \leq 2^{\lceil \lg N \rceil} -4 ,
\end{equation}
and let $ H $ be a heap on $ 2^{\lfloor \lg N \rfloor}  $ nodes produced by the first $ 2^{\lfloor \lg N \rfloor} - 1  $ moves of $ \mathsf{win}(N) $. Since $ H $ is also a heap  produced by the first $ 2^{\lfloor \lg N \rfloor} - 1  $ moves of $ \mathsf{par} $, by The 1$ ^{st} $ Optimality Theorem~\ref{thm:Spar}, $ H $ is a worst-case heap. Now, by the equality (\ref{eq:wincredits100}) in Theorem~\ref{thm:winscore} page \pageref{thm:winscore}, all moves $ 2^{\lfloor \lg N \rfloor}  $ through $ N - 1 $ of $ \mathsf{win}(N) $ are lossless. Therefore, the heap $ G $ that is produced by those moves is a worst-case heap, too. Since $ G $ is a heap produced by produced by the first $ N - 1  $ moves of $ \mathsf{win}(N) $, by virtue of Theorem~\ref{thm:win} page \pageref{thm:win}, $ \mathsf{win}(N) $ is a winning strategy for $ N $.
\hspace*{\fill} $\Box$

\subsection{A comment on the proof of optimality} \label{subsec:comment}

\medskip

Basically, all the trouble that we went through in this Section\footnote{Now, after I am done with all the details of my analysis, I can sense a trace of frustration, to which I can relate, in the comment that Donald Knuth wrote in \cite{knu:art}: ``Algorithm H is rather complicated, so it probably will never submit to a complete mathematical analysis [...].''.} was to prove the following deceitfully simple fact that entails the optimality of strategy $ \mathsf{win}(N) $  for any $ N \geq 2 $.

\begin{sith} \label{cor:parwin}
	No worst-case heap on $N = 2^{\lceil \lg N \rceil} - 4$ nodes admits a  lossless pull down.\footnote{The following stronger version of the Lemma~\ref{cor:parwin}  can be proved: \textit{No worst-case heap on $ N $ nodes admits a lossless pull down if, and only if,   $N = 2^{\lceil \lg N \rceil} - 4$.}} 
\end{sith}
{\bf Proof}. 
Assume to the contrary that $N = 2^{\lceil \lg N \rceil} - 4$ and $ H $ is a worst-case heap on $ N $ nodes that admits a lossless pull down. Let $ \tilde{H} $ be a residue of $ H $ on $M= 2^{\lfloor \lg N \rfloor} $ nodes, $ \tilde{G} $ be the heap on $ M $ nodes created by the first $ M-1 $ moves of strategy $ \mathsf{win}(N) $ and $ G $ be the heap on $ N $ nodes created by the first $ N-1 $ moves of strategy $ \mathsf{win}(N) $. 
By The 1$ ^{st} $ Optimality Theorem~\ref{thm:Spar}, $ \tilde{G} $ is a worst-case heap.
By the Diagram Lemma~\ref{lem:diag}~\ref{item:diag2} page \pageref{item:diag2}, all the $ N-M $ pull downs that reconstruct heap $ H $ from its residue $ \tilde{H} $ are lossless, thus $ \tilde{G} $ admits $ N-M+1 $
$ = $ $ M-1 $ lossless moves, contrary to Theorem~\ref{thm:parwin} page~\pageref{thm:parwin}. 
\hspace*{\fill} $\Box$
\medskip

\textit{Note}. 
The Singularity Theorem~\ref{cor:parwin}  can be expressed in the following equivalent form without any reference to \textit{ pull down}.

\begin{theorem} \label{cor:parwinVar}
	For every $ N = 2^{\lceil \lg (N-1) \rceil} - 3$ \footnote{That equality is equivalent to this slightly longer condition: $ N = 2^{\lceil \lg N \rceil} - 3$ and $ N \neq 5 $.} and every heap $ H $ on $ N $ nodes, the heap $ H.{\tt RemoveMax()}$ is not a worst-case heap or
	\begin{equation} \nonumber
	C_{{\tt RemoveMax()}} (H) < C^{max}_{{\tt RemoveMax()}} (N).
	\end{equation}
\end{theorem}
{\bf Proof of the equivalence of Theorems~\ref{cor:parwin} and \ref{cor:parwinVar}}. If $ H $ is a worst-case heap on $ N = 2^{\lceil \lg N \rceil} - 4$ nodes that admits a lossless pull down $ p $ then the heap $ H^{\prime} $ that is the result of application of $ p $ to $ H $ is a worst-case heap on $ N^{\prime} = 2^{\lceil \lg (N^{\prime}-1) \rceil} - 3$ nodes with $ C_{{\tt RemoveMax()}} (H^{\prime}) = C^{max}_{{\tt RemoveMax()}} (N) $. This would make Theorem~\ref{cor:parwinVar} false.

\medskip

If $ H $ is a heap on $ N = 2^{\lceil \lg (N-1) \rceil} - 3$ nodes and $ H^{\prime} =  H.{\tt RemoveMax()} $ is a worst-case heap with $ C_{{\tt RemoveMax()}} (H) = C^{max}_{{\tt RemoveMax()}} (N) $ then  $ H^{\prime} $ is a worst-case heap on  $ N^{\prime} = 2^{\lceil \lg N^{\prime} \rceil} - 4$ nodes that admits a lossless pull down.  This would make Theorem~\ref{cor:parwin} false. 
\hspace*{\fill} $\Box$
\medskip

One could try to find a strategy that makes a move with a loss of 1 credit only when the Singularity Theorem~\ref{cor:parwin} mandates a loss; such a strategy would automatically be optimal. Its loss of credit function $ \lambda ^* $ would be defined by the equality (\ref{eq:parStarDef}) page \pageref{eq:parStarDef} and visualized on Figure~\ref{fig:lambdaStar}  Unfortunately, such a strategy does not exist or otherwise, as a winning strategy for all $ N $, it would produce infinitely many hereditary worst-case heaps while there are only 1017 of them, as it has been demonstrated in \ref{Hereditary}. However, the Singularity Theorem~\ref{cor:parwin} dictates that $  \sum _{i=2} ^{N-1} \lambda ^*  (i) $ is a lower bound on the accumulated loss $ \sum _{i=2} ^{N-1} \lambda _{\mathsf s} (i) $ for the first $ N-1 $ moves of any strategy $ {\mathsf s} $. This fact, taking into account that, by virtue of Lemma~\ref{lem:sumwinstar} page~\pageref{lem:sumwinstar}, the strategy $ \mathsf{win}(N) $ actually reaches the said lower bound, leads to
a weaker criterion of optimality of a strategy, from which the optimality of  $ \mathsf{win}(N) $  for $ N $ follows.

\medskip

\begin{sLBlm} \label{lem:sumlambda*}
For every strategy $ {\mathsf s} $ and every $ N \geq 2 $,
	
	\begin{equation} \label{eq:sumlambda*}
	\sum _{i=2} ^{N} \lambda _{\mathsf s} (i) \geq \sum _{i=2} ^{N} \lambda ^*  (i) .
	\end{equation}
\end{sLBlm}

{\bf Proof} by induction on $ N $. For $ N \leq 12 $,  $ \sum _{i=2} ^{N} \lambda ^*  (i) = 0 $ thus the inequality (\ref{eq:sumlambda*}) holds. This completes the basis step.

\medskip

For the inductive step, let us assume that the inequality (\ref{eq:sumlambda*}) holds for some $ N \geq 12 $ and every strategy $ {\mathsf s} $. I am going to show that for every strategy $ {\mathsf s} $,
\begin{equation} \label{eq:sumlambda*N}
\sum _{i=2} ^{N+1} \lambda _{\mathsf s} (i) \geq \sum _{i=2} ^{N+1} \lambda ^*  (i) .
\end{equation}

If $N+1 \neq 2^{\lceil \lg (N+1) \rceil} - 4$ then, by (\ref{eq:parStarDef}) page~\pageref{eq:parStarDef}, $ \lambda ^*  (N+1) = 0 $ and
the inequality (\ref{eq:sumlambda*N})  follows from  (\ref{eq:sumlambda*}).

\medskip
If $N+1 = 2^{\lceil \lg (N+1) \rceil} - 4$ then, by (\ref{eq:parStarDef}), $ \lambda ^*  (N+1) = 1 $. In such a case,
let us assume to the contrary that there exists a strategy $ {\mathsf s} $ such that
\begin{equation} \label{eq:sumlambda*Nneg}
\sum _{i=2} ^{N+1} \lambda _{\mathsf s} (i) < \sum _{i=2} ^{N+1} \lambda ^*  (i)
=  \sum _{i=2} ^{N} \lambda ^*  (i) + 1.
\end{equation}
From (\ref{eq:sumlambda*Nneg}) and (\ref{eq:sumlambda*}) I infer $ \lambda _{\mathsf s} (N+1)=0 $, which implies that
\[ \sum _{i=2} ^{N} \lambda _{\mathsf s} (i) = \sum _{i=2} ^{N+1} \lambda _{\mathsf s} (i) <  \sum _{i=2} ^{N} \lambda ^*  (i) + 1, \]
or
\begin{equation} \label{eq:sumlambda*500}
\sum _{i=2} ^{N} \lambda _{\mathsf s} (i) \leq  \sum _{i=2} ^{N} \lambda ^*  (i) ,
\end{equation}
and, by virtue of the Singularity Theorem~\ref{cor:parwin}, that the heap \linebreak $ H = \mathscr{T} ({\mathsf s}_{1,N-1}) $ \footnote{The function $ \mathscr{T} $ has been defined on page~\pageref{def:createdT}.} on $ N $ nodes that has been created by $ {\mathsf s} $'s first $ N-1 $ moves is not a worst-case heap. Thus, by Theorem~\ref{thm:win} page~\pageref{thm:win}, $ {\mathsf s} $ is not optimal for $ N $ and, therefore, for some strategy  $ {\mathsf u} $,
\begin{equation} \label{eq:sumlambda*600}
\sum _{i=2} ^{N} \lambda _{\mathsf s} (i) > \sum _{i=2} ^{N} \lambda _{\mathsf u} (i).
\end{equation}
This, by virtue of (\ref{eq:sumlambda*500}), yields
\[  \sum _{i=2} ^{N} \lambda _{\mathsf u} (i) < \sum _{i=2} ^{N} \lambda ^*  (i)  \]
and, therefore, contradicts the inductive hypothesis.
\hspace*{\fill} $\Box$

\medskip

The $ \sum \lambda $ Lower Bound Lemma yields the following.

\begin{cropt} \label{cor:parwinlambda*}
	A strategy $ {\mathsf s} $ is optimal for $ N \geq 2 $ if, and only if,
	
	\begin{equation} \label{eq:parwinlambda*}
	\sum _{i=2} ^{N-1} \lambda _{\mathsf s} (i) = \sum _{i=2} ^{N-1} \lambda ^*  (i) .
	\footnote{Which, by virtue of Lemma~\ref{lem:Lambda*closed} page~\pageref{lem:Lambda*closed}, is equal to $ \max \{ {\lfloor \lg (N+3)  \rfloor} , 3 \} -3. $}
	\end{equation}
\end{cropt}

{\bf Proof}. By the Lemma~\ref{lem:winmin} page~\pageref{lem:winmin},
$ {\mathsf s} $ is optimal for $ N \geq 2 $ if, and only if, for every strategy $ {\mathsf u} $,
 \[	\sum _{i=2} ^{N-1} \lambda _{\mathsf s} (i) \leq  	\sum _{i=2} ^{N-1} \lambda _{\mathsf u} (i). \] Thus if (\ref{eq:parwinlambda*}) holds then, by the $ \sum \lambda $ Lower Bound Lemma~\ref{lem:sumlambda*}, $ {\mathsf s} $ is optimal. 

\medskip

If $ {\mathsf s} $ is optimal for $ N $ then, by the Lemma~\ref{lem:winmin},
\[  	\sum _{i=2} ^{N-1} \lambda _{\mathsf s} (i) \leq \sum _{i=2} ^{N-1} \lambda _{\mathsf{win}(N)}  (i) = \]
[by the equality (\ref{eq:sumwinStar1}) in the Theorem~\ref{lem:sumwinstar} page~\pageref{eq:sumwinStar1}]
\[ =  \sum _{i=2} ^{N-1} \lambda ^*  (i) ,\]
which, by the $ \sum \lambda $ Lower Bound Lemma~\ref{lem:sumlambda*}, yields (\ref{eq:parwinlambda*}).
\hspace*{\fill} $\Box$

\medskip

The Optimality Criterion~\ref{cor:parwinlambda*} may be used to prove the Singularity Theorem~\ref{cor:parwin}. Indeed, if $ H $ is a worst-case heap on $N = 2^{\lceil \lg N \rceil} - 4$ nodes that admits a lossless pull down, $ \tilde{H} $ is the heap produced from $ H $ by that lossless pull down, and $  {\mathsf s}  $ is a strategy that produces $ \tilde{H} $ in its first $ N $ moves then, by the Optimality Criterion, 
\begin{equation} \label{eq:parwinlambda*=600}
\sum _{i=2} ^{N-1} \lambda _{\mathsf s} (i) = \sum _{i=2} ^{N-1} \lambda ^*  (i).
\end{equation}
Since $ \lambda _{\mathsf s} (N)  = 0$ and $ \lambda ^*  (N) = 1 $, (\ref{eq:parwinlambda*=600}) implies that
\begin{equation} \nonumber 
\sum _{i=2} ^{N} \lambda _{\mathsf s} (i) < \sum _{i=2} ^{N} \lambda ^*  (i) ,
\end{equation}
thus contradicting the Optimality Criterion.

\medskip

Therefore, if one takes a proof Lemma~\ref{lem:sumwinstar} page~\pageref{lem:sumwinstar} for granted, the Optimality Criterion and the Singularity Theorem have proofs of roughly the same complexity since they can be easily derived one from another, as I have shown above. 

\medskip

Providing a straightforward proof of the Singularity Theorem~\ref{cor:parwin} that does not depend on The 1$ st $ Optimality Theorem~\ref{thm:Spar} (nor on Theorem \ref{thm:parwin} that was instrumental in proving it) would constitute a significant simplification of my proofs of the main results of this paper.

\section{The worst-case number of comparisons for ${\tt RemoveAll}$}
\label{W-cRemAll}

At this point, I have all the facts needed to conclude the fundamental result of this paper: a formula that
gives the exact number of comparisons of keys performed in the worst case by the ${\tt RemoveAll()}$, visualized on Figure~\ref{fig:cLB} page \pageref{fig:cLB}. 

\medskip

\begin{theorem} \label{thm:worstRemAll}
For every natural number $N \geq 2$, the number $ C_{\tt RemoveAll()} ^{\tt max} (N) $ 
of comparisons of keys performed in the worst case by the ${\tt H.RemoveAll()}$ on any heap ${\tt  H }$ of size $N$ is equal to:
\begin{equation} \label{eq:worstRemAll}
2(N-1)\lfloor \lg (N-1) \rfloor - 2 ^{\lfloor \lg (N-1) \rfloor +2} + \min (\lfloor \lg (N-1) \rfloor, 2) + 4 + c,
\end{equation}
where $ c $ is a binary function on the set of integers defined by:
\begin{equation} \label{eq:defc}
c = 
\left\{ \begin{array}{ll}
1 \mbox{ if } \;   N \leq 2 ^{\lceil \lg N \rceil} - 4   \\ \\
0 \mbox{ otherwise}.
\end{array} \right.
\end{equation}
\end{theorem}
{\bf Proof}. 
By The 2$ ^{nd} $ Optimality Theorem~\ref{cor:win(N)} page \pageref{cor:win(N)},  $ \mathsf{win}(N) $ is an optimal strategy for $ N $, so, by Theorem~\ref{thm:win} page \pageref{thm:win}, the payoff $ P _{\mathsf{win}(N)} (N-1)$, given by the formula (\ref{eq:wincredits}) in the Theorem~\ref{thm:wincredits} page \pageref{thm:wincredits}, for its first $ N-1 $ moves is equal to  
the number $C_{\tt RemoveAll()} ^{\tt max} (N)$ 
of comparisons of keys performed in the worst case by the ${\tt H.RemoveAll()}$ on any heap ${\tt  H }$ of size $N$. Since the formulas (\ref{eq:worstRemAll}) and (\ref{eq:wincredits}) are identical, the thesis of this theorem follows.
\hspace*{\fill} $\Box$

\medskip

Example of a 500-node worst-case heap for $ {\tt RemoveAll } $, created by my Java program, is included in \ref{sec:appExInput} page \pageref{sec:appExInput}.

\section{The worst-case number of comparisons for ${\tt Heapsort}$}
\label{W-cHeap}

\begin{figure}[h] 
\begin{center}
\includegraphics[scale=1]{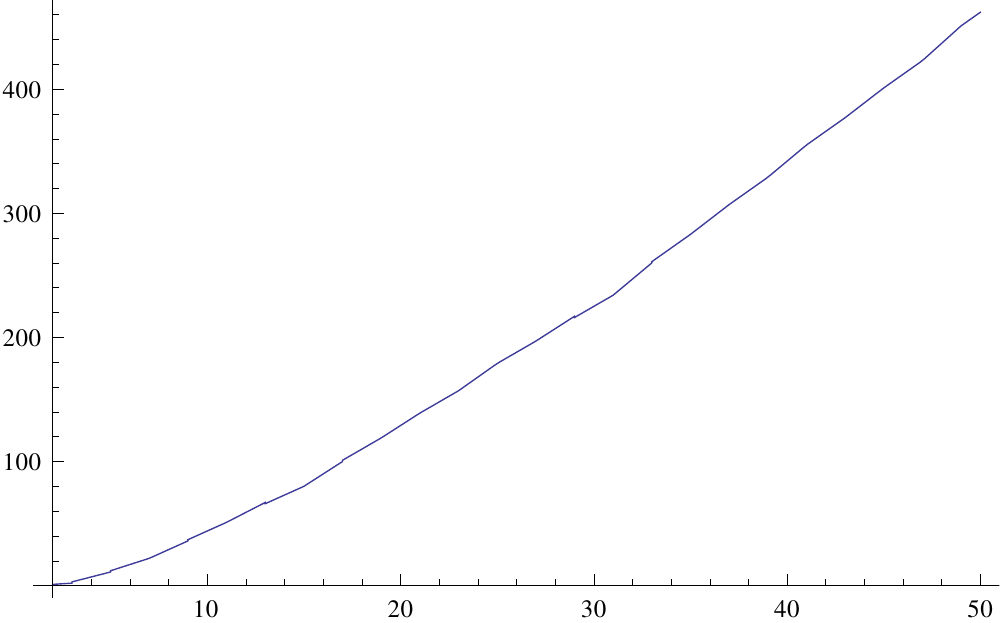}  
\end{center}
\caption{The worst-case number $ C_{\tt Heapsort} ^{\tt max}(N) $  of comparisons of keys by $ {\tt Heapsort} $. \label{fig:w-c_sort-50}}
\end{figure}

Adding the formulas for the exact numbers of comparisons of keys performed in the worst case by the $ {\tt MakeHeap} $ and ${\tt RemoveAll()}$ yields the exact numbers of comparisons of keys performed in the worst case by the $ {\tt Heapsort} $ visualized on Figure~\ref{fig:w-c_sort-50}.

\begin{theorem} \label{thm:worstSort}
For every $ N \geq 2 $,
\begin{equation} \label{eq:worstSort}
2(N-1)\lceil \lg N \rceil - 2 ^{\lceil \lg N \rceil +1} - 2  s_2(N) - e_2(N) + \min (\lceil \lg N \rceil, 3) + 5 + c,
\end{equation}
where $ c $ is a binary function on the set of integers defined by:
\begin{equation} \nonumber
c = 
\left\{ \begin{array}{ll}
1 \mbox{ if } \; N  \leq  2 ^{\lceil \lg N \rceil} - 4   \\ \\
0 \mbox{ otherwise}.
\end{array} \right.
\end{equation}
Moreover, if $ 8 \leq N = 2 ^{\lceil \lg N \rceil} $ then the formula
(\ref{eq:worstSort}) simplifies to
\begin{equation} \label{eq:worstSort2^k} 
  (2N-3) \lg \frac{N}{2}    + 3 . \footnote{If $ N=2 $ or $ N=4 $ then the the said formula simplifies to $ 2(N-1) \lg \frac{N}{2}    + 1 . $} 
\end{equation}
\end{theorem}
{\bf Proof} of (\ref{eq:worstSort}) is a direct application of Theorem~\ref{thm:worstRemAll} page~\pageref{thm:worstRemAll} to the equalities (\ref{eq:sum_maxes}) page~\pageref{eq:sum_maxes} and  (\ref{eq:MkHeapElem}) page~\pageref{eq:MkHeapElem}.

\medskip

If $ N =  2 ^{\lceil \lg N \rceil} $ then $ \lceil \lg N \rceil = \lg N $, $ s_2(N) = 1 $, $ e_2(N) = \lg N $, and $ c = 0 $, so that (\ref{eq:worstSort}) simplifies to
\[ 2(N-1) \lg N  - 2 N - 2  - \lg N + \min ( \lg N , 3) + 5 ,\]
or to
\[ (2N-3) \lg \frac{N}{2}  + \min ( \lg N , 3). \]
 If, moreover, $ N \geq 8 $ then $ \min ( \lg N , 3) = 3 $. Hence, in such a case, (\ref{eq:worstSort}) is equal to (\ref{eq:worstSort2^k}).
\hspace*{\fill} $\Box$

\medskip

\begin{figure}[h] 
\begin{center}
\includegraphics[scale=1]{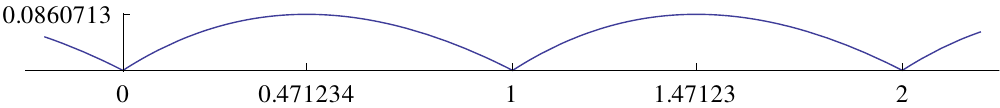} 
\end{center}
\caption{Graph of $ \varepsilon (x) $ $ = $ $ 1 + \lceil y \rceil -  y
- 2^{\lceil y \rceil -  y} $ as a function of $ y =  \lg x $. It assumes the maximum of $1 - \lg e + \lg \lg e \approx 0.0860713320559342$ for $ y = n + 1 - \lg\lg e $ $ \approx $ $ n + 0.4712336270551024 $ and any integer $ n $. \label{fig:epsilon} }
\end{figure}

The following theorem uses a version of (\ref{eq:worstSort}) that reduces the impact of the non-continuous function \textit{ceiling} in the formula (\ref{eq:worstSort}) to a small and continuous term $ \varepsilon $. 
It allows for easier evaluation of the \textit{exact} rate of growth of its terms as well as comparisons with some other sorting-related formulas that are expressible with $ \varepsilon $. 
\footnote{For instance, as indicated in \cite{knu:art}, p. 192, the minimum {external path length} $ epl_{\min} (m)$ in a finite binary tree with $ m $ leaves (treated as external nodes) is given by:
$ epl_{\min} (m) = m(\lg m + \varepsilon (m)) $,
where $ \varepsilon(x) $ is the function visualized on Figure~\ref{fig:epsilon} and defined in the proof 
of Theorem~\ref{thm:worstSortepsilon}, thus yielding this information-theoretic lower bound on the average number of comparisons of keys performed by any decision-tree-sorting algorithm: $ \lg N! + \varepsilon (N!) $ and this average number of comparisons of keys performed by successful binary search: 
$ \frac{N+1}{N}(\lg (N+1) + \varepsilon (N+1)) + 1 $.}

\begin{mth} \label{thm:worstSortepsilon}
For every natural number $N \geq 2$, the number 
of comparisons of keys performed in the worst case by the ${\tt Heapsort}$ on any array of size $N$ is equal to:
\begin{equation} \label{eq:worstSortepsilon}
2 (N-1)\, ( \,  \lg \frac{N-1}{2}  +\varepsilon  \, ) - 2s_2(N) - e_2(N) + \min (\lfloor \lg (N-1) \rfloor, 2) + 6 + c, 
\end{equation}
where $ \varepsilon $, given by:
\[\varepsilon = 1 + \theta
- 2^{\theta} \mbox{ and } \theta =  \lceil \lg \, (N-1) \rceil -  \lg \, (N-1),\]
is a continuous function of $ N $ 
on the set of reals $ >1 $, 
with the minimum value 0 and and the maximum (\textit{supremum} if 
$ \varepsilon $ is restricted to integers) value
\[\delta = 1 - \lg e + \lg \lg e \approx 0.0860713320559342 ,
\footnote{The constant
\[1 - \lg e + \lg \lg e \approx 0.0860713320559342\]
has been known as the {\em Erd\"{o}s constant} $ \delta $. Erd\"{o}s used it around 1955 in order to establish an asymptotic upper bound for the number $ M(k) $ of different numbers in a multiplication table of size $ k \times k $ by means of the following limit:
\[\lim _{k \rightarrow \infty} \frac{\ln \frac{k \times k}{M(k)}}{\ln \ln (k \times k)} =  \delta .
\]}
\]
$s_2(N)$ is the sum of all digits of the binary representation of $N$, $e_2(N)$ is the exponent of $2$ in the prime factorization of $N$, and $ c $ is a binary function on the set of integers defined by:
\begin{equation} \label{eq:cworstSortepsilon} 
c = 
\left\{ \begin{array}{ll}
1 \mbox{ if } \; N  \leq  2 ^{\lceil \lg N \rceil} - 4   \\ \\
0 \mbox{ otherwise}.
\end{array} \right.
\end{equation}
Moreover, if $ N \geq 5 $ then the formula (\ref{eq:worstSortepsilon}) simplifies to:
\begin{equation} \label{eq:worstSortepsilon2}
2 (N-1)\, ( \,  \lg \frac{N-1}{2}  +\varepsilon  \, ) - 2s_2(N) - e_2(N)  + 8 + c. 
\end{equation}
\end{mth}

\medskip

A graph of function $ c $ is visualized on Figure~\ref{fig:c3} page~\pageref{fig:c3}, while a graph of function $ \varepsilon $
is visualized on Figure~\ref{fig:epsilon}.

{\bf Proof}, given Theorem~\ref{thm:worstSort}, is elementarily algebraic and most of its components had been analyzed in works of others, for instance in \cite{knu:art}. I present it here in its entirety for the sake of completeness and reader's convenience.

\medskip

 First, I am going to show that
for every $x > 0$,
\begin{equation} \label{eq3} 
x \lceil \lg x \rceil
- 2^{\lceil \lg x \rceil} = x ( \lg x + \varepsilon ( x)-1),
\end{equation}
where $ \varepsilon $ is given by:
\[\varepsilon (x) = 1 + \theta
- 2^{\theta} \mbox{ and } \theta =  \lceil \lg \, x \rceil -  \lg \, x .\]

\smallskip

Indeed, substituting definition of $ \theta $ to the definition of $ \varepsilon $, we obtain:
\[\varepsilon (x)  = 1 + \lceil \lg \, x \rceil -  \lg \, x
- 2^{\lceil \lg \, x \rceil -  \lg \, x},\]
or
\[\lg \, x + \varepsilon (x) -1 = \lceil \lg \, x \rceil 
- \frac{2^{\lceil \lg \, x \rceil }}{2^{ \lg \, x}},\]
or
\[\lg \, x + \varepsilon (x) -1 = \lceil \lg \, x \rceil 
- \frac{2^{\lceil \lg \, x \rceil }}{x},\]
or
\[x(\lg \, x + \varepsilon (x) -1) = x \lceil \lg \, x \rceil 
- 2^{\lceil \lg \, x \rceil },\]
which proves (\ref{eq3}).

\smallskip

Now, putting $ x = N - 1 $, one gets
\[(N-1)(\lg \, (N-1) + \varepsilon  -1) = (N-1) \lceil \lg \, (N-1) \rceil 
- 2^{\lceil \lg \, (N-1) \rceil}.\]
Observation that
\begin{equation} \label{eq4} 
(N-1) \lceil \lg \, (N-1) \rceil 
- 2^{\lceil \lg \, (N-1) \rceil} = (N-1) \lceil \lg \, N \rceil 
- 2^{\lceil \lg \, N \rceil}
\end{equation}
[if $ N-1 = 2^{\lceil \lg \, (N-1) \rceil} $ 
then $ \lceil \lg \, N \rceil = \lceil \lg \, (N-1) \rceil + 1 =  \lg \, (N-1) + 1 $ 
and both sides of the equality (\ref{eq4}) reduce to $ (N-1)\lg (N-1)-N+1  $, 
while if $ N-1 \neq 2^{\lceil \lg \, (N-1) \rceil} $ 
then $ \lceil \lg \, N \rceil = \lceil \lg \, (N-1) \rceil $ and the equality (\ref{eq4}) is obviously true] 
allows one to derive:
\[2(N-1)\lceil \lg N \rceil - 2 ^{\lceil \lg N \rceil +1} =
2((N-1)\lceil \lg N \rceil - 2 ^{\lceil \lg N \rceil}) = \]
\[= 2((N-1)\lceil \lg (N-1) \rceil - 2 ^{\lceil \lg (N-1) \rceil }) =
(N-1)(\lg \, (N-1) + \varepsilon  -1) = \]
\[=(N-1)(\lg \, \frac{N-1}{2} + \varepsilon  ), \]
from which one concludes that (\ref{eq:worstSort}) is equal to (\ref{eq:worstSortepsilon}).

\smallskip

For $ N \geq 5 $, $ \min (\lfloor \lg (N-1) \rfloor, 2) + 6 = 8 $, so (\ref{eq:worstSortepsilon}) is equal to (\ref{eq:worstSortepsilon2}).
\hspace*{\fill} $\Box$

\medskip

Example of a 500-node worst-case array for $ {\tt Heapsort } $, created by my Java program, is included in \ref{sec:appExInput} page \pageref{sec:appExInput}.

\medskip

\textbf{\textit{Note}}.
Function $ \varepsilon (x) $ $ = $ $ 1 + \lceil \lg x \rceil -  \lg x
- 2^{\lceil \lg x \rceil -  \lg x} $ visualized on Figure~\ref{fig:epsilon} has been briefly analyzed in \cite{knu:art}. It assumes the maximum $1 - \lg e + \lg \lg e \approx 0.0860713320559342$ for $ \lg x = n + \lg\lg e $ and any integer $ n $, that is, for $ x $ $ = $ $ 2^n 2 ^{\lg\lg e} $ $ = $ $ 2^n \lg e $. Since $ \ln 2 $ and, therefore, $ \lg e $ are irrational numbers\footnote{Here, I only use the fact that $ \lg e $ does not have finite binary representation.}, so is $ x $. Therefore, function 
$ \varepsilon $ restricted to integers never reaches the value $1 - \lg e + \lg \lg e$. However, one can easily show that 
$ \lim _{n \rightarrow \infty} \varepsilon (\lfloor 2^n \lg e \rfloor) $ $ = $ $1 - \lg e + \lg \lg e$,\footnote{Indeed, $ 0 \leq  2^n \lg e - \lfloor 2^n \lg e \rfloor \leq 1 $, while $ \lim _{n \rightarrow \infty}  2^n \lg e = \infty $ and $ \varepsilon (x) $ is a continuous function differentiable on its domain minus the countable set of $ x = 2^{\lfloor \lg x \rfloor} $ and $ \lim _{2^{\lfloor \lg x \rfloor} \neq x \rightarrow \infty} \varepsilon ^{\prime} (x) = 0 $, so that
the $ \lim _{n \rightarrow \infty} (\varepsilon ( 2^n \lg e ) - \varepsilon (\lfloor 2^n \lg e \rfloor)) $ $ = $ $ 0 $. Thus $ \lim _{n \rightarrow \infty} \varepsilon (\lfloor 2^n \lg e \rfloor) $ $ = $ $ \lim _{n \rightarrow \infty} \varepsilon ( 2^n \lg e ) $ $ = $ 
$ 1 - \lg e + \lg \lg e $.} which makes $1 - \lg e + \lg \lg e$ the  \textit{supremum} of $ \varepsilon $ restricted to integers.

\section{Logarithm-based tight upper bounds on $  2 s_2(N) + e_2(N) $} \label{sec:s2e2} 

\begin{figure}[h]  
\begin{center}
\includegraphics[scale=1]{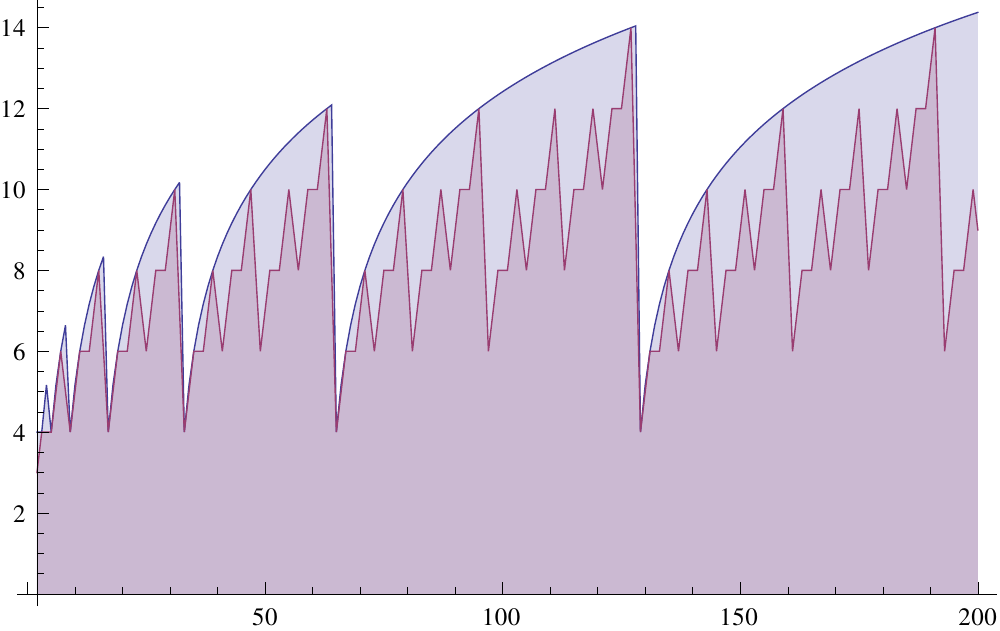} 
\end{center}
\caption{Graphs of $ 
2 s_2(N) + e_2(N) $ and its upper bound $ 2 \lg (N  - 2^{\lfloor \lg (N-1) \rfloor}+1) + 2 $. \label{fig:f(N)UB} }
\end{figure}

A jumpy function $ f(N) = 2 s_2(N) + e_2(N) $ that appears in formulas (\ref{eq:worstSort}) and (\ref{eq:worstSortepsilon}) oscillates between $ 4 $ and $ 2 \lg (N + 1) $ as $ N $ ranges between $ 3 $ and $ \infty $, assuming $ 4 $ for $ N = 2^{\lfloor \lg N \rfloor} + 1 $ and $ 2 \lg (N + 1) $ for $ N = 2^{\lceil \lg N \rceil} - 1 $. This yields:
\begin{equation} \label{eq:ineqf1} 
4 \leq 2 s_2(N) + e_2(N) \leq 2 \lg (N + 1).
\end{equation}
Using recurrence relations for $  e_2(N) $
\begin{equation} \label{eq:rece2} \nonumber
e_2(N) = \left\{ \begin{array}{ll}
e_2(\frac{N}{2}) + 1 \mbox{ if } 2 \mid N \\
0 \mbox{ otherwise}
\end{array} \right.
\end{equation}
 and $ s_2(N) $
\begin{equation} \label{eq:recs2} \nonumber
s_2(N) = \left\{ \begin{array}{ll}
1 \mbox{ if }  N = 1 \\
s_2(\lfloor \frac{N}{2} \rfloor) + N \% 2 \mbox{ otherwise}
\end{array} \right.
\end{equation}
one can derive a recurrence relation for $ f(N) $
\begin{equation} \label{eq:recf} 
f(N) = 2+ \left\{ \begin{array}{ll}
\lg N \mbox{ if }  N = 2^{\lfloor \lg N \rfloor} \\
f(N-2^{\lfloor \lg (N-1) \rfloor})  \mbox{ otherwise.}
\end{array} \right.
\end{equation}
This leads to another (tighter than (\ref{eq:ineqf1}), if
 $ 2^{\lfloor \lg N \rfloor} $ $ < $ $  N $ $ < $ $ 2^{\lceil \lg N \rceil} - 1 $, and continuous except for $ N = 2^{\lfloor \lg N \rfloor} + 1 $) upper bound on $ f(N) $ shown on Figure~\ref{fig:f(N)UB}.

\medskip

For every $ N \geq 2 $:
\begin{equation} \label{eq:fUB}
2 s_2(N) + e_2(N) \leq 2 \lg (N  - 2^{\lfloor \lg (N-1) \rfloor}+1) + 2 ,
\end{equation}
and for every $ N \neq 2^{\lceil \lg N \rceil} $:
\begin{equation} 
 2 \lg (N + 1 - 2^{\lfloor \lg (N-1) \rfloor}) + 2 \leq 2 \lg (N + 1).
\end{equation}
If $ N = 2^{\lceil \lg N \rceil} - 1 $ (the case considered in \cite{kru:heap}) then $ 2 \lg (N + 1 - 2^{\lfloor \lg (N-1) \rfloor}) + 2 $ $ = $ $ 2 \lg (N + 1) $; for all other $ N \neq 2^{\lceil \lg N \rceil} $, $ 2 \lg (N + 1 - 2^{\lfloor \lg (N-1) \rfloor}) + 2 $ $ < $ $ 2 \lg (N + 1) $.

\medskip

If $2 \leq N = 2^{\lceil \lg N \rceil} $ then $ 2 \lg (N  - 2^{\lfloor \lg (N-1) \rfloor}+1) + 2 $ $ = $ $ 2 \lg (N + 2) $ $ > $ $ 2 \lg (N + 1) $.

\begin{figure}[h]  
\begin{center}
\includegraphics[scale=1]{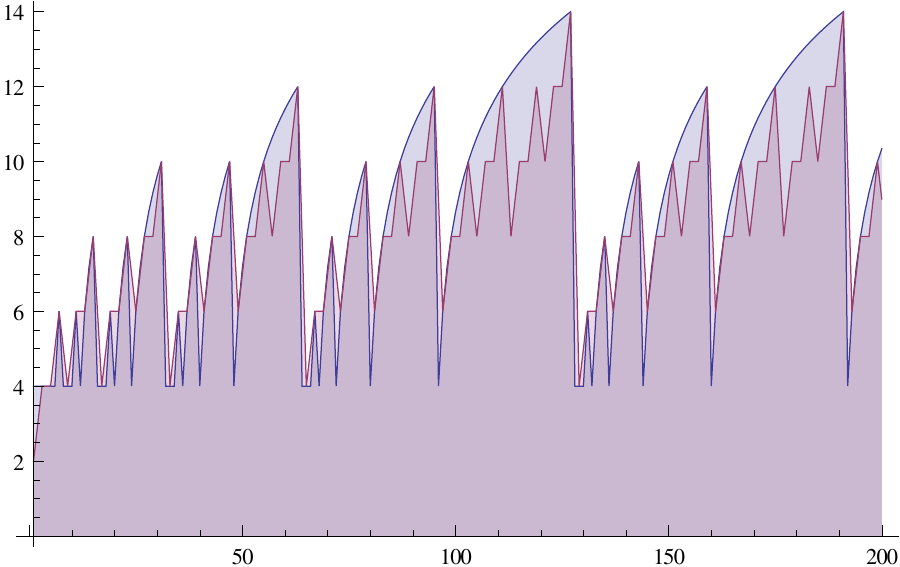} 
\end{center}
\caption{Graphs of $ 
2 s_2(N) + e_2(N) $ and its upper bound $ 2 \lg (N - 2^{\lfloor \lg (N-1) \rfloor} - 2^{\lfloor N  - 2^{\lfloor \lg (N-1) \rfloor} \rfloor }+1) + 2 $ for almost all $ N $.
\label{fig:f(N)UBmore}}
\end{figure}

\medskip
Let $ E = \{2^k \mid k \geq 3\} \cup \{2^m(2^k + 1) \mid k , m \geq 1 \} $.
For $ 2 \leq N \notin
E$,   (\ref{eq:recf}) may be further expanded to yield even tighter than (\ref{eq:fUB}) upper bound
\begin{equation} \label{eq:fUBmore} 
2 \lg (N - 2^{\lfloor \lg (N-1) \rfloor} - 2^{\lfloor N  - 2^{\lfloor \lg (N-1) \rfloor} \rfloor }+1) + 2
\end{equation} 
 on $ f(N) $,   visualized on Figure~\ref{fig:f(N)UBmore}.

\medskip

The size of the set $ E ^{<N} $ $ = $ $ E \cap \{0,...,N-1 \} $ of exceptions less than $ N $ is small relative to $ N $. It is clear that for every $ i \in E $, $ s_2(i) \leq 2 $. The number of numbers $ i < N $ with $ s_2(i) =1 $ is between 
$ \lfloor \lg N \rfloor  $ and $ \lceil \lg N \rceil $, and the number of numbers $ i < N $ with $ s_2(i) =2 $ is between 
$\frac{1}{2} (\lfloor \lg N \rfloor+1)(\lfloor \lg N \rfloor ) $ and $\frac{1}{2} (\lceil \lg N \rceil + 1)(\lceil \lg N \rceil) $, so that the number of numbers $ i < N $ with $ s_2(i) =1 $ or  $ s_2(i) =2 $ is between 
$\frac{1}{2} (\lfloor \lg N \rfloor+3)(\lfloor \lg N \rfloor ) $ and $\frac{1}{2} (\lceil \lg N \rceil + 3)(\lceil \lg N \rceil) $. Thus the size of $ E ^{<N} $ is $ \Theta (\log ^2 N) $.  In particular, the function given by (\ref{eq:fUBmore}) is an upper bound on $ f(N ) $ for almost all $ N $.\footnote{As 
$ \lim _{N \rightarrow \infty} \frac{\frac{1}{2} (\lceil \lg N \rceil + 3)(\lceil \lg N \rceil) }{N} = 0 $.
}
\begin{figure}[h]  
\begin{center}
\includegraphics[scale=1]{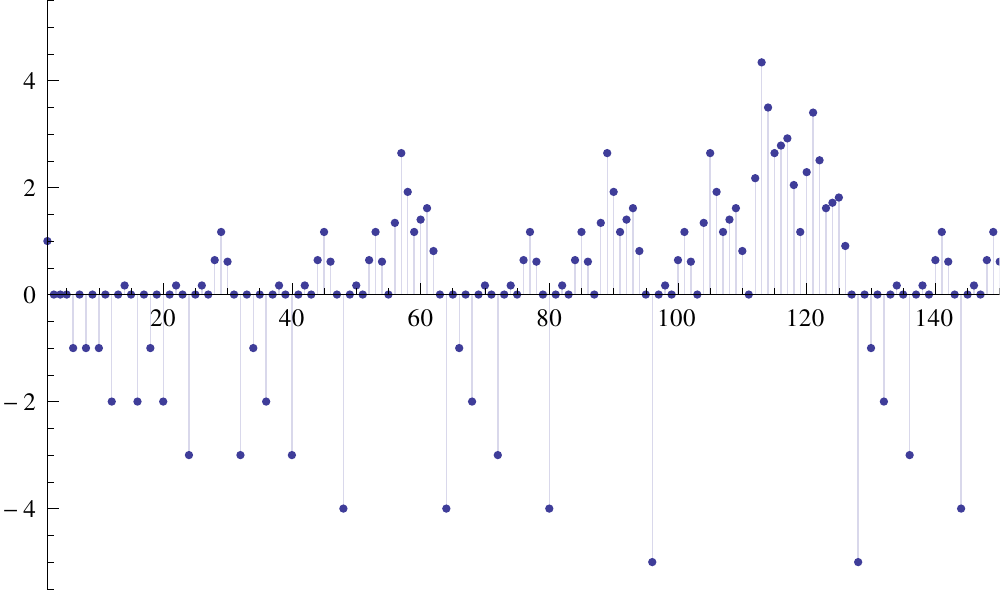} 
\end{center}
\caption{A graph of the difference between the upper bound $ 2 \lg (N - 2^{\lfloor \lg (N-1) \rfloor} - 2^{N  - 2^{\lfloor \lg (N-1) \rfloor}}+1) + 2 $ for almost all $ N $ and $ 
f(N) = 
2 s_2(N) + e_2(N) $. Points below the $ N $-axis indicate elements of the set $ E = \{2^k \mid k \geq 3\} \cup \{2^m(2^k + 1) \mid k , m \geq 1 \}$ for which the former is not an upper bound of the latter.
\label{fig:f(N)UBdif}}
\end{figure}

\medskip

A graph of the difference between the upper bound (\ref{eq:fUBmore}) and $ f(N) $ is visualized on Figure~\ref{fig:f(N)UBdif}.

\section{A note about the roots of this work} \label{Roots}

Ever since I learned $ {\tt Heapsort} $, I have always regretted that I had not been given a chance to invent it. I thought that a person commemorated with the epitaph
\begin{quote}
	\textit{``Here rests He who invented heaps''}
\end{quote}
should consider himself lucky. But, at last, I got my chance to precisely characterize the worst-case behavior of it, and I couldn't let it pass. And I didn't.

\medskip

I begun looking for a textbook on Analysis of Algorithms with an exact closed-form formula for the worst-case number of comparisons done by $ {\tt Heapsort} $ sometime in late 2000s, only to discover that no one seemed to know it. For instance, Cormen \textit{et al.}
\cite{clr:alg} had some close estimate of that number but not the exact formula.
I was able to derive such a formula for $ {\tt MakeHeap} $ in the Fall 2010. The paper with my derivation took a long path to print and appeared in the Summer 2012, more than a year after the same result (albeit with a totally different derivation) was published by Paparrizos~\cite{papa:tight}. I spent parts of the Summers 2012 and 2013 on pinpointing the formula for the entire $ {\tt Heapsort} $\footnote{It took me part of Spring and the Summer 2014 to clean up its derivation and simplify the optimality proof.}, and ``discovered'' a readable copy of Kruskal and Weixelbaum's old report with somewhat sketchy proof of a worst-case formula for a special case of $ N = 2^{\lceil \lg N \rceil}-1 $ in mid-July 2013 while waiting at a service station for a repair of my car and killing time by surfing the Internet, after I~had finished all the details of my early proofs of the said formula.

\medskip

 I~was amazed by striking similarities between their work and mine. Although we all learned from Knuth's writings or from writings of those who learned from Knuth's, so any similarities here are not totally coincidental 
even though my knowledge of The Art of Programming is rather spotty and often 
(like in the case the idea of running $ {\tt Heapsort} $ backwards) \textit{ex post facto}, 
the degree of the said similarities made me wonder if anyone of us who takes on certain kind of problems is destined to end up, eventually, on a similar path leading to similar results. It does feel, indeed, as if the proof of the formula for the worst-case number of comparisons done by $ {\tt Heapsort} $ was out there, somehow independently of our intellectual inquiries, like the gravity and the Sun storms, just waiting for somebody to discover it. 

\subsection{Comparison with Kruskal-Weixelbaum formulas} \label{CompK-W}

\medskip

\begin{figure}[h] 
\begin{center}
\includegraphics[scale=.6]{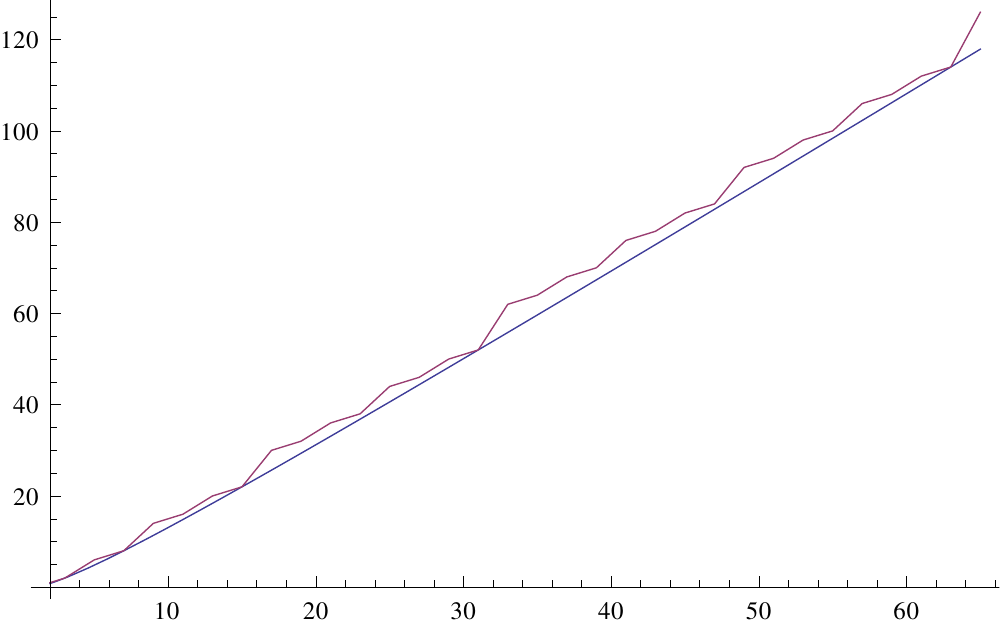} 
\includegraphics[scale=.6]{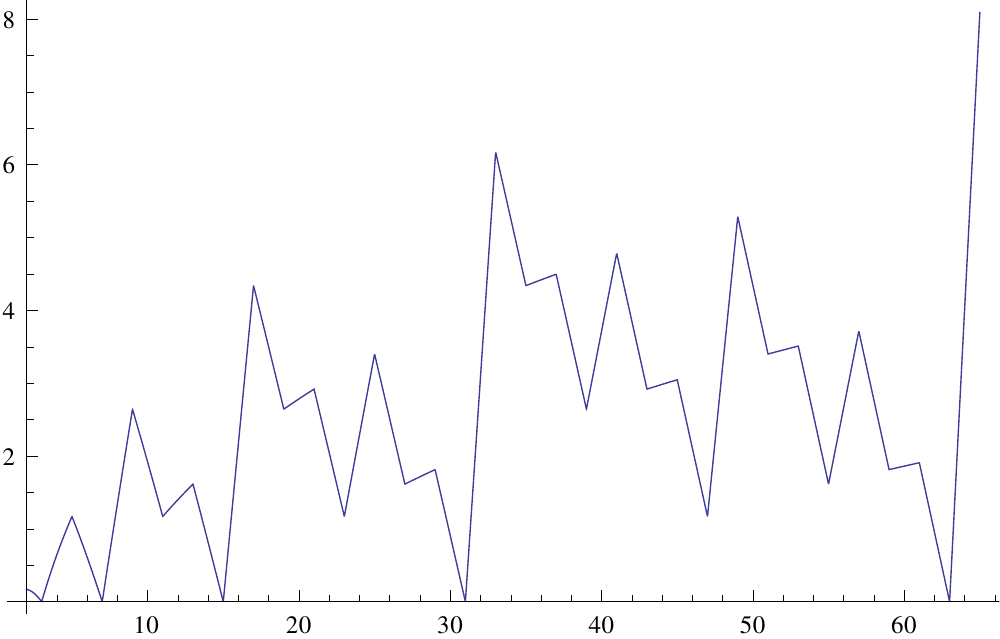} 
\end{center}
\caption{(Left:) Comparison of the actual worst-case for $ {\tt MakeHeap} $ (top line) with Kruskal-\-Weixelbaum's formula (bottom line). (Right:) The difference between the actual worst-case for $ {\tt MakeHeap} $ and Kruskal-Weixelbaum's lower bound. \label{fig:KWmakeheap}}
\end{figure}

\begin{figure}[h] 
\begin{center}
\includegraphics[scale=.6]{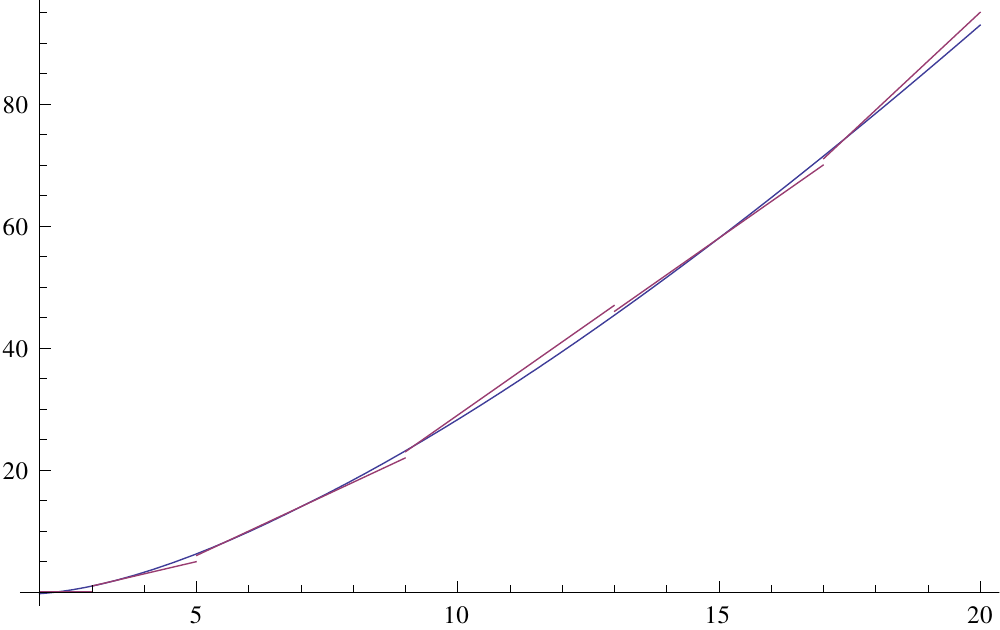} 
\includegraphics[scale=.6]{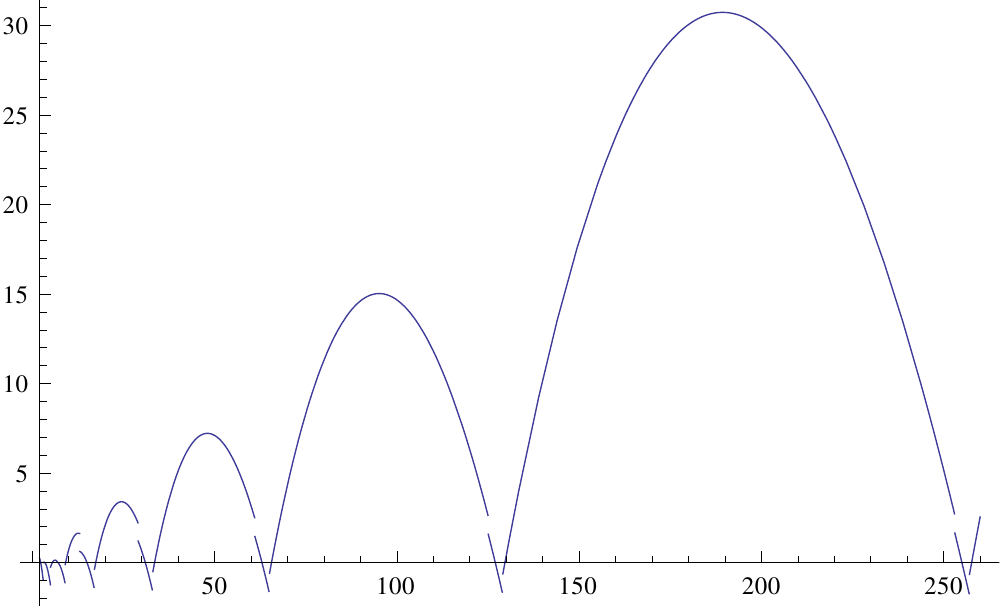} 
\end{center}
\caption{(Left:) Comparison of the actual worst-case for $ {\tt RemoveAll} $ (the crooked line) with Kruskal-Weixelbaum's formula (the smooth line). The latter overestimates the fromer for $ N=4,5,8,9,16 $ and $ 17 $. (Right:) The difference between the actual worst-case for $ {\tt RemoveAll} $ and Kruskal-Weixelbaum's lower bound. It is negative for for $ N=$ 4, 5, 8, 9, 16, 17, 32, 33, 64, 65, 128, 129, 256 and  257. \label{fig:KWremoveall}}
\end{figure}

Below are comparisons of my results with formulas published in \cite{kru:heap} that gave the actual worst-case numbers of comparisons for the special case of compete heaps (of size $ N = 2^{\lceil \lg N \rceil} -1 $, that is). As one can see, Kruskal-Weixelbaum's formulas, if extended over all $ N \geq 2 $,\footnote{Which was not the intention of their authors.} happen to give a lower bound for the number of comparisons done in the worst case by ${\tt MakeHeap}$ also for other cases of $ N $ (Figure~\ref{fig:KWmakeheap}, left), but not by ${\tt RemoveAll}$ (Figure~\ref{fig:KWremoveall}, left).

\medskip


The difference between the actual worst case for $ {\tt RemoveAll} $  and the Kru\-skal-\-Weixel\-baum's lower bound diverges and its limit superior is $ + \infty $, as Figure \ref{fig:KWremoveall} (right) illustrates it.

\medskip

The Kruskal-Weixelbaum's worst-case formula for $ {\tt RemoveAll} $ coincides with (\ref{eq:worstRemAll}) on page \pageref{eq:worstRemAll} (the exact value) for $ N = 2^{\lceil \lg N \rceil} -1 $ and
$ N \approx 2^{\lfloor \lg N \rfloor} + \xi (N) $  
for some $ \xi (N) \in ( 1.4, 1.8 ) $, where $ N \geq 7 $.

\medskip

\begin{figure}[h] 
\begin{center}
\includegraphics[scale=.6]{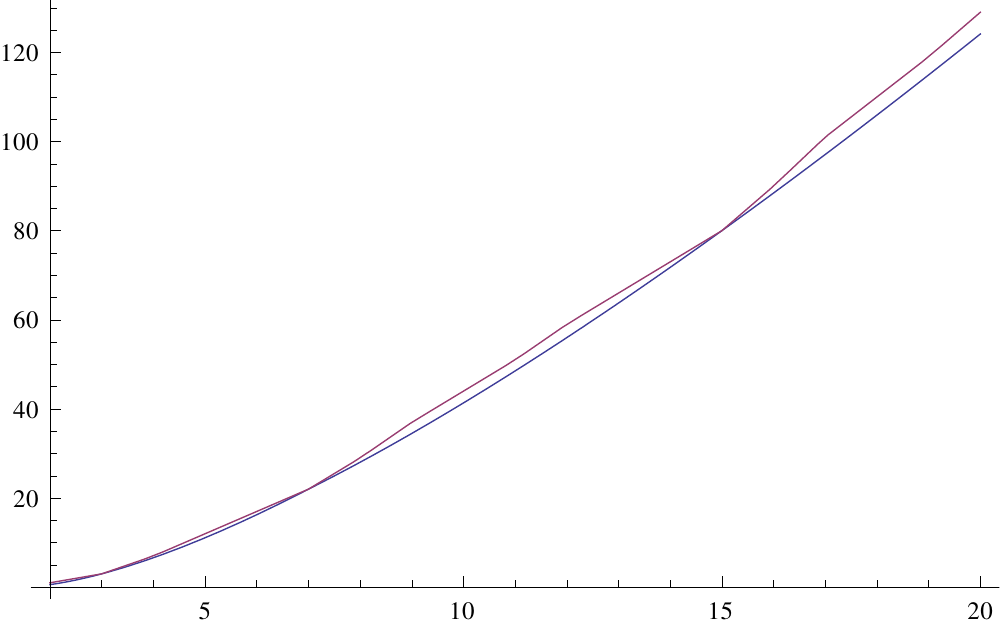} 
\includegraphics[scale=.6]{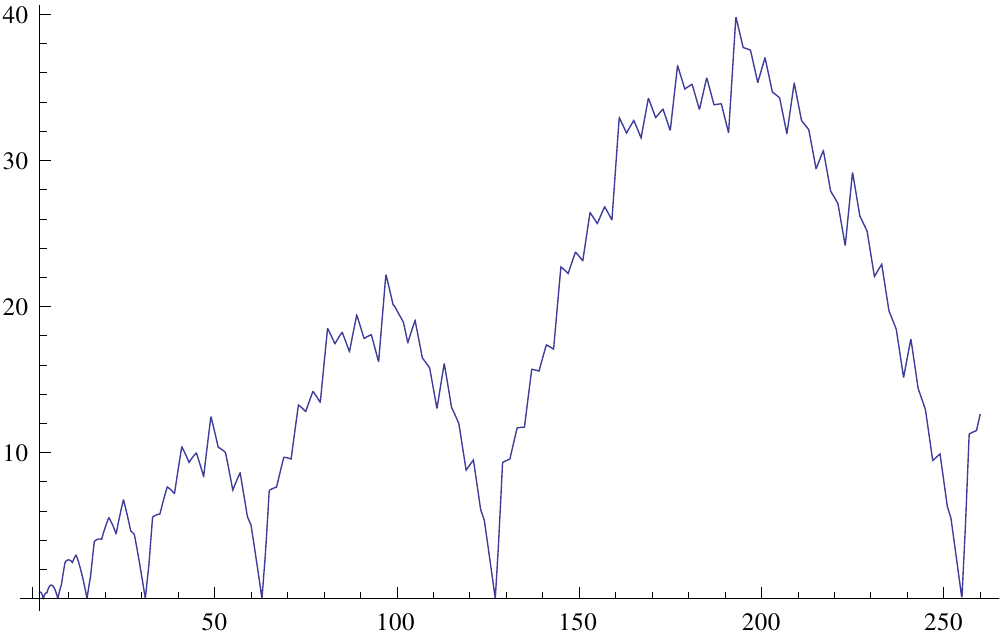}  
\end{center}
\caption{(Left:) Comparison of the actual worst-case for $ {\tt Heapsort} $ (top line) with Kruskal-Weixelbaum's formula (bottom line). (Right:) The difference between the actual worst-case for $ {\tt Heapsort} $ and  Kruskal-Weixelbaum's lower bound. \label{fig:KWheapsort}}
\end{figure}

Interestingly, Kruskal-Weixelbaum's formulas also give a lower bound for the number of comparisons in the worst case for the entire ${\tt Heapsort}$ (Figure~\ref{fig:KWheapsort}, left) although they overestimate that number for  ${\tt RemoveAll}$ for some $ N \neq 2^{\lceil \lg N \rceil} -1 $, for instance, for $ N = 2^{\lceil \lg N \rceil} $, and even if $ c = 1$, for instance, for $ N = 2^{\lfloor \lg N \rfloor} +1 $. One could speculate that this was a reason why they stopped short of deriving the formulas for arbitrary $ N \geq 2 $, not just for $ N = 2^{\lceil \lg N \rceil} -1 $.

\medskip

The difference between the actual worst case for $ {\tt Heapsort} $ and the Kru\-skal-Weixel\-baum's lower bound diverges and its limit superior is $ + \infty $, as Figure \ref{fig:KWheapsort} (right) illustrates it.

\medskip

 Moreover, Kruskal-Weixelbaum proof of decomposition of the worst-cases of ${\tt Heapsort}$ (Theorem 2 and Remark 1 in \cite{kru:heap}) works only for $ N = 2^{\lceil \lg N \rceil} -1$. Unlike my program $ {\tt unFixHeap} $ on Figure~\ref{fig:unfix} page~\pageref{fig:unfix}, the worst-case input generator for $ {\tt Makeheap} $ presented in \cite{kru:heap} works for input sizes $ N= 2^{\lceil \lg N \rceil} -1 $ but not for other sizes, except, incidentally.
 For instance, it doesn't work for $ N = 12 $ as their algorithm (Figure 2 in \cite{kru:heap}) does not generate a worst-case array (e.g.,  one visualized on Figure~\ref{fig:mkworstout} page \pageref{fig:mkworstout}) for the worst-case heap visualized on Figure~\ref{fig:ExHeap} page \pageref{fig:ExHeap}. As the first step, their algorithm will ``unsift'' the root 12 at index 1 of the said heap all the way down to the index 12 of the leaf 1, which happens to be the only node of the last level without a sibling. As a result, the corresponding ``sift'' will, by the equality (\ref{eq:credLeaf2}) page \pageref{eq:credLeaf2}, perform $ 2 \lfloor \lg 12 \rfloor - 1 $ comparisons while the maximum, given by (\ref{eq:creditUBform}) and  the right-hand side of the inequality (\ref{eq:creditUB}) page~\pageref{eq:creditUB}
 is $  \lfloor \lg 12 \rfloor +   \lfloor \lg 11 \rfloor$ $ = $ $ 2 \lfloor \lg 12 \rfloor $. Hence the ``reverse heap'' created of the said heap by their algorithm is not a worst-case array for the $ {\tt MakeHeap} $.

\medskip

The $ {\tt PROCEDURE \,\, \underline{UNSIFT} ( S, BOUND)} $ of their algorithm is functionally very close to Java method $ {\tt PullDown(i,j)} $ of Fig.~\ref{fig:PullDown} on page~\pageref{fig:PullDown}, substituting ${\tt BOUND} $ for $ {\tt i} $ and $ {\tt S} $ for $ {\tt j} $. Should they use ``{\tt the leftmost leaf in the tree rooted by P}'' rather than ``{\tt node containing smallest value in the
\linebreak tree rooted by P}'', their algorithm would be functionally equivalent to my program and would correctly generate worst-case arrays for any $ N \geq 2 $ for $ {\tt MakeHeap} $.

\bigskip

So, how close were they to discovering the general formulas for any $ N\geq 2 $? 

\medskip

The formula for the worst-case number of comparisons for $ {\tt RemoveAll} $ that they derived in \cite{kru:heap} (Theorem 3) for $ N = 2^{\lceil \lg N \rceil} -1 $ is equivalent to this one:
\[2(N-1) \lg (N+1) - 4(N-1) + \min (\lg(N+1)-1,2).\]
Should they try to derive a formula for $ N = 2^{\lfloor \lg N  \rfloor} +1 $, instead\footnote{Another possibility would be playing with the strategy $\mathsf{z}(N)$ of Example~\ref{ex:opt} for $ N = 2^{\lfloor \lg N  \rfloor} - 3 $.}, they would obtain a formula equivalent to this one:
\begin{equation} \label{eq:KW500}
2(N-1) \lg (N-1) - 4(N-1) + 4 + \min (\lg(N-1),2).
\end{equation}
\begin{figure}[h] 
\begin{center}
\includegraphics[scale=1]{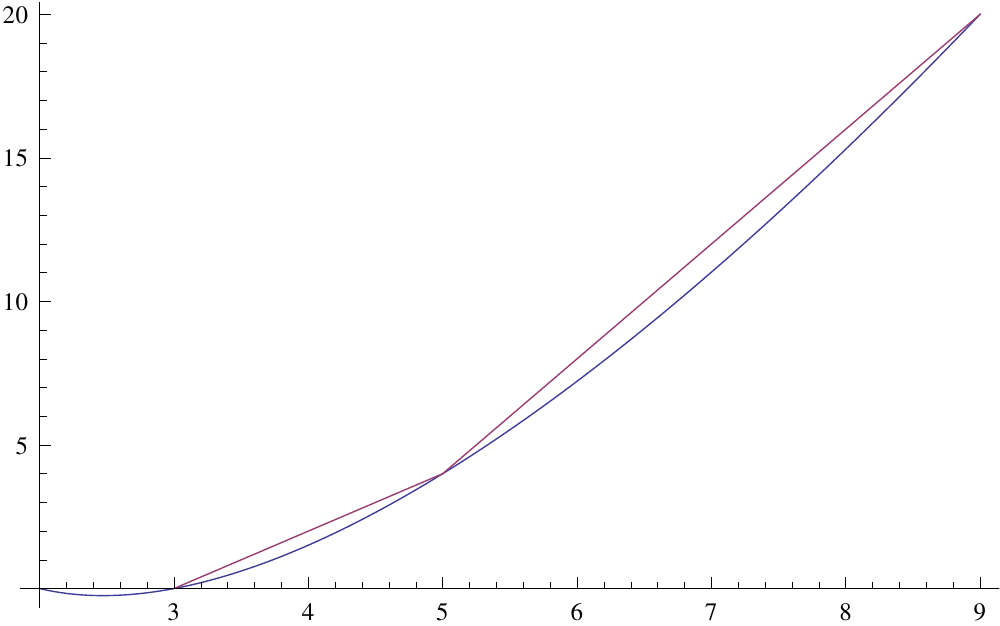} 
\end{center}
\caption{The function $ 2 (N - 1) \lfloor \lg (N - 1)  \rfloor - 
 2^{\lfloor \lg (N - 1)  \rfloor +2} +  4 $ (top line) interpolates the function $ 2(N-1) \lg (N-1) - 4(N-1) + 4$ (bottom line) between the points $ N = 2^{\lfloor \lg N  \rfloor} +1 $. \label{fig:KWinterpol}}
\end{figure}
Interpolating the first part $ 2(N-1) \lg (N-1) - 4(N-1) + 4$ of  (\ref{eq:KW500}) between the points $ N = 2^{\lfloor \lg N  \rfloor} +1 $, as it is visualized on Figure~\ref{fig:KWinterpol}, they could ``guess'' the exact value of the first part of the payoff $ P_{\mathsf{par}}(N-1) $ for the first $ N-1 $ moves of the strategy  $ \mathsf{par} $, that is,
\[2 (N - 1) \lfloor \lg (N - 1)  \rfloor - 
 2^{\lfloor \lg (N - 1)  \rfloor +2} +  4.\]
``Guessing'' that $ \lg(N-1) $ in $ \min (\lg(N-1),2) $ in (\ref{eq:KW500}) should really be $ \lfloor \lg (N - 1)  \rfloor $ would allow them to arrive at the correct formula for $ P_{\mathsf{par}}(N-1) $. After that, the only missing part of the actual formula for $C^{max}_{\tt RemAll}(N)$ is the function $ c  $ defined by the equation (\ref{eq:defc}) on page \pageref{eq:defc}, which does not seem like an obvious ``guess''. In order to compute $ c $, control of the index of node 4, which  Kruskal-Weixelbaum did not do, in construction of a general worst-case heap, as indicated by the Worst-case Heap Characterization 
 Lemma~\ref{lem:parwin4} page \pageref{lem:parwin4}, seems critical. And, of course, the proof that all the above are correct guesses would be nedded, too, and that would be, perhaps, the hardest part.

\medskip

Any ``guessing'' of the formula (\ref{eq:MkHeapElem}) on page \pageref{eq:MkHeapElem} 
\[ 2N - 2s_2(N) - e_2(N) \]
 for $ C_{{\tt MakeHeap}} ^{\tt max} (N) $ from $ 2N - 2 \lg(N+1) $ (formula derived in \cite{kru:heap}, correct for $ N = 2^{\lceil \lg N \rceil} -1 $) or from $ 2N - 4 $ (correct for $ N = 2^{\lfloor \lg N  \rfloor} +1 $) by interpolation or otherwise, seems out of the question if one takes into account the jumpy behavior of $ 2s_2(N) + e_2(N) $ as analyzed in Section~\ref{sec:s2e2} and visualized on Figure~\ref{fig:f(N)UB} page \pageref{fig:f(N)UB}.
 
\medskip

For $ 15 \leq N = 2^{\lceil \lg N \rceil} -1 $, Kruskal-Weixelbaum's 35-years old\footnote{At the time of publication of this paper.} formula gives a lesser (better, that is) value than one that I somewhat hastily\footnote{After ``exhaustively'' testing it for all permutations of up to 10 first positive integers; the test for all permutations of 13 would let me realize my mistake just in four months or so of running it on my laptop computer.} called,  in the Abstract and in the opening sentence of Section 9 of \cite{suc:elem}, the ``best-known upper bound'' on the
number of comparisons of ${\tt Heapsort}$.



\section*{APPENDIX}

\appendix

\section{Examples} \label{Examples}

\subsection{Example of construction of a worst-case 12-element heap using strategy $ \mathsf{win}(12) $}
\label{sec:ex12}

\medskip

The following is a compilation of excerpts from an output of my Java program that (among other things) visualizes construction of a worst-case heap of 12 elements of Figures~\ref{fig:ExHeap}, \ref{fig:ArrayHeap}, and \ref{fig:mkworstout} on pages~\pageref{fig:ExHeap} through \pageref{fig:mkworstout} using the strategy $ \mathsf{win}(12) $. It shows a sequence of 12 heaps (a game), each printed level-by-level, and the first 11 pull downs of $ \mathsf{win}(12) $ that are applied to the the first 11 of these heaps. All the pull downs are lossless. There is no lossless pull down after the last one in this sequence. The only two valid moves that are applicable to the last heap are pull down 1, which pulls down a leaf without a sibling, and pull down 2, which pulls down the parent of the former. They both are lossy because, by the equality (\ref{eq:credLeaf2}) on page \pageref{eq:credLeaf2}, they yield a credit
\[2 \lfloor \lg 12 \rfloor - 1  = \lfloor \lg 12 \rfloor + \lfloor \lg 11 \rfloor - 1\]
and this is less than the maximum credit
\[\lfloor \lg 12 \rfloor + \lfloor \lg 11 \rfloor\]
given by (\ref{eq:creditUBform}) and  the right-hand side of the inequality (\ref{eq:creditUB}) on page \pageref{eq:creditUB} for pulling down a node in a 12-element heap.
(After all, by virtue of Theorem \ref{thm:parwin} on page \pageref{thm:parwin}, substituting 3 for $ m $, every sequence of more than 11 consecutive pull downs the first of which is applied to the 1-element heap must be lossy.)

{ \footnotesize

\begin{verbatim}
 


Level 0: 1

H[1] -> H[2]

Level 0:  2
Level 1: 1^

H[2] -> H[3]

Level 0:  3
Level 1: 2^1

H[3] -> H[4]

Level 0:    4
Level 1:  2^^^3
Level 2: 1^

H[4] -> H[5]

Level 0:    5
Level 1:  4^^^3
Level 2: 2^1

H[4] -> H[6]

Level 0:    6
Level 1:  5^^^3
Level 2: 4^1 2^

H[5] -> H[7]

Level 0:    7
Level 1:  6^^^3
Level 2: 4^5 2^1

H[7] -> H[8]

Level 0:        8
Level 1:    6^^^^^^^7
Level 2:  4^^^5   2^^^3
Level 3: 1^

H[4] -> H[9]

Level 0:        9
Level 1:    8^^^^^^^7
Level 2:  6^^^5   2^^^3
Level 3: 1^4

H[8] -> H[10]

Level 0:               10
Level 1:        9^^^^^^^^^^^^^^^7
Level 2:    8^^^^^^^5       2^^^^^^^3
Level 3:  6^^^4   1^^

H[9] -> H[11]

Level 0:               11
Level 1:       10^^^^^^^^^^^^^^^7
Level 2:    9^^^^^^^5       2^^^^^^^3
Level 3:  6^^^8   1^^^4

H[10] -> H[12]

Level 0:               12
Level 1:       11^^^^^^^^^^^^^^^7
Level 2:    9^^^^^^10       2^^^^^^^3
Level 3:  6^^^8   5^^^4   1^^

The strategy was the first 11 pull downs of win(12): 
<1, 1, 1, 1, 2, 1, 1, 4, 1, 4, 1>

BUILD SUCCESSFUL (total time: 0 seconds)}

\end{verbatim}
}

\subsection{Example of construction of the last level a worst-case 30-element heap using strategy $ \mathsf{win}(30) $} \label{sec:win31} 

\medskip

The following is an annotated compilation of excerpts from an output of my Java program that visualizes construction of the last level of a worst-case heap of 30 elements.  It shows a sequence of 16 heaps (a subgame), each printed level-by-level, and the first 15 pull downs of $ \mathsf{win}(30) $ that are applied to the the first 15 of these heaps. All the pull downs, except for 14th pull down that incurs a loss of 1 credit, are lossless. 

{\footnotesize
\begin{verbatim}
Win(30) - annotated excerpt from the output of my Java program

Level 0:               15
Level 1:       12^^^^^^^^^^^^^^14
Level 2:    9^^^^^^11      13^^^^^^^3
Level 3:  4^^^6  10^^^5   7^^^8   2^^^1

Par ended. Greedy begins here.

Level 0:                               16
Level 1:               12^^^^^^^^^^^^^^^^^^^^^^^^^^^^^^15
Level 2:        9^^^^^^^^^^^^^^11              13^^^^^^^^^^^^^^14
Level 3:    4^^^^^^^6      10^^^^^^^5       7^^^^^^^8       2^^^^^^^3
Level 4:  1^^

Level 0:                               17
Level 1:               16^^^^^^^^^^^^^^^^^^^^^^^^^^^^^^15
Level 2:       12^^^^^^^^^^^^^^11              13^^^^^^^^^^^^^^14
Level 3:    9^^^^^^^6      10^^^^^^^5       7^^^^^^^8       2^^^^^^^3
Level 4:  1^^^4

Level 0:                               18
Level 1:               17^^^^^^^^^^^^^^^^^^^^^^^^^^^^^^15
Level 2:       16^^^^^^^^^^^^^^11              13^^^^^^^^^^^^^^14
Level 3:   12^^^^^^^6      10^^^^^^^5       7^^^^^^^8       2^^^^^^^3
Level 4:  9^^^4   1^^

Level 0:                               19
Level 1:               18^^^^^^^^^^^^^^^^^^^^^^^^^^^^^^15
Level 2:       17^^^^^^^^^^^^^^11              13^^^^^^^^^^^^^^14
Level 3:   16^^^^^^^6      10^^^^^^^5       7^^^^^^^8       2^^^^^^^3
Level 4:  9^^12   1^^^4

Level 0:                               20
Level 1:               19^^^^^^^^^^^^^^^^^^^^^^^^^^^^^^15
Level 2:       18^^^^^^^^^^^^^^11              13^^^^^^^^^^^^^^14
Level 3:   16^^^^^^17      10^^^^^^^5       7^^^^^^^8       2^^^^^^^3
Level 4:  9^^12   6^^^4   1^^

Level 0:                               21
Level 1:               20^^^^^^^^^^^^^^^^^^^^^^^^^^^^^^15
Level 2:       19^^^^^^^^^^^^^^11              13^^^^^^^^^^^^^^14
Level 3:   16^^^^^^18      10^^^^^^^5       7^^^^^^^8       2^^^^^^^3
Level 4:  9^^12   6^^17   1^^^4

Level 0:                               22
Level 1:               21^^^^^^^^^^^^^^^^^^^^^^^^^^^^^^15
Level 2:       19^^^^^^^^^^^^^^20              13^^^^^^^^^^^^^^14
Level 3:   16^^^^^^18      11^^^^^^^5       7^^^^^^^8       2^^^^^^^3
Level 4:  9^^12   6^^17  10^^^4   1^^

Level 0:                               23
Level 1:               22^^^^^^^^^^^^^^^^^^^^^^^^^^^^^^15
Level 2:       19^^^^^^^^^^^^^^21              13^^^^^^^^^^^^^^14
Level 3:   16^^^^^^18      20^^^^^^^5       7^^^^^^^8       2^^^^^^^3
Level 4:  9^^12   6^^17  10^^11   1^^^4

Level 0:                               24
Level 1:               23^^^^^^^^^^^^^^^^^^^^^^^^^^^^^^15
Level 2:       19^^^^^^^^^^^^^^22              13^^^^^^^^^^^^^^14
Level 3:   16^^^^^^18      20^^^^^^21       7^^^^^^^8       2^^^^^^^3
Level 4:  9^^12   6^^17  10^^11   5^^^4   1^^

Level 0:                               25
Level 1:               24^^^^^^^^^^^^^^^^^^^^^^^^^^^^^^15
Level 2:       19^^^^^^^^^^^^^^23              13^^^^^^^^^^^^^^14
Level 3:   16^^^^^^18      20^^^^^^22       7^^^^^^^8       2^^^^^^^3
Level 4:  9^^12   6^^17  10^^11   5^^21   1^^^4

Level 0:                               26
Level 1:               24^^^^^^^^^^^^^^^^^^^^^^^^^^^^^^25
Level 2:       19^^^^^^^^^^^^^^23              15^^^^^^^^^^^^^^14
Level 3:   16^^^^^^18      20^^^^^^22      13^^^^^^^8       2^^^^^^^3
Level 4:  9^^12   6^^17  10^^11   5^^21   7^^^4   1^^

Level 0:                               27
Level 1:               24^^^^^^^^^^^^^^^^^^^^^^^^^^^^^^26
Level 2:       19^^^^^^^^^^^^^^23              25^^^^^^^^^^^^^^14
Level 3:   16^^^^^^18      20^^^^^^22      15^^^^^^^8       2^^^^^^^3
Level 4:  9^^12   6^^17  10^^11   5^^21   7^^13   1^^^4

Level 0:                               28
Level 1:               24^^^^^^^^^^^^^^^^^^^^^^^^^^^^^^27
Level 2:       19^^^^^^^^^^^^^^23              26^^^^^^^^^^^^^^14
Level 3:   16^^^^^^18      20^^^^^^22      15^^^^^^25       2^^^^^^^3
Level 4:  9^^12   6^^17  10^^11   5^^21   7^^13   8^^^4   1^^

No lossless pull down is possible at this point

Level 0:                               29
Level 1:               24^^^^^^^^^^^^^^^^^^^^^^^^^^^^^^28
Level 2:       19^^^^^^^^^^^^^^23              26^^^^^^^^^^^^^^27
Level 3:   16^^^^^^18      20^^^^^^22      15^^^^^^25      14^^^^^^^3
Level 4:  9^^12   6^^17  10^^11   5^^21   7^^13   8^^^4   2^^^1

Level 0:                               30
Level 1:               24^^^^^^^^^^^^^^^^^^^^^^^^^^^^^^29
Level 2:       19^^^^^^^^^^^^^^23              26^^^^^^^^^^^^^^28
Level 3:   16^^^^^^18      20^^^^^^22      15^^^^^^25      27^^^^^^^3
Level 4:  9^^12   6^^17  10^^11   5^^21   7^^13   8^^^4  14^^^1   2^^


\end{verbatim}
}

\subsection{Example of a 500-element worst-case input for $ {\tt Heapsort} $ and heap for $ {\tt RemoveAll} $}
\label{sec:appExInput}

\medskip

Below is a 500-element worst-case input array $ A $ for the $ {\tt Heapsort} $ generated by my Java program $ unMakeHeap $ run on the worst-case heap $ H _{500} $ for $ {\tt RemoveAll} $. $ H_{500} $ had been created by my Java implementation of $ \mathsf{win}(N) $, and is shown below, after $ A $.  \medskip \\
65, 133, 7, 192, 10, 128, 14, 260, 17, 160, 5, 223, 8, 224, 29, 269, 24, 144, 6, 191, 11, 176, 21, 388, 13, 208, 25, 254, 15, 240, 60, 286, 39, 136, 9, 287, 18, 152, 36, 324, 20, 168, 40, 351, 22, 184, 44, 397, 16, 200, 32, 415, 26, 216, 52, 452, 28, 232, 56, 479, 30, 248, 123, 319, 70, 132, 12, 271, 33, 140, 67, 292, 35, 148, 71, 303, 37, 156, 75, 333, 23, 164, 47, 335, 41, 172, 83, 356, 43, 180, 87, 367, 45, 188, 91, 414, 27, 196, 55, 399, 49, 204, 99, 420, 51, 212, 103, 431, 53, 220, 107, 461, 31, 228, 63, 463, 57, 236, 115, 484, 59, 244, 119, 494, 1, 249, 2, 384, 34, 256, 19, 263, 48, 264, 130, 276, 66, 272, 134, 279, 68, 280, 138, 301, 38, 288, 78, 295, 72, 296, 146, 308, 74, 304, 150, 311, 76, 312, 154, 350, 42, 320, 86, 327, 80, 328, 162, 340, 82, 336, 166, 343, 84, 344, 170, 365, 46, 352, 94, 359, 88, 360, 178, 372, 90, 368, 182, 375, 92, 376, 186, 447, 50, 255, 102, 391, 64, 392, 194, 404, 98, 400, 198, 407, 100, 408, 202, 429, 54, 416, 110, 423, 104, 424, 210, 436, 106, 432, 214, 439, 108, 440, 218, 478, 58, 448, 118, 455, 112, 456, 226, 468, 114, 464, 230, 471, 116, 472, 234, 493, 62, 480, 126, 487, 120, 488, 242, 495, 122, 61, 246, 251, 124, 250, 3, 500, 96, 159, 142, 259, 79, 258, 257, 268, 129, 262, 261, 267, 131, 266, 265, 285, 69, 270, 141, 275, 135, 274, 273, 284, 137, 278, 277, 283, 139, 282, 281, 318, 73, 158, 149, 291, 143, 290, 289, 300, 145, 294, 293, 299, 147, 298, 297, 317, 77, 302, 157, 307, 151, 306, 305, 316, 153, 310, 309, 315, 155, 314, 313, 383, 81, 174, 165, 323, 95, 322, 321, 332, 161, 326, 325, 331, 163, 330, 329, 349, 85, 334, 173, 339, 167, 338, 337, 348, 169, 342, 341, 347, 171, 346, 345, 382, 89, 190, 181, 355, 175, 354, 353, 364, 177, 358, 357, 363, 179, 362, 361, 381, 93, 366, 189, 371, 183, 370, 369, 380, 185, 374, 373, 379, 187, 378, 377, 499, 97, 206, 197, 387, 111, 386, 385, 396, 193, 390, 389, 395, 195, 394, 393, 413, 101, 398, 205, 403, 199, 402, 401, 412, 201, 406, 405, 411, 203, 410, 409, 446, 105, 222, 213, 419, 207, 418, 417, 428, 209, 422, 421, 427, 211, 426, 425, 445, 109, 430, 221, 435, 215, 434, 433, 444, 217, 438, 437, 443, 219, 442, 441, 498, 113, 238, 229, 451, 127, 450, 449, 460, 225, 454, 453, 459, 227, 458, 457, 477, 117, 462, 237, 467, 231, 466, 465, 476, 233, 470, 469, 475, 235, 474, 473, 497, 121, 253, 245, 483, 239, 482, 481, 492, 241, 486, 485, 491, 243, 490, 489, 496, 125, 252, 4, 247.
\medskip
\\
Below is a 500-nodes worst-case heap $ H_{500} $, shown in the left-to-right level-by-level order, for $ {\tt RemoveAll } $. It was created by the first 499 moves of (my Java implementation) of $ \mathsf{win}(500) $. $ {H_{500}} $ is equal to the heap constructed by $ {\tt MakeHeap } $ run on the input array $ A $ visualized above. \medskip \\
500, 384, 499, 319, 383, 447, 498, 286, 318, 350, 382, 414, 446, 478, 497, 269, 285, 301, 317, 333, 349, 365, 381, 397, 413, 429, 445, 461, 477, 493, 496, 260, 268, 276, 284, 292, 300, 308, 316, 324, 332, 340, 348, 356, 364, 372, 380, 388, 396, 404, 412, 420, 428, 436, 444, 452, 460, 468, 476, 484, 492, 495, 251, 192, 259, 263, 267, 271, 275, 279, 283, 287, 291, 295, 299, 303, 307, 311, 315, 191, 323, 327, 331, 335, 339, 343, 347, 351, 355, 359, 363, 367, 371, 375, 379, 223, 387, 391, 395, 399, 403, 407, 411, 415, 419, 423, 427, 431, 435, 439, 443, 254, 451, 455, 459, 463, 467, 471, 475, 479, 483, 487, 491, 494, 247, 249, 250, 133, 159, 256, 258, 132, 262, 264, 266, 136, 270, 272, 274, 140, 278, 280, 282, 144, 158, 288, 290, 148, 294, 296, 298, 152, 302, 304, 306, 156, 310, 312, 314, 160, 174, 320, 322, 164, 326, 328, 330, 168, 334, 336, 338, 172, 342, 344, 346, 176, 190, 352, 354, 180, 358, 360, 362, 184, 366, 368, 370, 188, 374, 376, 378, 128, 206, 255, 386, 196, 390, 392, 394, 200, 398, 400, 402, 204, 406, 408, 410, 208, 222, 416, 418, 212, 422, 424, 426, 216, 430, 432, 434, 220, 438, 440, 442, 224, 238, 448, 450, 228, 454, 456, 458, 232, 462, 464, 466, 236, 470, 472, 474, 240, 253, 480, 482, 244, 486, 488, 490, 248, 252, 61, 246, 123, 124, 2, 3, 65, 96, 34, 142, 70, 79, 19, 257, 39, 129, 48, 261, 12, 131, 130, 265, 24, 69, 66, 141, 33, 135, 134, 273, 9, 137, 68, 277, 67, 139, 138, 281, 17, 73, 38, 149, 35, 143, 78, 289, 18, 145, 72, 293, 71, 147, 146, 297, 6, 77, 74, 157, 37, 151, 150, 305, 36, 153, 76, 309, 75, 155, 154, 313, 10, 81, 42, 165, 23, 95, 86, 321, 20, 161, 80, 325, 47, 163, 162, 329, 11, 85, 82, 173, 41, 167, 166, 337, 40, 169, 84, 341, 83, 171, 170, 345, 5, 89, 46, 181, 43, 175, 94, 353, 22, 177, 88, 357, 87, 179, 178, 361, 21, 93, 90, 189, 45, 183, 182, 369, 44, 185, 92, 373, 91, 187, 186, 377, 7, 97, 50, 197, 27, 111, 102, 385, 16, 193, 64, 389, 55, 195, 194, 393, 13, 101, 98, 205, 49, 199, 198, 401, 32, 201, 100, 405, 99, 203, 202, 409, 8, 105, 54, 213, 51, 207, 110, 417, 26, 209, 104, 421, 103, 211, 210, 425, 25, 109, 106, 221, 53, 215, 214, 433, 52, 217, 108, 437, 107, 219, 218, 441, 14, 113, 58, 229, 31, 127, 118, 449, 28, 225, 112, 453, 63, 227, 226, 457, 15, 117, 114, 237, 57, 231, 230, 465, 56, 233, 116, 469, 115, 235, 234, 473, 29, 121, 62, 245, 59, 239, 126, 481, 30, 241, 120, 485, 119, 243, 242, 489, 60, 125, 122, 4, 1.
\medskip
\\
The 499 pull downs that produced the above heap were: \\
$ \langle$~1, 1, 1, 1, 2, 1, 1, 2, 1, 2, 2, 1, 2, 1, 1, 2, 1, 2, 2, 1, 2, 1, 2, 1, 2, 1, 2, 1, 2, 1, 1, 2, 1, 2, 2, 1, 2, 1, 2, 1, 2, 1, 2, 1, 2, 1, 2, 1, 2, 1, 2, 1, 2, 1, 2, 1, 2, 1, 2, 1, 2, 1, 1, 2, 1, 2, 2, 1, 2, 1, 2, 1, 2, 1, 2, 1, 2, 1, 2, 1, 2, 1, 2, 1, 2, 1, 2, 1, 2, 1, 2, 1, 2, 1, 2, 1, 2, 1, 2, 1, 2, 1, 2, 1, 2, 1, 2, 1, 2, 1, 2, 1, 2, 1, 2, 1, 2, 1, 2, 1, 2, 1, 2, 1, 2, 1, 1, 2, 1, 2, 2, 1, 2, 1, 2, 1, 2, 1, 2, 1, 2, 1, 2, 1, 2, 1, 2, 1, 2, 1, 2, 1, 2, 1, 2, 1, 2, 1, 2, 1, 2, 1, 2, 1, 2, 1, 2, 1, 2, 1, 2, 1, 2, 1, 2, 1, 2, 1, 2, 1, 2, 1, 2, 1, 2, 1, 2, 1, 2, 1, 2, 1, 2, 1, 2, 1, 2, 1, 2, 1, 2, 1, 2, 1, 2, 1, 2, 1, 2, 1, 2, 1, 2, 1, 2, 1, 2, 1, 2, 1, 2, 1, 2, 1, 2, 1, 2, 1, 2, 1, 2, 1, 2, 1, 2, 1, 2, 1, 2, 1, 2, 1, 2, 1, 2, 1, 2, 1, 2, 1, 2, 1, 2, 1, 1, 4, 1, 4, 1, 4, 1, 4, 1, 4, 1, 4, 1, 4, 1, 4, 1, 4, 1, 4, 1, 4, 1, 4, 1, 4, 1, 4, 1, 4, 1, 4, 1, 4, 1, 4, 1, 4, 1, 4, 1, 4, 1, 4, 1, 4, 1, 4, 1, 4, 1, 4, 1, 4, 1, 4, 1, 4, 1, 4, 1, 4, 1, 4, 1, 4, 1, 4, 1, 4, 1, 4, 1, 4, 1, 4, 1, 4, 1, 4, 1, 4, 1, 4, 1, 4, 1, 4, 1, 4, 1, 4, 1, 4, 1, 4, 1, 4, 1, 4, 1, 4, 1, 4, 1, 4, 1, 4, 1, 4, 1, 4, 1, 4, 1, 4, 1, 4, 1, 4, 1, 4, 1, 4, 1, 4, 1, 4, 1, 4, 1, 4, 1, 4, 1, 4, 1, 4, 1, 4, 1, 4, 1, 4, 1, 4, 1, 4, 1, 4, 1, 4, 1, 4, 1, 4, 1, 4, 1, 4, 1, 4, 1, 4, 1, 4, 1, 4, 1, 4, 1, 4, 1, 4, 1, 4, 1, 4, 1, 4, 1, 4, 1, 4, 1, 4, 1, 4, 1, 4, 1, 4, 1, 4, 1, 4, 1, 4, 1, 4, 1, 4, 1, 4, 1, 4, 1, 4, 1, 4, 1, 4, 1, 4, 1, 4, 1, 4, 1, 4, 1, 4, 1, 4, 1, 4, 1, 4, 1, 4, 1, 4, 1, 4, 1, 4, 1, 4, 1, 4, 1, 4, 1, 4, 1~$\rangle $
\medskip
\\
It took $ {\tt MakeHeap } $ 986 comparisons to construct $ H_{500} $ from the worst-case array $ A $ shown above, and $ {\tt RemoveAll } $ 6,967 comparisons to deconstruct it, for a total of 7,953 comparisons to ${\tt Heapsort}$ the input array $ A $. The total time my Java program took for creation of the input array $ A $, which included creation of heap $ H_{500} $, and sorting it, was less than 1 second  under Netbeans IDE 6.9 on a Dell Lattitude E5510 laptop computer with Intel$^{\mbox{\textregistered}}$  Core$\texttrademark$ i5 2.40GHz processor, running Ubuntu 10.10 operating system. With all the diagnostics, dumping all the subheaps fixed and constructed (the time spent on which was $ \Theta (N^2 \log N) $), and ornamental overhead, the total time was 24 seconds

\section{Hereditary worst-case heaps} \label{Hereditary}

Hereditary worst-case heaps  for $ {\tt RemoveAll()} $ are defined as worst-case heaps whose all \textit{residua} are also worst-case.
For example, every worst-case heap on 12 or less nodes is hereditary worst-case. 

\begin{figure}[h] 
\begin{center}
\includegraphics[scale=.5]{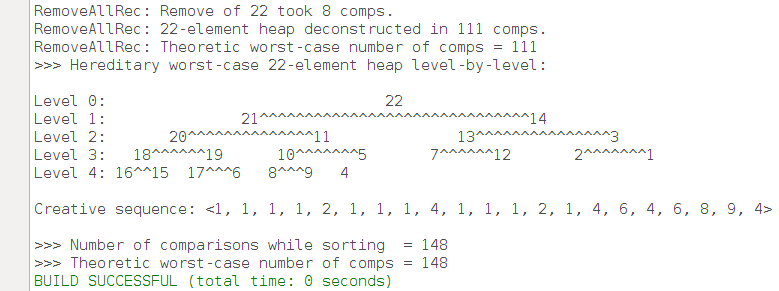} 
\end{center}
\caption{A hereditary worst-case heap of 22 nodes produced by creative sequence $ \langle 1, 1, 1, 1, 2, 1, 1, 1, 4, 1, 1, 1, 2, 1, 4, 6, 4, 6, 8, 9, 4 \rangle $  whose 12th move looses 1 credit to the upper bound
$ \lfloor \lg 12 \rfloor + \lfloor \lg 11 \rfloor $ $ = $ $ 6 $ and all other moves are lossless. \label{fig:hereditary22}}
\end{figure}

\medskip

For any hereditary worst-case heap of size $ N $, its creative sequence yields a substrategy $\mathsf{sub}$  that is optimal for any $ i $ with $ 1 \leq i < N $ (not just for $ N-1 $). The loss of credit function $ \lambda_{\mathsf{sub}}(i) $ for its moves $ i $ coincides with the delayed loss function $ \lambda^*(i) $ defined by (\ref{eq:parStarDef}) page~\pageref{eq:parStarDef} for all $ 1 \leq i < N $, and, therefore, is minimal for every move. Thus the creative sequence of any hereditary worst-case heap yields a greedy substrategy. 

\medskip

For example, one can take any 12-element worst-case heap (one created by $ {\mathsf{win}}(12) $ will do) and apply to it any greedy strategy of pull-downs. With relatively straightforward experimentations, one can find this way a 20-element worst-case heap
$ [ 20,19,15,18,10,13,14,16,17,4,5,7,12,2,3,9,11,8,6,1 ] $ created with a greedy substrategy
$ \langle 1, 1, 1, 1, 2, 1, 1, 1, 4, 1, 1, 1, 2, 1, 1, 6, 1, 6, 1 \rangle $  whose 12th move looses 1 credit to the upper bound
$ \lfloor \lg 12 \rfloor + \lfloor \lg 11 \rfloor $ $ = $ $ 6 $ and all other moves are lossless.

\medskip

\begin{figure}[h] 
\begin{center}
\includegraphics[scale=.5]{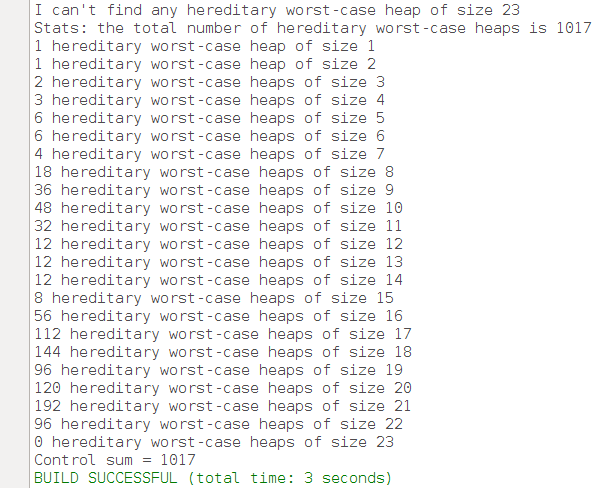} 
\end{center}
\caption{A statistics of all hereditary worst-case heaps produced by a Java program. \label{fig:all_heredit}}
\end{figure}

Unfortunately, one can only go so far playing greedy as the largest hereditary worst-case heap has only 22 nodes. For instance, a hereditary worst-case heap of 22 nodes is visualized on Figure~\ref{fig:hereditary22}. Given the function credit $ cr(i,k) $ defined by (\ref{eq:creditFH}) page~\ref{eq:creditFH} and the discussion of cases of maximal credit on the following pages, one can write a simple Java program (as I did) that generates all greedy substrategies and, by the Mapping Theorem~\ref{thm:1-1heapPatch} page~\pageref{thm:1-1heapPatch}, finds all 1017 hereditary worst-case heaps by means of pre-order traversal of a tree of their creative sequences\footnote{For instance,
the above mentioned 20-node hereditary worst-case heap appears as $\#$~698 on the list generated by my Java program.}. An excerpt of an output of such a program is shown on 
Figure~\ref{fig:all_heredit}. The complete output has been posted at: \\
{\small
\begin{verbatim}
http://csc.csudh.edu/suchenek/Papers/Hereditary_worst-case_heaps.pdf
\end{verbatim}
}

\newpage






\bibliographystyle{siam}
\bibliography{ref}
\bigskip
\bigskip
\copyright \textit{2015 Marek A. Suchenek. All rights reserved by the author. \newline A non-exclusive license to distribute this article is granted to arXiv.org}.

\end{document}